\documentclass [11pt] {article}

\usepackage[dvips]{graphicx,color}

\usepackage[hyperindex,pdfmark,backref,letterpaper]{hyperref}

\usepackage{verbatim}

\usepackage{enumerate}

\newcommand{\smallfrac}[2]{\frac{\mbox{\scriptsize $#1$}}{\mbox{\scriptsize $#2$}}}
\newcommand{\bfc}{{\bf c}}
\newcommand{\bfC}{{\bf C}}
\newcommand{\bfN}{{\bf N}}
\newcommand{\bfx}{{\bf x}}
\newcommand{\bfy}{{\bf y}}
\newcommand{\bfzero}{{\bf 0}}
\newcommand{\bfalpha}{{\mbox{\boldmath $\alpha$}}}
\newcommand{\bfbeta}{{\mbox{\boldmath $\beta$}}}
\newcommand{\bfgamma}{{\mbox{\boldmath $\gamma$}}}
\newcommand{\bfvarepsilon}{{\mbox{\boldmath $\varepsilon$}}}

\newcommand{\bflambda}{{\mbox{\boldmath $\lambda$}}}
\newcommand{\bfpi}{{\mbox{\boldmath $\pi$}}}
\newcommand{\bfrho}{{\mbox{\boldmath $\rho$}}}
\newcommand{\calL}{{\cal L}}
\newcommand{\calO}{{\cal O}}
\newcommand{\dt}{{\Delta t}}
\newcommand{\dx}{{\Delta x}}
\newcommand{\bra}[1]{\left\langle #1\right|}
\newcommand{\ket}[1]{\left| #1\right\rangle}
\newcommand{\braket}[2]{\left\langle #1\mid #2\right\rangle}

\newcommand{\Neq}[1]{{N^{\mbox{\scriptsize eq}}_{#1}}}
\newcommand{\Neqj}{{\Neq{j}}}
\newcommand{\bfNeq}{{\bfN^{\mbox{\scriptsize eq}}}}
\newcommand{\veps}{\varepsilon}
\newcommand{\bfveps}{\bfvarepsilon}
\newcommand{\alphabar}{{\bar{\alpha}}}
\newcommand{\betabar}{{\bar{\beta}}}
\newcommand{\gammabar}{{\bar{\gamma}}}
\newcommand{\vepsbar}{{\bar{\varepsilon}}}
\newcommand{\pibar}{{\bar{\pi}}}

\newtheorem{CollisionProperty}{Collision Property}

\begin{document}

\title{
  \bf Entropic Lattice Boltzmann Methods}
\author{
  Bruce M. Boghosian\\
  {\footnotesize Department of Mathematics, Tufts University,
    Bromfield-Pearson Hall, Medford, MA 02155, U.S.A.}\\
  {\footnotesize{\tt bruce.boghosian@tufts.edu}}\\[0.3cm]
  Jeffrey Yepez\\
  {\footnotesize Air Force Research Laboratory, Hanscom AFB,
    MA, 01731, U.S.A.,}\\
  {\footnotesize {\tt yepez@plh.af.mil}}\\[0.3cm]
  Peter V. Coveney\\
  {\footnotesize Centre for Computational Science,
    Queen Mary and Westfield College,}\\
  {\footnotesize University of London, Mile End Road,}\\
  {\footnotesize London E1 4NS, U.K.}\\
  {\footnotesize{\tt p.v.coveney@qmw.ac.uk}}\\[0.3cm]
  Alexander Wagner\\
  {\footnotesize Department of Materials Science, Massachusetts
    Institute of Technology, Cambridge, MA 02138, U.S.A.}\\
  {\footnotesize {\tt awagner@mit.edu}}\\[0.3cm]
  }
\date{\today}
\maketitle

\begin{abstract}
  We present a general methodology for constructing lattice Boltzmann
  models of hydrodynamics with certain desired features of statistical
  physics and kinetic theory.  We show how a methodology of linear
  programming theory, known as Fourier-Motzkin elimination, provides an
  important tool for visualizing the state space of lattice Boltzmann
  algorithms that conserve a given set of moments of the distribution
  function.  We show how such models can be endowed with a Lyapunov
  functional, analogous to Boltzmann's $H$, resulting in unconditional
  numerical stability.  Using the Chapman-Enskog analysis and numerical
  simulation, we demonstrate that such entropically stabilized lattice
  Boltzmann algorithms, while fully explicit and perfectly conservative,
  may achieve remarkably low values for transport coefficients, such as
  viscosity.  Indeed, the lowest such attainable values are limited only
  by considerations of accuracy, rather than stability.  The method thus
  holds promise for high-Reynolds number simulations of the
  Navier-Stokes equations.
\end{abstract}

\vspace{0.5truein}


\newpage
\tableofcontents
\newpage

\section{Introduction}

Though lattice models have been used to study equilibrium systems since
the 1920's, their application to hydrodynamic systems is a much more
recent phenomenon.  Lattice-gas models of hydrodynamics were first
advanced in the late 1980's, and the related lattice Boltzmann method
was developed in the early 1990's.  This paper describes a particularly
interesting and useful category of lattice Boltzmann models, which we
call {\it Entropic Lattice Boltzmann models}, and provides a number of
mathematical tools for constructing and studying them.  This
introductory section describes and contrasts lattice methods for
hydrodynamics, traces their development, and endeavors to place the
current work in historical context.

\subsection{Lattice Gases}

The first isotropic lattice-gas models of hydrodynamics were introduced
in the late 1980's~\cite{bib:fhp,bib:swolf,bib:fchc}.  Such models
consist of discrete particles moving and colliding on a lattice,
conserving mass and momentum as they do.  If the lattice has sufficient
symmetry, it can be shown that the density and hydrodynamic velocity of
the particles satisfy the Navier-Stokes equations in the appropriate
scaling limit.

This method exploits an interesting hypothesis of kinetic theory: The
Navier-Stokes equations are the dynamic renormalization group fixed
point for the hydrodynamic behavior of a system of particles whose
collisions conserve mass and momentum.  We refer to this assertion as a
hypothesis, because it is notoriously difficult to demonstrate in any
rigorous fashion.  Nevertheless, it is very compelling because it
explains why a wide range of real fluids with dramatically different
molecular properties -- such as air, water, honey and oil -- can all be
described by the Navier-Stokes equations.  A lattice gas can then be
understood as a ``minimalist'' construction of such a set of interacting
particles.  Viewed in this way, it is perhaps less surprising that they
satisfy the Navier-Stokes equations in the scaling limit.

For a time, lattice-gas models were actively investigated as an
alternative methodology for computational fluid dynamics
(CFD)~\cite{bib:lga}.  Unlike all prior CFD methodologies, they do not
begin with the Navier-Stokes equations; rather, these equations are an
{\it emergent} property of the particulate model.  Consequently, the use
of such models to simulate the Navier-Stokes equations tends to parallel
theoretical and experimental studies of natural fluids.  One derives a
Boltzmann equation for the lattice-gas particles, and applies the
Chapman-Enskog analysis to determine the form of the hydrodynamic
equations and the transport coefficients.  In fact, numerical
experimentalists often circumvent this theory by simply {\it measuring}
lattice-gas transport coefficients in a controlled setting in advance of
using them to study a particular flow problem, just as do laboratory
experimentalists.

One often overlooked advantage of lattice-gas models is their
unconditional stability.  By insisting that lattice-gas collisions obey
a detailed-balance condition~\footnote{This is certainly possible for an
  ideal single-phase viscous fluid.  For complex fluids with
  finite-range interaction potentials, it is an outstanding problem.},
we are ensured of the validity of Boltzmann's $H$ theorem, the
fluctuation-dissipation theorem, Onsager reciprocity, and a host of
other critically important properties with macroscopic consequences.  By
contrast, when the microscopic origins of the Navier-Stokes equations
are cavalierly ignored, and they are ``chopped up'' into
finite-difference schemes, these important properties can be lost.  The
discretized evolution equations need no longer possess a Lyapunov
functional, and the notion of underlying fluctuations may lose meaning
altogether.  As the first practioneers of finite-difference methods
found in the 1940's and 1950's, the result can be the development and
growth of high-wavenumber {\it numerical instabilities}, and indeed
these have plagued essentially all CFD methodologies in all of the
decades since.  Such instabilities are entirely unphysical because they
indicate the absence of a Lyapunov functional, analogous to Boltzmann's
$H$.  Numerical analyists have responded to this problem with methods
for mitigating these anomalies -- such as upwind differencing -- but
from a physical point of view it would have been much better if the
original discretization process had retained more of the underlying
kinetics, so that these problems had not occurred in the first place.
Lattice gases represent an important first step in this direction.  As
was shown shortly after their first applications to hydrodynamics,
lattice-gas models for single-phase ideal fluids can be constructed with
an $H$ theorem~\cite{bib:henon} that rigorously precludes any kind of
numerical instability.  More glibly stated, lattice gases avoid
numerical instabilities in precisely the same way that Nature herself
does so.

\subsection{Lattice Boltzmann Methods}

In spite of these appealing features, the presence of intrinsic kinetic
fluctuations makes lattice-gas models less than ideal as a CFD
methodology.  Accurate values for the velocity field at selected
locations, or even for bulk coefficients such as drag and lift, have an
intrinsic statistical error that can be reduced only by extensive
averaging.  On the other hand, the presence of such fluctuations in a
simple hydrodynamic model make lattice gases an ideal tool for those
studying the statistical physics of fluids, molecular hydrodynamics, and
complex-fluid hydrodynamics; not surprisingly, these remain the method's
principal application areas~\cite{bib:me2d3d}.  Consequently, many CFD
researchers who appreciated the emergent nature of lattice-gas
hydrodynamics but wanted to eliminate (or at least control) the level of
fluctuations, turned their attention to the direct simulation of the
Boltzmann equation for lattice gases.  Such simulations are called
lattice Boltzmann models~\cite{bib:higueraa,bib:higuerab,bib:mcn,bib:succi,bib:lbe}.
These models evolve the real-valued single-particle distribution function,
rather than the discrete particles themselves, and in this manner eliminate
the kinetic fluctuations.

Early attempts along these lines restricted attention to Boltzmann
equations for actual lattice gases.  It was soon
realized, however, that these quickly become unwieldy as the number of
possible particle velocities increases.  In order for lattice Boltzmann
models to become practical tools, it was necessary to develop simplified
collision operators that did not necessarily correspond to an underlying
lattice-gas model.  The most successful collision operators of this type
are those of the form due to Bhatnager, Gross and
Krook~\cite{bib:liboff}, and these have given rise to the so-called
lattice BGK models~\cite{bib:qianbgk}.

\subsection{Lattice BGK Models}

Lattice BGK collision operators allow the user to specify the form of
the equilibrium distribution function to which the fluid should relax.
For lattice gases obeying a detailed balance~\footnote{Actually, a
  weaker condition called {\it semi-detailed balance} suffices for this
  purpose.} condition, this is known to be a Fermi-Dirac distribution.
Having abandoned lattice-gas collision operators, however, it seemed
unnecessary to continue to use lattice-gas equilibria, and practitioners
exploited the freedom of choosing lattice BGK equilibria to achieve
certain desiderata, such as Galilean invariance, and the correct form of
the compressible Navier-Stokes equations.

The alert reader will have noticed, however, that the evolution to
lattice-BGK methods has jettisoned the last vestiges of kinetic-level
physics that might have been left over from the original lattice-gas
models.  The move to a Boltzmann description of the lattice gas
eliminated kinetic fluctuations, but at least it retained an
$H$-theorem; more precisely, its global equilibrium still extremized a
Lyapunov functional of the dynamics.  The move to lattice-BGK operators
and the arbitrary legislation of the equilibrium distribution function,
however, completely abandoned even the concept of detailed balance and
the $H$-theorem.  Without a Lyapunov functional, lattice-BGK methods
became susceptible to a wide variety of numerical instabilities which
are ill-understood and remain the principal obstacle to the wider
application of the technique to the present day.

\subsection{Motivation for Entropic Lattice Boltzmann Methods}

In this paper, we argue that the most natural way to eliminate numerical
instabilities from lattice Boltzmann models is to return to the method
its kinetic underpinnings, including the notion of a Lyapunov
functional.  We present a general program for the construction of
``entropically stabilized'' lattice Boltzmann models, and illustrate its
application to several example problems.

The crux of the idea is to encourage the model builder to specify an
appropriate Lyapunov functional for the model, rather than try to
blindly dictate an equilibrium state.  Of course, specifying a Lyapunov
functional determines the equilibrium distribution, but it also governs
the {\it approach} to this equilibrium.  It can therefore be used to
control the stability properties of the model.  It should be emphasized
that the presence of a Lyapunov functional guarantees the {\it
  nonlinear} stability of the model, which is a much stronger condition
than linear stability.

As we shall show, collision operators that are constructed in this way
are similar in form to the lattice BGK collision operators, except that
their relaxation parameter may depend on the current state.  As a
result, the transport coefficients may have a certain minimum value in
models of this type but, as we shall see, this value may actually be
zero.  In fact, the ultimate limitation to the application of these
algorithms for very small transport coefficients will come from
considerations of accuracy, rather than stability.

\subsection{Structure of Paper}

The plan of this paper is as follows: In Section~\ref{sec:elbm} we
present the general program of construction of entropic lattice
Boltzmann models.  This includes the introduction of a {\it conservation
  representation} of the lattice Boltzmann distribution function, by
means of which the collision process is most naturally described.  This
representation also provides an interesting geometric interpretation, as
it allows us to describe a collision process as a mathematical map from
a certain polytope into (and perhaps onto) itself.  We then show a way
of constructing Lyapunov functionals on this polytope, contrast these
with Boltzmann's $H$, and use them to construct collision operators for
absolutely stable models.

Section~\ref{sec:diff} carries out this program for a very simple
lattice Boltzmann model of diffusion in one dimension.  The model is
useful from a pedagogical point of view, since an entropically
stabilized collision operator can be written for it in closed form.
Moreover, it is possible to determine the transport coefficient
of this model analytically.  The result
is a fully explicit, perfectly conservative, absolutely stable
method of integrating the diffusion equation.  We present numerical
simulations for various values of the diffusivity in order to probe
the limitations of the technique.  In the course of presenting this
model, we show that a linear algebraic procedure, called the
Fourier-Motzkin algorithm, is very useful for constructing and
visualizing the conservation representation.

Section~\ref{sec:fluid} applies the method to a simple five-velocity
model of fluid dynamics in one dimension, first considered by Renda et
al. in 1997~\cite{bib:renda}.  Here the geometric picture involves a
four dimensional polytope, and the Fourier-Motzkin algorithm is shown to
be very useful in describing it.  Indeed, Appendix~\ref{app:tour} gives
the reader a ``tour'' of this polytope.

Finally, Section~\ref{sec:fhpfchc} applies the method to the
two-dimensional FHP model and the three-dimensional FCHC model of fluid
dynamics.  For these examples, we are able to provide appropriate
conservation representations.  Though the latter example has too many
degrees of freedom to allow geometric visualization of the collision
dynamics, it is nevertheless possible to develop and present a Lyapunov
functional and an entropic model.  We use these more complicated models
to demonstrate that entropic lattice Boltzmann models are not
computationally prohibitive, even when the number of velocities involved
is large.  We describe the computational aspects of such models in some
detail, and end with a discussion of Galilean invariance.

\newpage
\section{Entropic Lattice Boltzmann Models}
\label{sec:elbm}

In this section, we present the general program of construction of
entropic lattice Boltzmann models.  We note that the lattice Boltzmann
distribution function admits a variety of representations, and we show
how to transform between these.  In particular, we introduce a {\it
  conservation representation}, by means of which the collision process
has a very natural description.  We also examine the geometric aspects
of this representation, and use these insights to show how to construct
lattice Boltzmann collision operators possessing a Lyapunov functional.

\subsection{Representation}

Lattice Boltzmann models are constructed on a $D$ dimensional regular
lattice $\calL$, and evolve in discrete time intervals $\dt$.  We
imagine that a population of particles exists at each site
$\bfx\in\calL$, and that these particles can have one of a finite number
$n$ of velocities, $\bfc_j/\dt$ where $j\in\{1,\ldots,n\}$.  The
displacement vectors $\bfc_j$ are required to be combinations of lattice
vectors with integer coefficients, so that if $\bfx\in\calL$, then
$\bfx+\bfc_j\in\calL$.  We note that there is nothing precluding some of
the $\bfc_j$ from being the same, or from being zero; indeed, this
latter choice is often made to incorporate so-called ``rest particles''
into the model.

Mathematically, the state of the system of particles is represented by
$n$ real numbers at each site $\bfx\in\calL$ and time step $t$.  These
are denoted by $N_j(\bfx,t)\in\Re$, and can be thought of as the
expected number of particles with mass $m_j$ and velocity $\bfc_j/\dt$
at site $\bfx$ and time $t$.  As such, we expect them to be nonnegative,
$N_j(\bfx,t)\geq 0$.  Taken together, these quantities constitute the
single-particle distribution function of the system.  This is all the
information that is retained in a Boltzmann description of interacting
particles.

In what follows, we shall think of the set of single-particle
distribution functions as a vector space.  That is, we shall regard the
$n$ quantities $N_j(\bfx,t)$, for $i\in\{1,\ldots,n\}$, as the
components of an $n$-vector or ``ket,'' $\ket{\bfN(\bfx,t)}\in\Re^n$.
If no ambiguity is likely to result, we shall often omit the explicit
dependence on $t$ and write simply $\ket{\bfN(\bfx)}$.  If discussion is
restricted to a single lattice site, we may further abbreviate this as
$\ket{\bfN}$.

In one discrete time step $\dt$, the state of the system is modified in
a manner that is intended to model collisions between the particles at
each site, followed by propagation of the collided particles to new
sites.  The collisions are modelled by modifications to the distribution
function
\begin{equation}
  \ket{\bfN(\bfx)}\rightarrow\ket{\bfN'}.
  \label{eq:modifyN}
\end{equation}
Here and henceforth, we use a prime to denote the postcollision state.
It is usually the case that $\ket{\bfN'(\bfx)}$ depends only on
$\ket{\bfN(\bfx)}$, though in more sophisticated lattice Boltzmann
models it may also depend on the $\ket{\bfN}$'s in a local neighborhood
of sites about $\bfx$.  Following convention, we define the collision
operator $C_j(\ket{\bfN(\bfx)})$ as the difference between the old and
new values of the single-particle distribution,
\begin{equation}
  C_j(\ket{\bfN(\bfx)}) = N'_j(\bfx) - N_j(\bfx).
  \label{eq:cdef}
\end{equation}
As this is the difference of two kets, it can also be thought of as a
ket,
\begin{equation}
  \ket{\bfC} = \ket{\bfN'} - \ket{\bfN}.
  \label{eq:cket}
\end{equation}

\subsection{Conservation Laws}

The collision process specified in Eq.~(\ref{eq:modifyN}) is usually
required to conserve some number of locally defined quantities.
Usually, these quantities are additive over the lattice sites and
directions.  For example, we may require that there be a conserved mass
\begin{equation}
M = \sum_{\bfx\in\calL}\rho(\bfx,t),
\end{equation}
where we have defined the mass density
\begin{equation}
  \rho(\bfx,t)
  = \sum_{j=0}^n m_j N_j(\bfx,t)
  = \sum_{j=0}^n m_j N'_j(\bfx,t).
\end{equation}
To streamline the notation for this, we can define a covector or ``bra''
by the prescription
\begin{equation}
  \bra{\bfrho}_j = m_j,
  \label{eq:masscov}
\end{equation}
and write simply
\begin{equation}
  \rho = \braket{\bfrho}{\bfN} = \braket{\bfrho}{\bfN'}.
  \label{eq:rhodef}
\end{equation}
Thus, to the extent that we think of the single-particle distribution
function as a vector, to each conserved quantity there corresponds a
covector such that the value of the conserved quantity is the
contraction of the two.  Because collisions are required to preserve
this value, the contraction of the covector with the collision operator
must vanish,
\begin{equation}
  \braket{\bfrho}{\bfC} = 0,
\end{equation}
as can be seen from Eqs.~(\ref{eq:cket}) and (\ref{eq:rhodef}).

Lattice Boltzmann models may also conserve momentum,
\begin{equation}
  \bfpi(\bfx,t)
  = \sum_{j=0}^n m_j\frac{\bfc_j}{\dt} N_j(\bfx,t),
  = \sum_{j=0}^n m_j\frac{\bfc_j}{\dt} N'_j(\bfx,t).
\end{equation}
This is still linear in the $N_j$, so that we can define
\begin{equation}
  \bra{\bfpi}_j = m_j\frac{\bfc_j}{\dt},
\end{equation}
and write
\begin{equation}
  \bfpi(\bfx,t) = \braket{\bfpi}{\bfN} = \braket{\bfpi}{\bfN'}.
\end{equation}
Here we must be a bit careful to define the meaning of the quantities
involved: Note that $\bra{\bfpi}_j$ is a covector in its $j$ index, but
a vector in its spatial index~\footnote{The pedant will note that
  momentum is more properly thought of as a covector in its spatial
  index, but we do not bother to distinguish between {\it spatial}
  vectors and covectors in this paper.}.  Thus, the contraction with
$\ket{\bfN}_j$ is over the $j$ index, and results in the vector $\bfpi$.
Once again, this covector annihilates the collision operator,
\begin{equation}
  \braket{\bfpi}{\bfC} = \bfzero,
\end{equation}
where the right-hand side is a null vector.

The subsequent propagation process is described mathematically by the
prescription
\begin{equation}
  N_j(\bfx+\bfc_j,t+\dt) = N'_j(\bfx,t).
  \label{eq:prop}
\end{equation}
Note that this operation must be carried out {\it simultaneously} over
the entire lattice.  This may alter the values of the conserved
quantities at each site, but because it is nothing more than a
permutation of the values of the distribution function about the
lattice, it clearly leaves unaltered the global values of the conserved
quantities,
\begin{equation}
  \begin{array}{ccccc}
    \sum_{\bfx\in\calL}\rho(\bfx,t) & &
    \mbox{and} & &
    \sum_{\bfx\in\calL}\bfpi(\bfx,t).
  \end{array}
\end{equation}
Combining Eqs.~(\ref{eq:cdef}) and (\ref{eq:prop}), we find the general
dynamical equation for a lattice Boltzmann model,
\begin{equation}
  N_j(\bfx+\bfc_j,t+\dt) - N_j(\bfx) = C_j(\ket{\bfN(\bfx)}).
\end{equation}

For compressible fluids, it is also necessary to pay attention to
conservation of energy.  Conservation of kinetic energy can be expressed
using the bra,
\begin{equation}
  \bra{\bfveps}_j = \frac{m_j}{2}\left|\bfc_j\right|^2.
  \label{eq:kincov}
\end{equation}
Indeed, the right-hand side could be generalized without difficulty to
anything that depends only on $j$.  The problem of including a potential
energy function between particles at different sites, however, is much
more difficult.  If we suppose that there is a potential
$V(|\bfx_j-\bfx_k|)$ between particles at $\bfx_j,\bfx_k\in\calL$, then
two problems arise: First, the total potential energy will depend on the
pair distribution function -- that of finding two particles a certain
distance apart.  This is outside the scope of Boltzmann methods, which
retain only the single-particle distribution.  We can avoid this problem
by making the {\it mean-field} approximation, in which the probability
of having one particle at $\bfx_j$ and another at $\bfx_k$ is simply the
product of the two single-particle probabilities, but then the potential
energy is quadratic, rather than linear, in the $N_j$'s.  Second, and
perhaps more distressingly, the potential energy is not preserved by the
propagation step~\cite{bib:rkpart}.  These considerations make it very
difficult to add potential interaction to lattice Boltzmann models.  If
we are willing to give up the idea of an exactly conserved energy, and
instead consider isothermal systems, then methods of skirting these
difficulties have been known and actively investigated for some time
now~\cite{bib:shanchen,bib:yeomans}.  More recently, a method of
incorporating interaction that maintains exact (kinetic plus potential)
energy conservation has also been proposed~\cite{bib:martys}.  In any
case, this paper will be restricted to systems with kinetic energy only.
The possibility of extending our methods to models with nontrivial
interaction potentials is discussed briefly in the Conclusions section.

\subsection{Geometric Viewpoint}

The requirements of maintaining the conserved quantities and the
nonnegativity of the distribution function place very important
constraints on the collision process.  Much of this paper will be
devoted to describing -- algebraically and geometrically -- the most
general set of collisional alterations of the $N_j(\bfx)$ that meet
these requirements.

Suppose that there are $n_c < n$ independent conserved quantities.  As
described above, these define $n_c$ linearly independent covectors or
bras, whose contraction with the collision operator vanishes.  We shall
sometimes adopt special names for these; for example, that corresponding
to mass conservation shall be called $\bra{\bfrho}$, that for momentum
conservation shall be called $\bra{\bfpi}$, etc.  Generically, however,
let us refer to them as $\bra{\bflambda_\sigma}$, where
$\sigma=1,\ldots,n_c$.  These covectors are not necessarily orthogonal
in any sense.  Note, for example, that the covectors for mass and
kinetic energy in Eqs.~(\ref{eq:masscov}) and (\ref{eq:kincov}) are not
orthogonal with respect to the Euclidean metric.  In fact, there is no
reason to insist on any kind of metric in this covector space.

The $n_c$ covectors are not uniquely defined, since a linear combination
of two conserved quantities is also conserved.  The $n_c$ dimensional
subspace defined by the covectors, on the other hand, is uniquely
defined, and we shall call it the {\it hydrodynamic
  subspace}~\footnote{Note that we are using the term ``hydrodynamic''
  here to describe degrees of freedom that will result in macroscopic
  equations of motion, whether or not they are those of a fluid.  If
  mass is the only conserved quantity, a diffusion equation generically
  results; if momentum and/or energy are conserved as well, a set of
  fluid equations generically results.  We use the term ``hydrodynamic''
  in either case.  This terminology is standard in kinetic theory and
  statistical physics, but often seems strange to hydrodynamicists.}.
It is always possible to construct $n-n_c$ more covectors that are
linearly independent of each other, and of the $n_c$ corresponding to
the conserved quantities.  For example, this may be done by the
Gram-Schmidt procedure.  Let us generically call these
$\bra{\bfalpha_\eta}$, where $\eta=1,\ldots,n-n_c$; if there are only a
few of these and we want to avoid excessive subscripting, we shall use
the successive Greek letters $\bra{\bfalpha}, \bra{\bfbeta},
\bra{\bfgamma}, \ldots$ for this purpose.  Once again, these are not
uniquely defined or orthogonal in any sense, but they do span an $n-n_c$
dimensional subspace which we shall call the {\it kinetic subspace}.
The union of all $n$ of our covectors, namely the
$\bra{\bflambda_\sigma}$'s and the $\bra{\bfalpha_\eta}$'s, thus
constitute a basis for the full $n$ dimensional covector space.

The ket $\ket{\bfN}$ whose components are the single-particle
distribution function is thus seen to be but one representation of the
dynamical variable.  As has been recognized for some time
now~\cite{bib:dhum}, we are free to make a change of basis in the
lattice Boltzmann equation.  In fact, one other basis suggests itself
naturally.  Given a basis of covectors or bras, it is always possible to
construct a dual basis of vectors or kets.  Indeed, our decomposition of
the covector space into hydrodynamic and kinetic subspaces is naturally
mirrored by a corresponding decomposition of the vector space.  Thus we
construct $\ket{\bflambda_\sigma}$ where $\alpha=1,\ldots,n_c$ and
$\ket{\bfalpha_\eta}$ where $\beta=1,\ldots,n-n_c$, such that
\begin{equation}
  \begin{array}{ll}
    \braket{\bflambda_\sigma}{\bflambda_\tau} = \delta_{\sigma\tau} &
    \braket{\bflambda_\sigma}{\bfalpha_\eta} = 0 \\
    \braket{\bfalpha_\eta}{\bflambda_\sigma} = 0 &
    \braket{\bfalpha_\eta}{\bfalpha_\theta} = \delta_{\eta\theta}.
  \end{array}
  \label{eq:ortho}
\end{equation}
Thus, we can expend $\ket{\bfN}$ in terms of these {\it hydrodynamic and
  kinetic basis kets}.
\begin{equation}
  \ket{\bfN} =
  \sum_{\sigma=1}^{n_c}\lambda_\sigma\ket{\bflambda_\sigma} +
  \sum_{\eta=1}^{n-n_c}\alpha_\eta\ket{\bfalpha_\eta}.
  \label{eq:expansion}
\end{equation}
We shall call the set of coefficients $\lambda_\sigma$ and $\alpha_\eta$
the {\it conservation representation}.  The explicit construction of
this representation is best illustrated by example, and we give several
of these in the following sections.

The advantage of the conservation representation for describing
collisions is clear: When $\ket{\bfN}$ is expanded in terms of
hydrodynamic and kinetic kets, collisions may change only the
coefficients of the latter.  From Eqs.~(\ref{eq:ortho}) and
(\ref{eq:expansion}), we see that the coefficients of the hydrodynamic
kets, namely the $\lambda_\sigma$'s, are the conserved quantities
themselves.  These are precisely what must remain unchanged by a
collision.  So, in the conservation representation, the collision
process is simply an alteration of the $n-n_c$ coefficients of the
kinetic kets, namely the $\alpha_\eta$'s.  Thus, this representation
effectively reduces the dimensionality of the space needed to describe
the collision process to $n-n_c$.  Again, we shall illustrate the
construction of and transformations between these alternative
representations of the single-particle distribution function for several
examples of entropic lattice Boltzmann models.

\subsection{Nonnegativity}

While the conservation representation of $\ket{\bfN}$ is most natural
for describing collisions, the original representation is more natural
for describing the constraint of nonnegativity of the distribution
function components.  In the original representation, we had a
restriction to $N_j\geq 0$ for $j=1,\ldots,n$.  In the conservation
representation, these $n$ inequalities transform to a corresponding set
of linear inequalities on the hydrodynamic and kinetic parameters,
$\lambda_\sigma$ and $\alpha_\eta$.  As is well known, such a set of
linear inequalities define a convex polytope in the parameter space.
This construction, in the full $n$ dimensional space of hydrodynamic and
kinetic parameters, shall be called the {\it master polytope}.
Specification of the hydrodynamic parameters then defines the cross
section of the master polytope that bounds the kinetic parameters,
$\alpha_\eta$; we shall call these cross sections the {\it kinetic
  polytopes}, and it is clear that they must also be convex.

We shall construct the master and kinetic polytopes for simple entropic
lattice Boltzmann models later in this paper.  We shall see that they
become very difficult to visualize when the number of particle
velocities becomes large.  This difficulty raises the question of
whether or not there is a general method to describe the shape of
polytopes defined by linear equalities in this way.  It turns out that
such a method is well known in linear programming and optimization
theory, and is called {\it Fourier-Motzkin elimination}~\cite{bib:fm}.
It is constructive in nature, and works for any set of inequalities in
any number of unknowns.  If the inequalities cannot be simultaneously
satisfied, the method will indicate that.  Otherwise, it will yield an
ordered sequence of inequalities for each variable, the bounds of which
depend only on the previously bounded variables in the sequence.  If it
were desired to perform a multiple integral over the polytopic region,
for example, the Fourier-Motzkin method would provide a systematic
procedure for setting up the limits of integration.

\subsection{Collisions}

Once we are able to construct and visualize these polytopes, it is
straightforward to describe the constraints that conservation imposes on
the collision process: Because the propagation process is nothing more
than a permutation of the values of the $N_j(\bfx)$ on the lattice, it
is clear that it will maintain nonnegativity.  That is, if the
$N_j(\bfx)$ were positive prior to the propagation, they will be so
afterward.  This means that the post-propagation state of each site will
lie within the allowed master polytope.  Given such a post-propagation
state $\ket{\bfN}$, we transform to the conservation represenation.  The
coefficients $\lambda_\sigma$ of the hydrodynamic kets are the conserved
quantities and must remain unchanged by the collision.  Geometrically,
these determine a cross section of the master polytope within which the
state must remain.  This cross section is the kinetic polytope of the
pre-collision state.  The essential point is that the post-collision
state must also lie within this kinetic polytope to preserve
nonnegativity.  To the extent that the postcollision state is determined
by the precollision state, this means that:
\begin{CollisionProperty}
  \label{cp:nonnegativity}
  The collision process is a map from the kinetic polytope into itself.
\end{CollisionProperty}

We note that this requirement is not without some controversy.  It may
be argued that lattice Boltzmann algorithms are ficticious kinetic
models from which realistic hydrodynamics are emergent.  Since the
details of the kinetics are ficticious anyway, why not also dispense
with the requirement that the single-particle distribution function be
positive?  As long as the conserved densities of positive-definite
quantities, such as mass and kinetic energy, are positive, why should we
care if the underlying lattice Boltzmann distribution function is
likewise?  There are two reasons: The first is that, even if the system
is initialized with nonnegative physical densities, the propagation step
may give rise to negative physical densities if negative values of the
distribution are allowed.  To see this, imagine a postcollision state in
which all of the neighbors of site $\bfx$ have a single negative
distribution function component in the direction heading toward $\bfx$.
Even if all these neighboring sites had positive mass density, site
$\bfx$ will have a negative mass density after one propagation step.

The second reason for demanding nonnegativity of the distribution
function is, to some extent, a matter of taste.  We like to think that
the kinetic underpinnings of the lattice Boltzmann algorithm are more
than just a mathematical trick to yield a desired set of hydrodynamic
equations.  Though there may be no physical system with such a collision
operator (not to mention dynamics constrained to a lattice), we feel
that the more properties of real kinetics that can be maintained, the
more useful the algorithm is likely to be.  This is particularly
important for complex fluids, for which the form of the hydrodynamic
equations is often unknown, and for which we must often appeal to some
kinetic level of description.  Nevertheless, we shall revisit this
question in Subsection~\ref{ssec:galileo}.

Collision Property~\ref{cp:nonnegativity} is simple in statement and
motivation, but in fact it weeds out many putative lattice Boltzmann
collision operators, including those most commonly used in computational
fluid dynamics research today -- namely, the overrelaxed lattice BGK
operators.  For sufficiently large overrelaxation parameter (collision
frequency), such operators are well known to give rise to negative
values of the $N_j(\bfx)$.  This is usually symptomatic of the onset of
a numerical instability in the lattice Boltzmann algorithm.  We shall
discuss such instabilities in more detail later in this paper.  For the
present, we emphasize that we are {\it not} claiming that Collision
Property~\ref{cp:nonnegativity} will eliminate such instabilities.  The
property does mandate a lower bound of zero on the $N_j(\bfx)$, and
hence it restricts the manner in which such instabilities might grow and
saturate.  Nevertheless, it is generally still possible for collision
operators that obey Collision Property~\ref{cp:nonnegativity} to exhibit
numerical instability if they lack a Lyapunov functional.

\subsection{Reversibility}

Collision Property~\ref{cp:nonnegativity} is the minimum requirement
that we impose on our lattice Boltzmann collision operators.  It is
possible to define more stringent requirements for them, and we shall
continue to do exactly that to satisfy various desiderata.  For example,
we might want to demand that our lattice Boltzmann algorithm be {\it
  reversible}.  A reversible algorithm could be run backwards in time
from any final condition by alternately applying the inverse propagation
operator,
\begin{equation}
  N_j(\bfx-\bfc_j)\leftarrow N_j(\bfx),
  \label{eq:invp}
\end{equation}
followed by an inverse collision operator to recover an initial
condition at an earlier time.  For such an inverse collision operator to
exist, the map from the allowed polytope to itself must be one-to-one.
Since we have already demanded that the map be {\it into}, this means
that it must also be {\it onto}, hence:
\begin{CollisionProperty}
  \label{cp:reversibility}
  A collision process is reversible if it is a map from the allowed
  polytope onto itself.
\end{CollisionProperty}

\subsection{Imposition of a Lyapunov Functional}

The criteria that we have set out thusfar assure that the conservation
laws -- and hence the First Law of Thermodynamics -- will be obeyed.  To
ensure stability and thermodynamic consistency, however, it is necessary
to also incorporate the requirements of the Second Law.

We suppose that our system has a function $H$ of the state variables
$N_j$ that is additive over sites $\bfx$,
\begin{equation}
  H = \sum_{\bfx\in\calL} h\left(\bfN(\bfx)\right),
\end{equation}
and additive over directions $j$,
\begin{equation}
  h\left(\bfN(\bfx)\right) =
  \sum_{j=1}^n \theta_j\left(N_j\left(\bfx\right)\right),
  \label{eq:hadditive}
\end{equation}
where the functions $\theta_j$ are defined on the nonnegative real
numbers.  One might wonder if Eq.~(\ref{eq:hadditive}) could be
generalized, but in fact it has been shown to be a necessary condition
for an $H$-theorem~\cite{bib:ajw}.  It is clear from this construction
that $H$ is preserved under the propagation operation of
Eq.~(\ref{eq:prop}).  If we require that our collisions never decrease
the contribution to $H$ from each site -- that is, that
$h\left(\bfN(\bfx)\right)$ can only be increased by a collision -- then
$H$ is a Lyapunov functional for our system, and the existence and
stability of an equilibrium state is guaranteed.

We note that $h$ is a function of the $N_j$ for all $N_j\geq 0$, and can
therefore be thought of as a scalar function on the master polytope.
Specification of the coefficients $\lambda_\sigma$ of the hydrodynamic
kets then determine the cross section of allowed collision outcomes, or
the kinetic polytope, parametrized by the coefficients of the kinetic
kets $\alpha_\eta$.  Thus, for a given incoming state, $h$ can be
thought of as a scalar function on the kinetic polytope.  We denote this
by $h(\alpha_\eta)$, and we demand that the collision process increase
this function:
\begin{CollisionProperty}
  \label{cp:htheorem}
  To ensure the existence and stability of an equilibrium state, a
  collision at site $\bfx$ must not decrease the restriction of the
  function $h$ to the kinetic polytope.
\end{CollisionProperty}

From a more geometric point of view, note that the (hyper)surfaces of
constant $h(\alpha_\eta)$ provide a codimension-one foliation of the
kinetic polytope.  These (hyper)surfaces degenerate to a point where $h$
reaches a maximum.  An incoming state lies on such a (hyper)surface.  A
legitimate collision is required to map such an incoming state to an
outgoing one that lies inside, or at least on, this (hyper)surface.
Clearly, the point with maximal $h$ must be mapped to itself.

Indeed, we can subsume Collision Property~\ref{cp:nonnegativity} into
Collision Property~\ref{cp:htheorem} by constructing the function $h$ so
that it takes a minimum value on the boundary of the polytope, and
increases to a single maximum somewhere inside.  The master polytope is
defined as the region for which all of the $N_j(\bfx)$'s are
nonnegative.  The boundary of the master polytope is therefore the place
where at least one of the $N_j(\bfx)$'s vanishes.  It follows that
$\prod_j^n N_j(\bfx)$ is constant -- in fact, it is zero -- on the
polytope boundary.  More generally, if $\zeta_j(x)$ for $j=1,\ldots,n$
are such that
\begin{equation}
  \zeta_j(0) = 0,
  \label{eq:zatz}
\end{equation}
then $\prod_j^n \zeta_j\left(N_j(\bfx)\right)$ also goes to zero on the
polytope boundary.  To ensure that there is only one maximum inside the
convex (master or kinetic) polytope, we also require that the $\zeta_j$
be strictly increasing,
\begin{equation}
  \zeta'_j(x) > 0,
  \label{eq:increasing}
\end{equation}
for nonnegative $x$.  Indeed, from Eqs.~(\ref{eq:zatz}) and
(\ref{eq:increasing}) it follows that
\begin{equation}
  \zeta_j(x) \geq 0,
  \label{eq:positivity}
\end{equation}
for nonnegative $x$.  Thus, the simplest choice would be to make
$h\left(\bfN(\bfx)\right)$ a function of $\prod_j^n
\zeta_j\left(N_j(\bfx)\right)$.  To make this consistent with
Eq.~(\ref{eq:hadditive}), however, we see that the functions $\theta_j$
should be chosen to be the logarithms of the functions $\zeta_j$, so
\begin{equation}
  h\left(\bfN(\bfx)\right) =
  \sum_{j=1}^n \ln\left[\zeta_j\left(N_j(\bfx)\right)\right].
  \label{eq:hdef}
\end{equation}
This goes to negative infinity on the boundaries of the master (and
hence kinetic) polytopes, and it has a unique maximum in the interior.
\begin{CollisionProperty}
  \label{cp:hfunction}
  A valid functional form for $h$ is given by Eq.~(\ref{eq:hdef}), where
  the functions $\zeta_j$ obey Eq.~(\ref{eq:zatz}),
  Eq.~(\ref{eq:increasing}), and Eq.~(\ref{eq:positivity}).
\end{CollisionProperty}
This may well be the most controversial of the four collision properties
that we have presented.  Indeed, it is {\it not} necessary to assure that $H$
have a single maximum within the polytope.  We shall discuss an $H$ function
that violates Collision Property~\ref{cp:hfunction} in Subsection~\ref{ssec:galileo}.

The equilibrium distribution is the point within the kinetic polytope
where $h$ has its maximum value.  If we denote the kinetic parameters by
$\alpha_j$, then we can find this point by demanding that the gradient
of $h$ vanish,
\begin{equation}
  0 =
  \frac{\partial h}{\partial\alpha_\eta} =
  \sum_{j=1}^n
  \frac{\zeta'_j\left(N_j\right)}{\zeta_j\left(N_j\right)}
  \frac{\partial N_j}{\partial\alpha_\eta} =
  \sum_{j=1}^n
  \frac{\zeta'_j\left(N_j\right)}{\zeta_j\left(N_j\right)}
  \ket{\bfalpha_\eta}_j.
\end{equation}
If we denote the equilibrium distribution function by $\Neqj$, it
follows that the covector with components
$\zeta'_j\left(\Neqj\right)/\zeta_j\left(\Neqj\right)$ lies in the
hydrodynamic subspace,
\begin{equation}
  \frac{\zeta'_j\left(\Neqj\right)}{\zeta_j\left(\Neqj\right)} =
  \sum_{\sigma=1}^{n_c} Q_\sigma \bra{\bflambda_\sigma}_j.
  \label{eq:equilibrium}
\end{equation}
The $n_c$ coefficients $Q_\sigma$ must be chosen so that the values of
the conserved quantities are as desired,
\begin{equation}
  \braket{\bflambda_\tau}{\bfNeq} = \lambda_\tau
  \label{eq:lagmult}
\end{equation}
for $\tau=1,\ldots,n_c$.  Eqs.~(\ref{eq:equilibrium}) and
(\ref{eq:lagmult}) can be regarded as $n+n_c$ equations for the $n+n_c$
unknowns, $\Neqj$ (for $j=1,\ldots,n$) and $Q_\sigma$ (for
$\sigma=1,\ldots,n_c$).  In general, these equations are nonlinear and
it is not always possible to write the $\Neqj$ in closed form.

The equilibrium distribution is something over which the model builder
would like to retain as much control as possible, since it is often used
to tailor the form of the resultant hydrodynamic equations.  For
example, a judicious choice of the equilibrium distribution function is
required to obtain Galilean invariant hydrodynamic equations for lattice
Boltzmann models of fluids.  Customarily, lattice Boltzmann model
builders have simply dictated the form of the equilibrium distribution
function.  To the extent that this is necessary, however, it is more in
keeping with the philosophy espoused in this paper to (try to) dictate
the form of the Lyapunov functional, rather than that of the equilibrium
distribution.  The challenge is to find a set of functions $\zeta_j$,
subject to the requirements of Eqs.~(\ref{eq:zatz}),
(\ref{eq:increasing}) and (\ref{eq:positivity}), such that
Eq.~(\ref{eq:equilibrium}) yields the desired equilibrium distribution.
We shall return to this problem in Subsection~\ref{ssec:galileo}.

While this formulation of an entropic principle for lattice Boltzmann
models seems reasonable, note that it is rather different from the usual
one encountered in kinetic theory.  The usual choice there would be that
of Boltzmann, $h=\sum_{j=1}^n N_j\ln N_j$.  A moment's examination of
Eq.~(\ref{eq:hdef}) indicates that this would correspond to the choice
$\zeta_j(x)=x^x$, but this function decreases for $x\in (0,1/e)$ and
hence violates Eq.~(\ref{eq:increasing}).  Thus, while it is possible to
construct lattice Boltzmann models with Maxwell-Boltzmann
equilibria~\cite{bib:ajw}, the Lyapunov functionals of the models
described in this paper need to be rather different from those commonly
used in kinetic theory.  With this caveat in mind, we shall henceforth
abuse terminology by taking ``$H$ function'' and ``Lyapunov functional''
to be synonymous.

\subsection{Collision Operator}

Finally, we turn our attention to the construction of a collision
operator that is in keeping with the collision properties described
above.  Obviously, there are many ways to describe maps from the kinetic
polytopes into (and perhaps onto) themselves, that do not decrease an
$H$ function.  For the purposes of this paper, however, we restrict our
attention to the BGK form of collision operator in which the outgoing
state is a linear combination of the incoming state and the equilbrium,
\begin{equation}
  \ket{\bfC} = \frac{1}{\tau}\left(\ket{\bfNeq} - \ket{\bfN}\right),
  \label{eq:cbgk}
\end{equation}
where $\tau$ is called the relaxation time.  Usually $\tau$ is taken to
be constant, but more generally it may depend on the conserved
quantities, and most generally on all of the $N_j$.

If we combine Eqs.~(\ref{eq:cbgk}) and (\ref{eq:cket}), we get the BGK
equation,
\begin{equation}
  \ket{\bfN'} = \ket{\bfN} + \frac{1}{\tau}
  \left(\ket{\bfNeq} - \ket{\bfN}\right).
  \label{eq:bgk}
\end{equation}
From a geometric point of view, this equation tells us to draw a line in
the kinetic polytope from the position of the incoming state
$\ket{\bfN}$ through the equilibrium state $\ket{\bfNeq}$.  The final
state is a weighted combination of these two states and hence lies on
this line.  The incoming state is weighted by $1-1/\tau$, and the
equilibrium state is weighted by $1/\tau$.  Thus, for $\tau\geq 1$ the
outgoing state lies somewhere on the segment between the incoming state
and the equilibrium.  Since both of these states are in the kinetic
polytope, and since this polytope is convex, the outgoing state must lie
within it as well.  Thus Collision Property~\ref{cp:nonnegativity} is
satisfied.  Moreover, since the restriction of $h$ to this segment
increases as one approaches the equilibrium, Collision
Property~\ref{cp:htheorem} is also satisfied.

The instabilities associated with lattice BGK operators arise because
practitioners try to {\it overrelax} them.  A large class of lattice BGK
models for the Navier-Stokes equations have shear viscosity $\nu \propto
(\tau - 1/2)$.  In an effort to achieve lower viscosity (and hence
higher Reynolds number), practitioners set $\tau$ to values between
$1/2$ and unity.  In this situation, the outgoing state ``overshoots''
the equilibrium and lies on the other side of the polytope, opposite the
equilibrium state.  For sufficiently small $\tau$, the outgoing state
will lie on a contour of $h$ that is lower than that of the incoming
state, thereby violating Collision Property~\ref{cp:htheorem}.  Still
smaller values of $\tau$ might cause the outgoing state to lie outside
the kinetic polytope altogether, thereby violating Collision
Property~\ref{cp:nonnegativity}.  In either case, numerical instability
is likely to result.

\subsection{Entropically Stabilized BGK Operators}
\label{ssec:esbgko}

A potential solution to this problem is suggested by our geometrical
viewpoint.  For $\tau=1$ the outgoing state is the equilibrium, for
which $h$ is maximal.  As $\tau$ is decreased from unity, the final
value of $h$ decreases from its maximal value.  At some value of $\tau$
less than unity, the outgoing value of $h$ will be equal to the incoming
one.  In order to respect Collision Property~\ref{cp:htheorem}, we must
not make $\tau$ any lower than this value, given by the solution to the
equation
\begin{equation}
  h\left(\bfN'\right) = h\left(\bfN\right),
  \label{eq:entstab}
\end{equation}
where $\ket{\bfN'}$ is given by Eq.~(\ref{eq:bgk}).  This may be
regarded as a nonlinear algebraic equation for the single scalar unknown
$\tau$.  Indeed, this limitation on $\tau$ was suggested independently
by Karlin {\it et al.}~\cite{bib:karlin}, and by Chen and
Teixeira~\cite{bib:hudong}.  It is straightforward to find the solution
to this equation numerically, since it is easily bounded: We know that
the desired solution has an upper bound of unity.  The lower bound will
be that for which the solution leaves the kinetic polytope; this happens
when one of the $N'_j$'s vanish, or equivalently when $\tau$ is equal to
the largest value of $1-\Neqj/N_j$, for $j\in\{1,\ldots,n\}$, that lies
between zero and one.  Given these two bounds on $\tau$, the regula
falsi algorithm will reliably locate the desired solution.  Call this
solution $\tau_*(\bfN)$.  This will generally be a function of the
incoming state.  A useful way to parametrize $\tau$ is then to write
\begin{equation}
  \tau(\bfN) = \frac{\tau_*(\bfN)}{\kappa},
\end{equation}
where $0 < \kappa < 1$ is a constant parameter.  The case
$\kappa\rightarrow 0$ corresponds to $\tau\rightarrow\infty$ so that the
collision operator vanishes; the case $\kappa\rightarrow 1$ corresponds
to $\tau\rightarrow\tau_*$, which is the largest value possible that
respects Collision Property~\ref{cp:htheorem}.  Thus, the entropically
stabilized version of the lattice BGK equation is
\begin{equation}
  \ket{\bfN'} = \ket{\bfN} + \frac{\kappa}{\tau_*(\bfN)}
  \left(\ket{\bfNeq} - \ket{\bfN}\right),
  \label{eq:bgkstab}
\end{equation}
where $\tau_*$ is the solution to Eq.~(\ref{eq:entstab}), and
$0\leq\kappa\leq 1$.

Of course, making $\tau$ a nontrivial function of the incoming state
will impact the hydrodynamic equations derived from the model.  The
challenge to the designer of entropic lattice Boltzmann models is then
to choose the $\zeta_j$ very judiciously, so that Eq.~(\ref{eq:bgkstab})
yields the desired hydrodynamic equations, while stability is guaranteed
by keeping $0<\kappa<1$.

In constructing a lattice Boltzmann model in this fashion, as opposed to
using the simpler prescription of specifying $\tau$, it may be argued
that we are relinquishing some control over the transport coefficients.
After all, if $\nu\propto (\tau-1/2)$, it appears that we can specify
arbitrarily small viscosity by reducing $\tau$.  In fact, this is not
the case since uncontrolled instabilities are known to set in for $\tau$
well above $1/2$.  The objective of entropic lattice Boltzmann models is
to allow the user some ability to overrelax the collision process,
without sacrificing stability.  The price that one may have to pay for
this stable overrelaxation is living with a bounded transport
coefficient.  Thus, entropic lattice Boltzmann models for
fluid flow may be restricted to minimum values of
viscosity.  As we shall show, however, for certain
entropically stabilized lattice Boltzmann models, this minimum value can
actually be zero.  This allows for fully explicit, perfectly
conservative, absolutely stable algorithms at arbitrarily small
transport coefficient.

Finally, we note that this particular prescription is only one way of
creating a stable lattice Boltzmann algorithm.  In fact, any map obeying
Collision Properties~\ref{cp:nonnegativity} and \ref{cp:htheorem} will
work.  More general mappings of polytopes to themselves can and should
be considered, and we leave this to future study.

\newpage
\section{One-Dimensional Diffusion Model}
\label{sec:diff}

Having discussed entropic lattice Boltzmann models in general terms, we
now apply the formalism to mass diffusion in one dimension.  In
elementary books on numerical analysis, it is demonstrated that the
fully explicit finite-difference approximation to the one-dimensional
diffusion equation is stable only if the Courant
condition~\cite{bib:courant}, $\dt\leq(\dx)^2/(2D)$, is satisfied; note
that this places an {\it upper bound} on the transport coefficient.  It
is also shown that this condition may be removed by adopting an implicit
differencing scheme, such as that of Crank and Nicolson~\cite{bib:cn},
or an alternating-direction implicit scheme, such as that of Peaceman
and Rachford~\cite{bib:pr}.  The DuFort-Frankel algorithm~\cite{bib:df}
is fully explicit and unconditionally stable, but it achieves this by a
differencing scheme that involves three time steps, even though the
diffusion equation is first-order in time.

For the problem of achieving high Reynolds number, one would like the
transport coefficient to be as {\it small} as possible.  We shall show
that the entropic lattice Boltzmann algorithm provides a fully explicit,
perfectly conservative, two-time-step algorithm that is absolutely
stable for arbitrarily small transport coefficient.  While this result
seems significant in and of itself, part of our purpose is pedagogical.
In the course of our development of this model, we shall discuss optimal
conservation representations and present the Fourier-Motzkin method for
visualizing the master and kinetic polytopes.  Nongeneric features of
the example are noted, in preparation for more sophisticated examples in
subsequent sections.

\subsection{Description of Model}

We suppose that we have a regular one-dimensional lattice $\calL$ whose
sites $x\in\calL$ are occupied by particles whose velocities may take on
one of only $n=3$ discrete values, namely $-1$, $0$ and $+1$.  We abuse
notation slightly by letting $j$ take its values from the set of symbols
$\{-,0,+\}$, so that we may write the single-particle distribution as
$N_j(x)$.  Thus, the state of a given site is captured by the ket,
\begin{equation}
  \ket{\bfN}
  =
  \left(
    \begin{array}{l}
      N_{-}\\
      N_{0}\\
      N_{+}
    \end{array}
  \right),
\end{equation}
where we have suppressed the dependence on the coordinate $x$ for
simplicity.

Diffusion conserves mass, so we suppose that the mass per unit lattice
site (mass density in lattice units),
\begin{equation}
  \rho
  =
  N_{-}+N_{0}+N_{+}
  =
  \braket{\bfrho}{\bfN},
\end{equation}
is conserved by the collisions.  Here we have introduced the
hydrodynamic ``bra,''
\begin{eqnarray}
  \bra{\bfrho}
  &=&
  \left(
    \begin{array}{rrr}
      +1 & +1 & +1
    \end{array}
  \right).\\
  & & \\
  \noalign{\hbox{\parbox{6.5truein}{
        This is the one and only hydrodynamic degree of freedom in this
        example.  Because there are a total of three degrees of freedom, the
        other two must be kinetic in nature.  To span these kinetic degrees of
        freedom and thereby make the bra basis complete, we introduce the
        linearly independent bras,}}}
  & & \nonumber\\
  \bra{\bfalpha}
  &=&
  \left(
    \begin{array}{rrr}
      -1 & +2 & -1
    \end{array}
  \right)
  \label{eq:dbra1}\\
  \bra{\bfbeta}
  &=&
  \left(
    \begin{array}{rrr}
      +1 & \phantom{+}0 & -1
    \end{array}
  \right).
  \label{eq:dbra2}
\end{eqnarray}
Next, we form a matrix of the three bras and invert it to get the dual
basis of kets,
\begin{equation}
  \left(
    \begin{array}{rrr}
      \ket{\bfrho} & \ket{\bfalpha} & \ket{\bfbeta}
    \end{array}
  \right)
  =
  \left(
    \begin{array}{r}
      \bra{\bfrho}\\
      \bra{\bfalpha}\\
      \bra{\bfbeta}
    \end{array}
  \right)^{-1}
  =
  \left(
    \begin{array}{rrr}
      +1 & +1 & +1\\
      -1 & +2 & -1\\
      +1 &  0 & -1
    \end{array}
  \right)^{-1}
  =
  \frac{1}{6}
  \left(
    \begin{array}{rrr}
      +2 & -1 & +3\\
      +2 & +2 &  0\\
      +2 & -1 & -3
    \end{array}
  \right).
  \label{eq:dbasis}
\end{equation}
From Eq.~(\ref{eq:dbasis}) we identify the linearly independent basis
kets
\begin{equation}
  \begin{array}{ccccc}
    \ket{\bfrho}
    =
    \frac{1}{3}
    \left(
      \begin{array}{r}
        +1\\
        +1\\
        +1
      \end{array}
    \right) &
    \phantom{aaa} &
    \ket{\bfalpha}
    =
    \frac{1}{6}
    \left(
      \begin{array}{r}
        -1\\
        +2\\
        -1
      \end{array}
    \right) &
    \phantom{aaa} &
    \ket{\bfbeta}
    =
    \frac{1}{2}
    \left(
      \begin{array}{r}
        +1\\
        0\\
        -1
      \end{array}
    \right).
  \end{array}
\end{equation}
The first of these is a hydrodynamic basis ket, while the last two are
kinetic basis kets.

One nongeneric feature should be noted: In this example, we were able to
choose $\bra{\bfalpha}$ and $\bra{\bfbeta}$ so that each ket is
proportional to the transpose of a corresponding bra.  Such a choice is
convenient but unnecessary.  The only real requirement in choosing the
kinetic bras is that they be linearly independent of each other and of
the hydrodynamic bras.  The next example will illustrate a situation in
which there is no obvious correspondence between the individual bras and
kets.

We can expand the state ket in this basis as follows
\begin{equation}
  \ket{\bfN}
  =
  \rho\ket{\bfrho} +
  \alpha\ket{\bfalpha} +
  \beta\ket{\bfbeta}
  =
  \frac{\rho}{3}
  \left(
    \begin{array}{c}
      1-\frac{\alphabar}{2}+\frac{3\betabar}{2}\\
      1+\alphabar\\
      1-\frac{\alphabar}{2}-\frac{3\betabar}{2}\\
    \end{array}
  \right),
  \label{eq:dN}
\end{equation}
where we have defined $\alphabar\equiv\alpha/\rho$ and
$\betabar\equiv\beta/\rho$.  The coefficients $\rho$, $\alpha$ and
$\beta$ (or equivalently $\rho$, $\alphabar$ and $\betabar$) constitute
the conservation representation.  The inverse of this transformation is
seen to be
\begin{eqnarray}
  \rho  &=& \braket{\bfrho}{\bfN}  = +N_{-}+ N_{0}+N_{+}\nonumber\\
  \alpha &=& \braket{\bfalpha}{\bfN} = -N_{-}+2N_{0}-N_{+}\label{eq:dinv}\\
  \beta &=& \braket{\bfbeta}{\bfN} = +N_{-}       -N_{+}.\nonumber
\end{eqnarray}

\subsection{Nonnegativity}

\begin{figure}
  \center{
    \mbox{
      \includegraphics[bbllx=72,bblly=242,bburx=540,bbury=556,width=5.0truein]{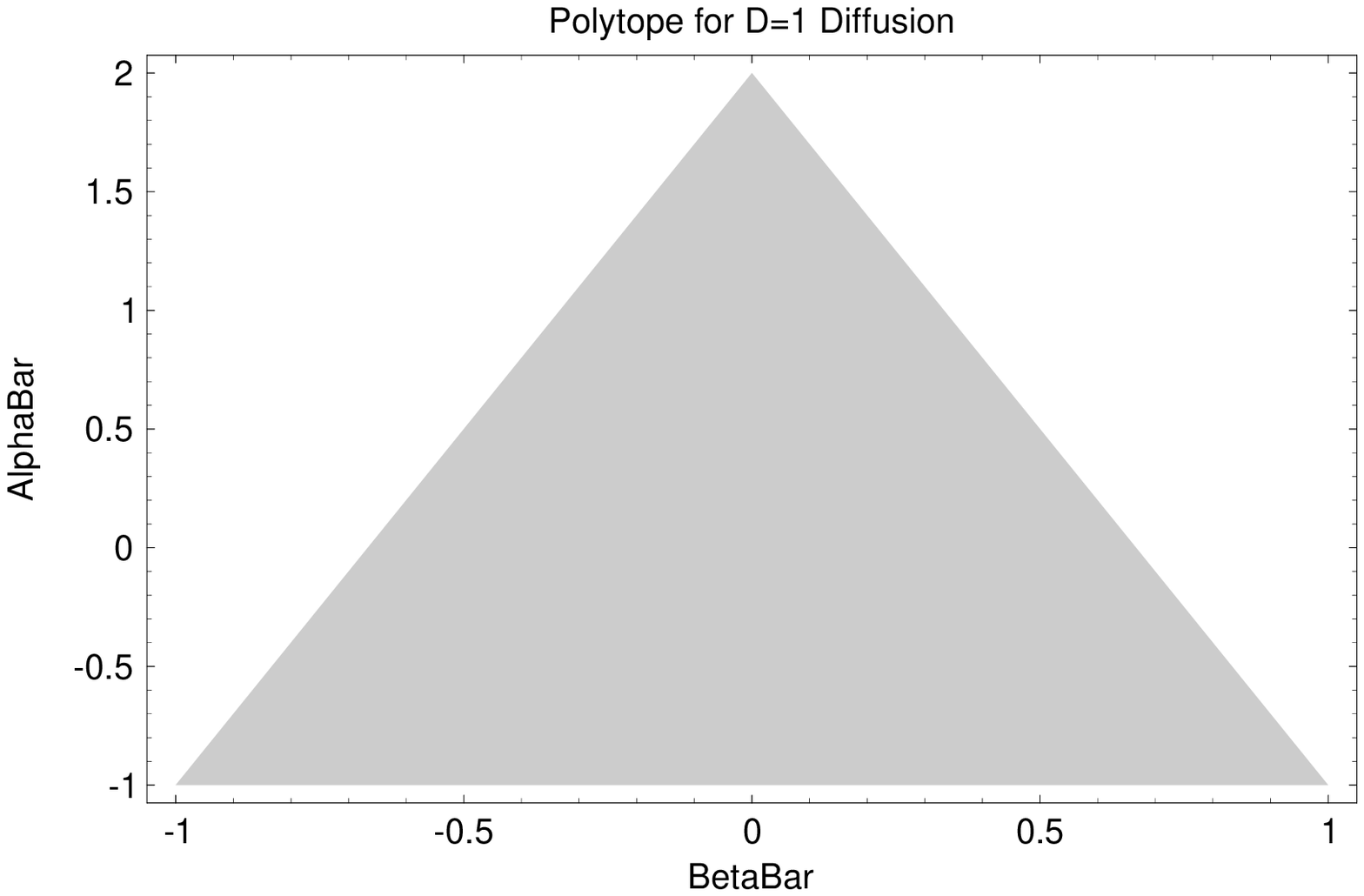}
      }}
  \caption{{\bf The Kinetic Polytope:}  Nonnegativity of the
    distribution function requires that the kinetic parameters $\alphabar$
    and $\betabar$ lie in the shaded triangular region.}
  \label{fig:dpolytope}
\end{figure}

The nonnegativity of the components of the single-particle distribution,
$\ket{\bfN}$, places inequality constraints on the parameters $\rho$,
$\alpha$ and $\beta$.  For example, it is clear that $\rho\geq 0$.
Referring to Eq.~(\ref{eq:dN}), we see that we must also demand
\begin{eqnarray}
  0 &\leq& 1-\frac{\alphabar}{2}+\frac{3\betabar}{2}\nonumber\\
  0 &\leq& 1+\alphabar\label{eq:dineqs}\\
  0 &\leq& 1-\frac{\alphabar}{2}-\frac{3\betabar}{2}.\nonumber
\end{eqnarray}
It is easy to see that all three of these inequalities may be subsumed
by the single statement
\begin{equation}
  -1\leq\alphabar\leq 2-3\left|\betabar\right|.
  \label{eq:tribounds}
\end{equation}
This restricts the kinetic parameters to a triangular region in the
$(\alphabar,\betabar)$ plane, as is shown in Fig.~\ref{fig:dpolytope}.

Note that the effect of varying $\rho$ at constant $\alphabar$ and
$\betabar$ is to simply scale the components of the distribution
function.  The triangular region bounding the parameters $\alphabar$ and
$\betabar$ is then independent of $\rho$.  That is, the bounds on the
barred kinetic parameters do not depend on the hydrodynamic parameter,
so there is really no need for distinction between the master and
kinetic polytopes.  It should be noted that this is another nongeneric
feature of the present model.  While it is always possible to scale out
by a single nonnegative-definite hydrodynamic parameter, as we have done
with $\rho$ in this case, more sophisticated models will have several
hydrodynamic parameters.  In such cases, the shape of the region
bounding the kinetic parameters will depend on the remaining
hydrodynamic parameters.  We shall see this more clearly in the example
of Section~\ref{sec:fluid}.

\subsection{Optimality of Representation}

In this example, there is clearly a great deal of latitude in the choice
of the kinetic bras.  The only requirement that we have placed on these
is that they be linearly independent of the hydrodynamic bra and of each
other.  This raises the question of whether there is some optimal choice
that might be made for these.  Of course, this depends entirely on what
is meant by ``optimal'' in this context.

It has been noted that the original representation of $\ket{\bfN}$ is
more natural for expressing the constraint of nonnegativity, while the
conservation representation is more natural for expressing the collision
process.  For this reason, any computer implementation of entropic
lattice Boltzmann methods will require frequent transformations between
the two representations.  This transformation is precisely what we have
worked out in Eqs.~(\ref{eq:dN}) and (\ref{eq:dinv}) above.  Thus, one
natural figure of merit is the number of arithmetic operations required
to perform such transformations.

To investigate this question, rather than requiring that the kinetic
bras be given by Eqs.~(\ref{eq:dbra1}) and (\ref{eq:dbra2}), we leave
them in the general form
\begin{eqnarray}
  \bra{\bfalpha}
  &=&
  \left(
    \begin{array}{rrr}
      \alpha_1 & \alpha_2 & \alpha_3
    \end{array}
  \right)
  \\
  \bra{\bfbeta}
  &=&
  \left(
    \begin{array}{rrr}
      \beta_1 & \beta_2 & \beta_3
    \end{array}
  \right).
\end{eqnarray}
When we invert to get the kets, the results are
\begin{equation}
  \begin{array}{ccccc}
    \ket{\bfrho}
    =
    \frac{1}{\Delta}
    \left(
      \begin{array}{r}
        \alpha_2\beta_3-\alpha_3\beta_2\\
        \alpha_3\beta_1-\alpha_1\beta_3\\
        \alpha_1\beta_2-\alpha_2\beta_1
      \end{array}
    \right) &
    \phantom{aaa} &
    \ket{\bfalpha}
    =
    \frac{1}{\Delta}
    \left(
      \begin{array}{r}
        \beta_2-\beta_3\\
        \beta_3-\beta_1\\
        \beta_1-\beta_2
      \end{array}
    \right) &
    \phantom{aaa} &
    \ket{\bfbeta}
    =
    \frac{1}{\Delta}
    \left(
      \begin{array}{r}
        \alpha_3-\alpha_2\\
        \alpha_1-\alpha_3\\
        \alpha_2-\alpha_1
      \end{array}
    \right),
  \end{array}
\end{equation}
where we have defined the determinant
\begin{equation}
  \Delta\equiv
  \alpha_1\left(\beta_2-\beta_3\right) +
  \alpha_2\left(\beta_3-\beta_1\right) +
  \alpha_3\left(\beta_1-\beta_2\right).
\end{equation}
One way to simplify the transformation process would be to find a
representation in which as many components as possible of the kinetic
bras and kets vanish, without making the transformation singular.  This
means that we want to choose the $\alpha_j$'s and $\beta_j$'s such that
$\Delta\neq 0$, while making vanish as many as possible of the following
twelve quantities:
\begin{eqnarray*}
  \alpha_1, & \alpha_2, & \alpha_3,\\
  \beta_1, & \beta_2, & \beta_3,\\
  \alpha_3-\alpha_2, & \alpha_1-\alpha_3, & \alpha_2-\alpha_1,\\
  \beta_2-\beta_3, & \beta_3-\beta_1, & \beta_1-\beta_2.
\end{eqnarray*}
In this example, it is straightforward to see that there are a number
of ways to make six of these quantities equal to zero.  For example,
we could choose
\begin{eqnarray}
  \alpha_1 \neq 0, & \alpha_2 = 0, & \alpha_3 = 0,
  \nonumber\\
  \beta_1 = 0, & \beta_2 \neq 0, & \beta_3 = 0,
  \label{eq:dchoice}\\
  \noalign{\hbox{\parbox{6.5truein}{resulting in $\Delta=\alpha_1\beta_2\neq 0$ and}}}
  & &\nonumber\\
  \alpha_3-\alpha_2 = 0, & \alpha_1-\alpha_3 \neq 0, & \alpha_2-\alpha_1 \neq 0,
  \nonumber\\
  \beta_2-\beta_3 \neq 0, & \beta_3-\beta_1 = 0, & \beta_1-\beta_2 \neq 0.
\end{eqnarray}
This choice also has the added virtue of making the hydrodynamic ket
equal to
\begin{equation}
  \ket{\bfrho} =
  \left(
    \begin{array}{c}
      0\\
      0\\
      1
    \end{array}
  \right),
\end{equation}
which also has two vanishing components.

The computation of the $\ket{\bfN}$ components from the parameters
$\rho$, $\alphabar$ and $\betabar$ is thus reduced to
\begin{eqnarray*}
  N_{-} &=& \rho\alphabar\\
  N_{0} &=& \rho\betabar\\
  N_{+} &=& \rho\left(1-\alphabar-\betabar\right),
\end{eqnarray*}
involving a total of two multiplications and two additions/subtractions;
this may be contrasted with Eqs.~(\ref{eq:dN}) which involve six
multiplications and five additions/subtractions.  Likewise, using the
choice of Eqs.~(\ref{eq:dchoice}), the computation of the parameters
$\rho$, $\alphabar$ and $\betabar$ from the $\bfN$ is reduced to
\begin{eqnarray*}
  \rho &=& N_{-}+N_{0}+N_{+}\\
  \alphabar &=& N_{-}/\rho\\
  \betabar &=& N_{0}/\rho,\\
\end{eqnarray*}
involving two additions/subtractions and two divisions; this may be
contrasted to Eqs.~(\ref{eq:dinv}) which involve five
additions/subtractions, one multiplication, and two divisions.

Clearly, more sophisticated figures of merit could be devised to
optimize the transformations used in lattice Boltzmann computations.  As
we have noted, vanishing components of the kinetic bras and kets
eliminate the addition/subtraction of terms.  Likewise, components equal
to $\pm 1$ do require an addition/subtraction, but not a multiplication,
and this fact could be taken into account in a more refined figure of
merit.  The computation of the collision outcome is the principal
``inner loop'' of a lattice Boltzmann computation, insofar as it must be
performed at each site of a spatial grid at each time step.  Such
considerations may be especially important for lattice Boltzmann models
with large numbers of velocities.

\subsection{Fourier-Motzkin Elimination}

For this simple example, we had no difficulty visualizing the master and
kinetic polytopes.  In preparation for the succeeding sections where the
task will be substantially more difficult, however, we take this
opportunity to introduce the Fourier-Motzkin elimination method for this
purpose.  The algorithm consists of the following sequence of steps:
\begin{enumerate}
\item We rewrite the set of inequalities, Eq.~(\ref{eq:dineqs}) in
  matrix format,
  \begin{equation}
    \left(
      \begin{array}{rrr}
        -\smallfrac{1}{2} & +\smallfrac{3}{2} & +1\\
        +1 & 0 & +1\\
        -\smallfrac{1}{2} & -\smallfrac{3}{2} & +1
      \end{array}
    \right)
    \left(
      \begin{array}{c}
        \alphabar\\
        \betabar\\
        1
      \end{array}
    \right)
    \geq 0
    \label{eq:dineqsm}
  \end{equation}
  We have adopted the convention of including constant terms in an extra
  column of the matrix, using the device of appending $1$ to the column
  vector of unknowns.  In general, there will be $m$ inequalities for $n$
  unknowns, and the matrix will be of size $m\times (n+1)$.
  
\item We scale each inequality by a positive factor so that the
  pivot~\footnote{As in discussions of Gaussian elimination, the term
    ``pivot'' refers to the first nonzero entry in a row.} is either
  $-1$, $0$ or $+1$.  (Recall that scaling by a positive factor
  preserves the sense of an inequality.)  We then reorder the
  inequalities, sorting by their (scaled) pivots so that the zero pivots
  are last.  Beginning with Eq.~(\ref{eq:dineqsm}), this yields
  \begin{equation}
    \left(
      \begin{array}{rrr}
        -1 & +3 & +2\\
        -1 & -3 & +2\\
        +1 &  0 & +1
      \end{array}
    \right)
    \left(
      \begin{array}{c}
        \alphabar\\
        \betabar\\
        1
      \end{array}
    \right)
    \geq 0.
  \end{equation}
  Note that there were no zero pivots in this first step.  We did,
  however, reorder the inequalities so that those with pivot $-1$ are
  together, and preceed that with pivot $+1$.

\item We now add all pairs of inequalities, such that the first member
  of the pair has pivot $-1$ and the second member of the pair has pivot
  $+1$, and we append the new zero-pivot inequalities thus obtained to
  the system.  If there are $m_-$ inequalities with pivot $-1$, and
  $m_+$ inequalities with pivot $+1$, this results in the addition of
  $m_- m_+$ new inequalities to the system.  Since there were
  $m-m_--m_+$ zero-pivot inequalities to begin with, the new system will
  have a total of $m_0=m-m_--m_++m_-m_+$ zero-pivot inequalities.  For
  our above example, $m=3$, $m_-=2$ and $m_+=1$, so we get $m_0=2$
  zero-pivot inequalities in the system, which can now be written
  \begin{equation}
    \left(
      \begin{array}{rrr}
        -1 & +3 & +2\\
        -1 & -3 & +2\\
        +1 &  0 & +1\\
        0 & +3 & +3\\
        0 & -3 & +3\\
      \end{array}
    \right)
    \left(
      \begin{array}{c}
        \alphabar\\
        \betabar\\
        1
      \end{array}
    \right)
    \geq 0.
  \end{equation}

\item Finally, we recurse by returning to step 2 for the $m_0\times n$
  submatrix obtained by taking only the zero-pivot inequalities and
  neglecting the first unknown (since the zero-pivot inequalities do not
  involve it anyway).  We continue in this fashion until all of the
  first $n$ elements of the rows of the matrix are zero.
\end{enumerate}

For our example, the recursion step asks us to return to step 2 for the
submatrix indicated below
\begin{equation}
  \left(
    \mbox{
      \begin{tabular}{r|rr}
        \multicolumn{1}{r}{-1} & \multicolumn{1}{r}{+3} & \multicolumn{1}{r}{+2}\\
        \multicolumn{1}{r}{-1} & \multicolumn{1}{r}{-3} & \multicolumn{1}{r}{+2}\\
        \multicolumn{1}{r}{+1} & \multicolumn{1}{r}{ 0} & \multicolumn{1}{r}{+1}\\
        \cline{2-3}\\
        0 & +3 & +3\\
        0 & -3 & +3
      \end{tabular}
      }
  \right)
  \left(
    \begin{array}{c}
      \alphabar\\
      \betabar\\
      1
    \end{array}
  \right)
  \geq 0.
\end{equation}
As before, we begin by normalizing the rows and sorting them by pivot,
\begin{equation}
  \left(
    \mbox{
      \begin{tabular}{r|rr}
        \multicolumn{1}{r}{-1} & \multicolumn{1}{r}{+3} & \multicolumn{1}{r}{+2}\\
        \multicolumn{1}{r}{-1} & \multicolumn{1}{r}{-3} & \multicolumn{1}{r}{+2}\\
        \multicolumn{1}{r}{+1} & \multicolumn{1}{r}{ 0} & \multicolumn{1}{r}{+1}\\
        \cline{2-3}\\
        0 & -1 & +1\\
        0 & +1 & +1
      \end{tabular}
      }
  \right)
  \left(
    \begin{array}{c}
      \alphabar\\
      \betabar\\
      1
    \end{array}
  \right)
  \geq 0.
\end{equation}
In this example there is now one row of the submatrix with pivot $-1$
and one with pivot $+1$, so we append the sum of these to the system to
get
\begin{equation}
  \left(
    \mbox{
      \begin{tabular}{r|rr}
        \multicolumn{1}{r}{-1} & \multicolumn{1}{r}{+3} & \multicolumn{1}{r}{+2}\\
        \multicolumn{1}{r}{-1} & \multicolumn{1}{r}{-3} & \multicolumn{1}{r}{+2}\\
        \multicolumn{1}{r}{+1} & \multicolumn{1}{r}{ 0} & \multicolumn{1}{r}{+1}\\
        \cline{2-3}\\
        0 & -1 & +1\\
        0 & +1 & +1\\
        0 &  0 & +2
      \end{tabular}
      }
  \right)
  \left(
    \begin{array}{c}
      \alphabar\\
      \betabar\\
      1
    \end{array}
  \right)
  \geq 0.
\end{equation}
At this point, the first $n$ elements of the last row of the matrix
are zero, so we stop the recursion and examine the last row.  It
states a true inequality, namely $+2\geq 0$, so we can conclude that
the original inequalities are not mutually exclusive; that is, a
nonnull polytope of solutions will exist.  If the $n+1$ element of the
final row had been negative, we would have concluded that no such
solution was possible.

Once we have stopped the recursion and established the consistency of
the inequalities, we can get the final set of inequalities by ``back
substituting'' up the matrix.  The last two nontrivial inequalities
tell us that $-1 \leq \betabar \leq +1$, or
\begin{equation}
  \left|\betabar\right|\leq 1.
\end{equation}
The first three equations then yield $-1 \leq \alphabar \leq
\min\left(2-3\betabar, 2+3\betabar\right)$ which is equivalent to
Eq.~(\ref{eq:tribounds}), describing the triangular region of
Fig.~\ref{fig:dpolytope}.

More generally, each variable will have a lower bound that is the
maximum of all the lower bounds determined by the inequalities with
pivot $+1$, and an upper bound that is the minimum of all the upper
bounds determined by the inequalities with pivot $-1$.  The arguments of
the maximum and minimum functions will depend only on those variables
that have already been bounded.  If we were using the technique to find
limits of integration for a multiple integral, the inequalities at the
bottom of the Fourier-Motzkin-eliminated matrix would correspond to the
outermost integrals.  We shall revisit this technique in the next
section.

\subsection{Collision Operator}

\begin{figure}
  \center{
    \mbox{
      \includegraphics[bbllx=72,bblly=160,bburx=540,bbury=625,width=3.0truein]{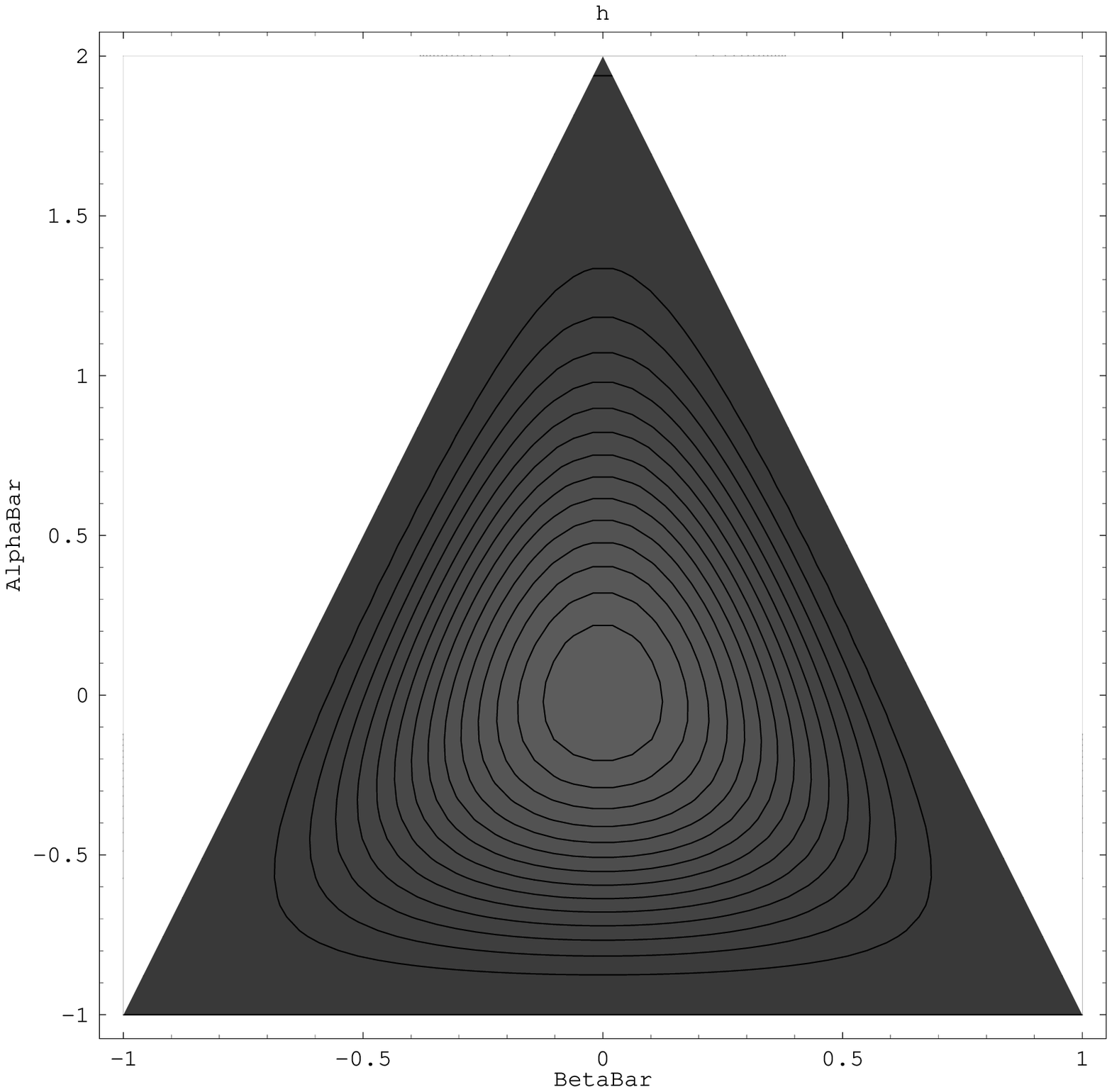}
      }}
  \caption{{\bf $H$ Function on the Kinetic Polytope:}  Note that $H$
    reaches a maximum when $\alphabar=\betabar=0$, and goes to $-\infty$
    on the boundary of the polytope.}
  \label{fig:entropy}
\end{figure}

To illustrate the construction of an entropically stable collision
operator for this model, we adopt the simplest possible $H$ function by
taking $\zeta_j(x)=x$.  Since there is only one hydrodynamic degree of
freedom, Eqs.~(\ref{eq:equilibrium}) reduce to
\begin{equation}
  \Neq{-} = \Neq{0} = \Neq{+} = \frac{1}{Q}.
\end{equation}
Eq.~(\ref{eq:lagmult}) then tells us that $Q=3/\rho$, so
\begin{equation}
  \bfNeq = \frac{\rho}{3}
  \left(
    \begin{array}{c}
      1\\
      1\\
      1
    \end{array}
  \right).
\end{equation}
Comparing this with Eq.~(\ref{eq:dN}), we see that, within the kinetic
polytope, the equilibrium point is the origin,
$\alpha^{\mbox{\scriptsize eq}}=\beta^{\mbox{\scriptsize eq}}=0$.  A
contour plot of the $H$ function on the kinetic polytope is presented in
Fig.~\ref{fig:entropy}.

The lattice BGK equation in the conservation representation,
Eq.~(\ref{eq:bgk}), is then equivalent to the prescription
\begin{equation}
  \rho' = \rho
\end{equation}
and
\begin{equation}
\left(
\begin{array}{l}
\alpha'\\
\beta'
\end{array}
\right) =
\left(
\begin{array}{l}
\alpha\\
\beta
\end{array}
\right) +
\frac{1}{\tau}
\left[
\left(
\begin{array}{l}
\alpha^{\mbox{\scriptsize eq}}\\
\beta^{\mbox{\scriptsize eq}}
\end{array}
\right) -
\left(
\begin{array}{l}
\alpha\\
\beta
\end{array}
\right)
\right] = z
\left(
\begin{array}{l}
\alpha\\
\beta
\end{array}
\right),
\label{eq:alcol}
\end{equation}
where $z\equiv 1-1/\tau$.  These can also be written as the single equation
\begin{equation}
\left(
\begin{array}{l}
\rho'\\
\alpha'\\
\beta'
\end{array}
\right) =
\left(
\begin{array}{ccc}
1 & 0 & 0\\
0 & z & 0\\
0 & 0 & z
\end{array}
\right)
\left(
\begin{array}{l}
\rho\\
\alpha\\
\beta
\end{array}
\right).
\label{eq:becol}
\end{equation}
As promised, the collision is dramatically simplified in this
representation.  The hydrodynamic parameter is unchanged; only the two
kinetic parameters are altered by the collision.  Transformation back to
the original representation yields
\begin{equation}
  \ket{\bfN'}
  =
  \frac{\rho}{3}
  \left(
    \begin{array}{c}
      1-\frac{z\alphabar}{2}+\frac{3z\betabar}{2}\\
      1+z\alphabar\\
      1-\frac{z\alphabar}{2}-\frac{3z\betabar}{2}\\
    \end{array}
  \right),
  \label{eq:dNp}
\end{equation}
or equivalently
\begin{equation}
  \ket{\bfN'}
  =
\frac{1}{3}
\left(
\begin{array}{ccc}
1+2z & 1-z & 1-z \\
1-z & 1+2z & 1-z \\
1-z & 1-z & 1+2z
\end{array}
\right)
  \ket{\bfN}.
\end{equation}

\subsection{Entropic Stabilization}

\begin{figure}
  \center{
    \mbox{
      \includegraphics[bbllx=72,bblly=160,bburx=540,bbury=625,width=3.0truein]{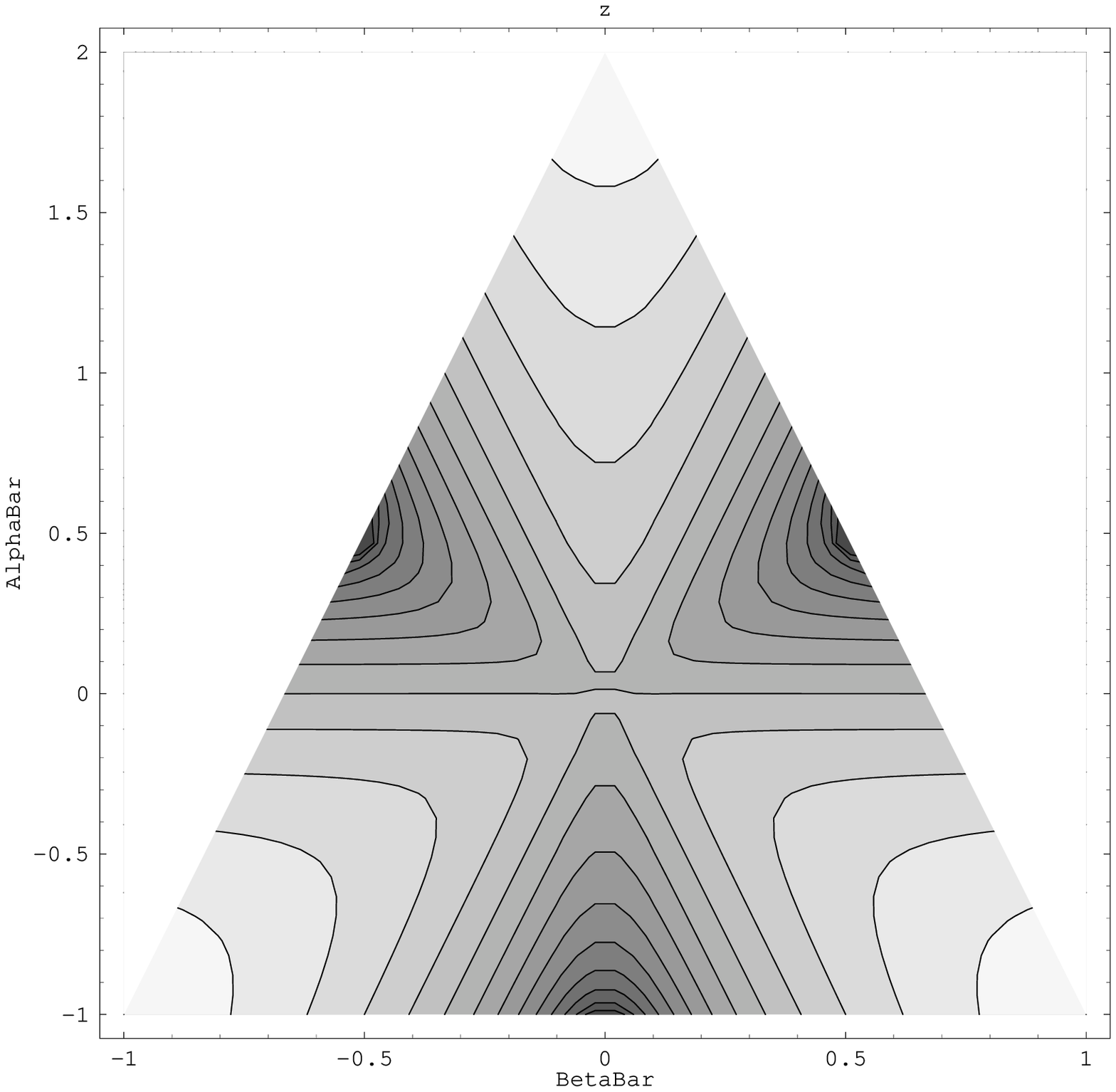}
      }}
  \caption{{\bf Contours of $z_*(\alphabar,\betabar)$ on the Kinetic
      Polytope:} This function approaches $-1$ when
    $\alphabar=\betabar=0$, is $-2$ on the midpoints of the sides of the
    triangle, and is $-1/2$ at the vertices.}
  \label{fig:z}
\end{figure}

The condition for marginal entropic stabilization,
Eq.~(\ref{eq:entstab}), for this model is then
\begin{equation}
  \ln N'_+ + \ln N'_0 + \ln N'_- = \ln N_+ + \ln N_0 + \ln N_-.
\end{equation}
This needs to be solved for the limiting value $z=z_*$, where
$z_*=1-1/\tau_*$.  Using Eqs.~(\ref{eq:dN}) and (\ref{eq:dNp}), this
becomes
\begin{equation}
  \left(1-\frac{z_*\alphabar}{2}+\frac{3z_*\betabar}{2}\right)
  \left(1+z_*\alphabar\right)
  \left(1-\frac{z_*\alphabar}{2}-\frac{3z_*\betabar}{2}\right) =
  \left(1-\frac{\alphabar}{2}+\frac{3\betabar}{2}\right)
  \left(1+\alphabar\right)
  \left(1-\frac{\alphabar}{2}-\frac{3\betabar}{2}\right).
\label{eq:diffes}
\end{equation}
This appears to be a cubic equation for $z_*$, but in fact $z_*=1$
(corresponding to $\tau_*\rightarrow\infty$) is clearly a root.  It is
an uninteresting root since it corresponds to no collision.  Once it is
removed, we are left with a quadratic for $z_*$, so the relevant
solution is easily written in closed form,
\begin{equation}
  z_* =
  \frac{3\alphabar^2 - \alphabar^3 + 9\betabar^2 + 9\alphabar\betabar^2 -
    \sqrt{3(\alphabar^2 + \alphabar^3 + 3\betabar^2 - 9\alphabar\betabar^2)
      (3\alphabar^2 - \alphabar^3 + 9\betabar^2 + 9\alphabar\betabar^2)}}
  {2(\alphabar^3 - 9\alphabar\betabar^2)}.
\label{eq:zstar}
\end{equation}
A contour plot of the function $z_*(\alphabar,\betabar)$ on the kinetic
polytope is displayed in Fig.~\ref{fig:z}.  There it can be seen that
$z_*$ has minimum value of $-2$ at the midpoints of the sides of the
triangle, has maximum value of $-1/2$ at the vertices, and approaches
$-1$ at the equilibrium $\alphabar=\betabar=0$.

The entropically stabilized collision operator is then given by
Eq.~(\ref{eq:dNp}), with $\tau\rightarrow\tau_*/\kappa$, or equivalently
\begin{equation}
z\rightarrow 1-\kappa (1-z_*).
\label{eq:z}
\end{equation}
It is useful to discuss three possible choices for $\kappa$:
\begin{itemize}
\item $\kappa\rightarrow 0$ means that $z\rightarrow 1$ and
  $\tau\rightarrow\infty$.  This is the uninteresting limit in which the
  collision operator vanishes and hydrodynamic behavior is lost.
\item $\kappa=1/2$ means that $z=(1+z_*)/2$ and $\tau=2/(1-z_*)$.  This
  is interesting because near equilibrium, where $z_*\sim -1$, it
  coincides with the prescription $\tau=1$.  This is the limit,
  mentioned in Subsection~\ref{ssec:esbgko}, wherein the outgoing state
  is the local equilibrium.  This is the smallest value of $\tau$ for
  which the standard lattice BGK algorithm is guaranteed to be stable.
  (In fairness, the standard algorithm is likely to be stable for
  smaller $\tau$; it's just that this is not guaranteed.)
\item $\kappa\rightarrow 1$ means that $z\rightarrow z_*$ and
  $\tau\rightarrow\tau_*$.  This is the largest value of $\tau$ for
  which the entropic lattice BGK algorithm is guaranteed to be stable.
\end{itemize}

\subsection{Attainable Transport Coefficient}

The complicated form of Eq.~(\ref{eq:zstar}) may lead one to believe
that the hydrodynamic equation (in this case, a diffusion equation)
would be very difficult to derive for the entropically stabilized
collision operator.  In fact, this is not the case at all.  The
Chapman-Enskog analysis that yields the hydrodynamic equations requires
only that one linearize the collision operator about the local
equilibrium.  Since the local equilibrium has $\alpha=\beta=0$, it is
clear from Eqs.~(\ref{eq:alcol}) and (\ref{eq:becol}), that only the
value of $z$ at equilibrium will enter.  Since $z_*\rightarrow -1$ at
the equilibrium, Eq.~(\ref{eq:z}) indicates that the transport
coefficient will be precisely that obtained by the standard lattice BGK
algorithm with $\tau=1/(2\kappa)$.

The Chapman-Enskog analysis of the standard lattice BGK algorithm is an
elementary exercise~\cite{bib:chapensk}.  The result for the diffusivity
in natural lattice units is
\begin{equation}
D = \frac{1}{3}\left(\tau-\frac{1}{2}\right).
\label{eq:dtau}
\end{equation}
Hence, by the above argument, the result for the diffusivity of the
entropic lattice BGK algorithm is
\begin{equation}
D = \frac{1}{6}\left(\frac{1}{\kappa}-1\right).
\label{eq:dkappa}
\end{equation}
From this, we can clearly see the benefit of the entropic algorithm.
The standard lattice BGK algorithm is not guaranteed stable for $\tau<1$
or $\kappa>1/2$, as discussed above.  From Eq.~(\ref{eq:dtau}) or
(\ref{eq:dkappa}), respectively, we see that this corresponds to a
minimum diffusivity of $1/6$.  By contrast, the entropic lattice BGK
algorithm loses stability only when $\kappa=1$, corresponding to $D=0$.

It is remarkable that we have found a stable, conservative, explicit
algorithm that {\it seems} to allow the transport coefficient to become
arbitrarily small.  After all, the analysis leading to the shear
viscosity of a fluid model is not very different from that leading to
the diffusivity above, and arbitrarily small shear viscosity would allow
for arbitrarily large Reynolds number.  In the field of computational
fluid dynamics, however, results that seem too good to be true usually
are just that.  For one thing, stability does not imply accuracy.  A
perfectly stable algorithm is not terribly useful if it does not
converge to the correct answer.  Flows at higher Reynolds number involve
ever smaller eddies, and at some point the lattice spacing becomes
insufficient to resolve these.  While it is comforting to have an
algorithm that does not lose stability in this situation, it is probably
a mistake to attach much physical significance to its results.

Another problem is that the time required for the system to come to
equilibrium goes to infinity as $\kappa\rightarrow 1$.  In fact, when
$\kappa=1$ the entropic lattice BGK collision operator is its own
inverse: If incoming state $\ket{\bfN}$ yields outgoing state
$\ket{\bfN'}$, then incoming state $\ket{\bfN'}$ would yield outgoing
state $\ket{\bfN}$.  This means that if the entire lattice were
initialized slightly away from equilibrium with no spatial gradients
whatsoever, it would simply thrash back and forth between two states,
without ever converging to the desired equilibrium.  When the time
required to achieve local equilibrium exceeds time scales of interest,
hydrodynamic behavior breaks down.

To illustrate this problem, we simulated the entropically stabilized
diffusion model with the initial density profile
\begin{equation}
\rho=\rho_0+\rho_1\sin\left(\frac{2\pi\ell x}{L}\right)
\label{eq:profile}
\end{equation}
on a periodic lattice.  We used lattice size $L=32$, and wavenumber
$\ell=3$ in all the simulations reported below.  We initialized the
state of each site in a local equilibrium $N_-=N_0=N_+=\rho/3$, without
the Chapman-Enskog correction due to gradient.  Now
Eq.~(\ref{eq:profile}) is a solution of the diffusion equation, if
$\rho_1$ decays in time as $\exp(-\gamma t)$, where
\begin{equation}
\gamma = D\left(\frac{2\pi\ell}{L}\right)^2.
\label{eq:diffcoef}
\end{equation}
So we fit our results to the functional form of Eq.~(\ref{eq:profile}),
measured $\rho_1(t)$, and did a least-squares fit of its logarithm to a
linear function of time in order to determine $\gamma$.  We then used
Eq.~(\ref{eq:diffcoef}) to get the diffusion coefficient, and we
compared this to the theoretical value provided by Eq.~(\ref{eq:dkappa})
for several different values of $\kappa$, approaching unity.

\begin{figure}
  \center{
    \mbox{
      \includegraphics[bbllx=0,bblly=53,bburx=288,bbury=235,width=3.0truein]{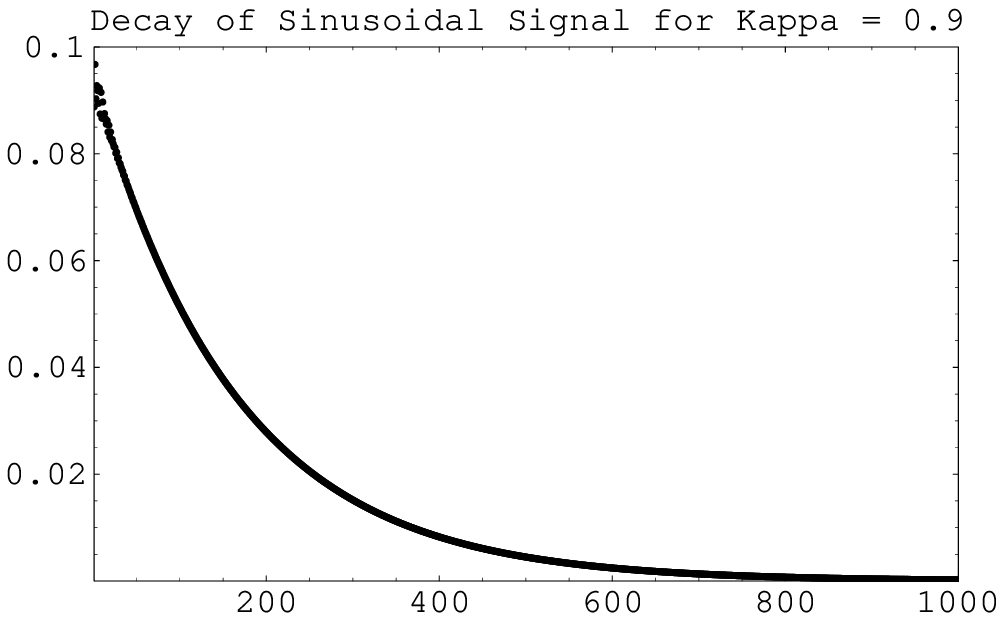}
      \includegraphics[bbllx=0,bblly=54,bburx=288,bbury=234,width=3.0truein]{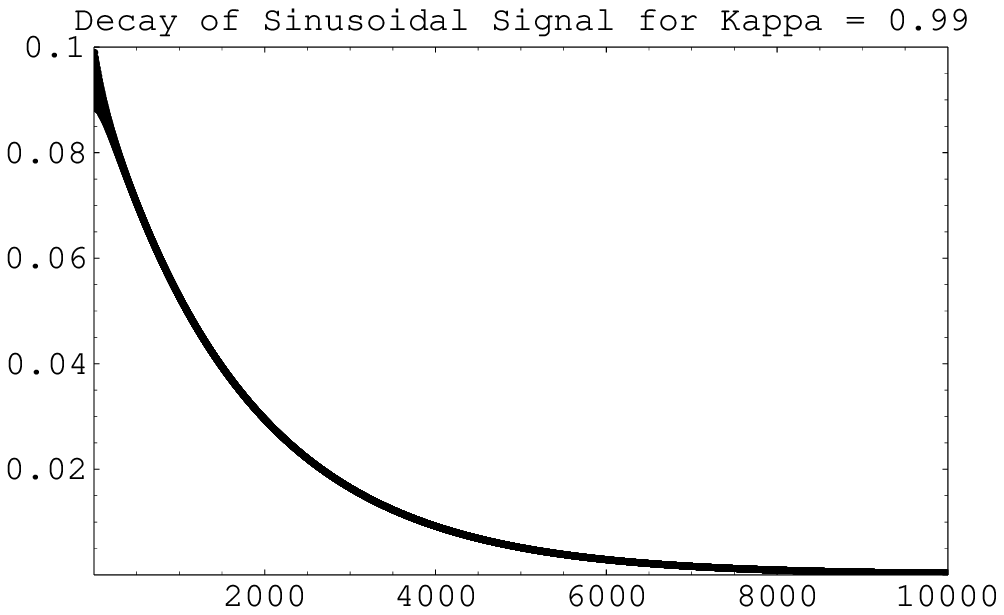}
      }}
  \caption{{\bf Decay of sinusoidal density profile} for $\kappa=0.9$ (left) and $\kappa=0.99$ (right).}
  \label{fig:decay90000}
\end{figure}

\begin{figure}
  \center{
    \mbox{
      \includegraphics[bbllx=0,bblly=52,bburx=288,bbury=236,width=3.0truein]{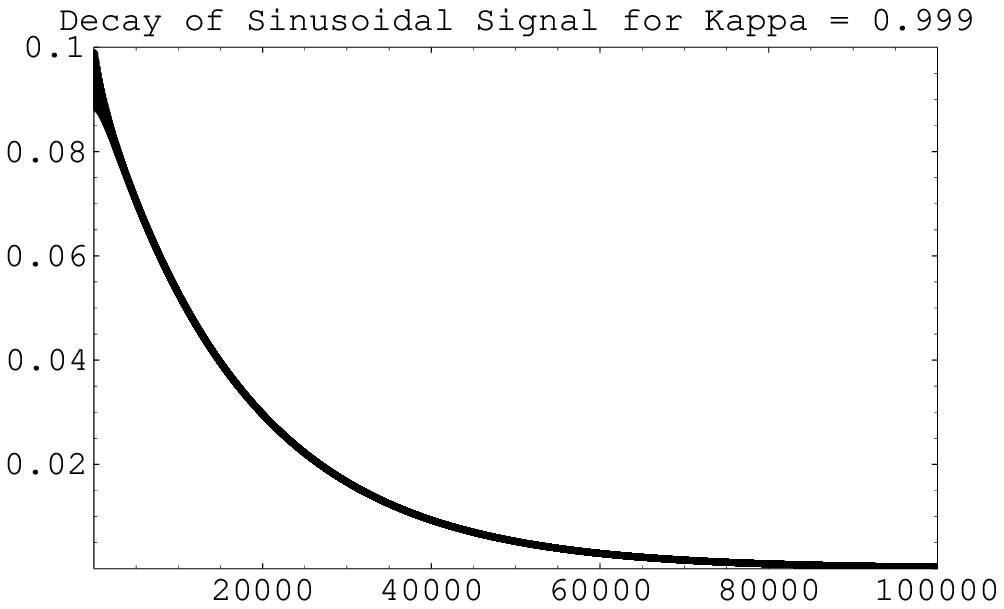}
      \includegraphics[bbllx=0,bblly=56,bburx=288,bbury=232,width=3.0truein]{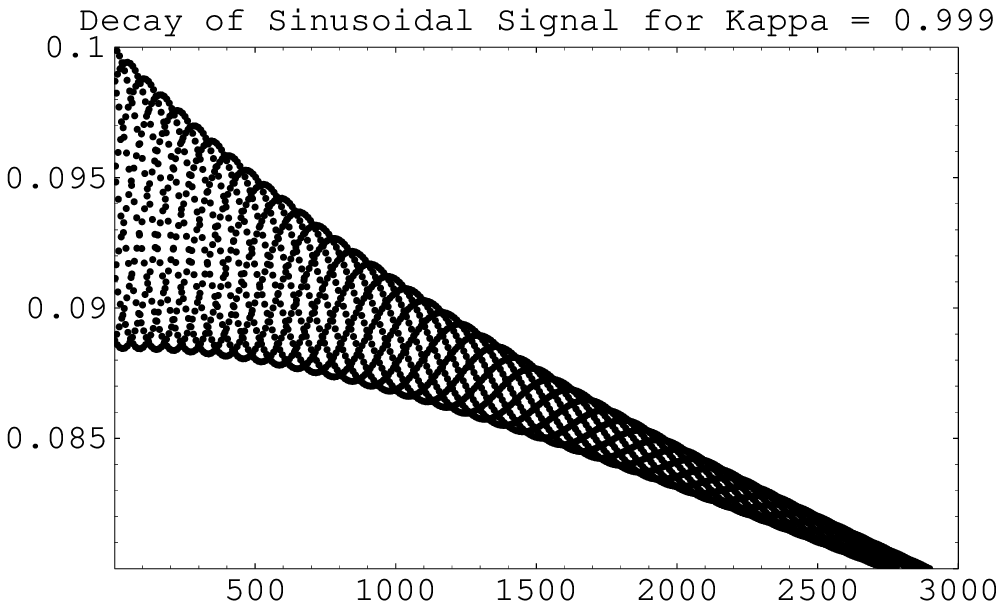}
      }}
  \caption{{\bf Decay of sinusoidal density profile} for $\kappa=0.999$}
  \label{fig:decay99900}
\end{figure}

\begin{figure}
  \center{
    \mbox{
      \includegraphics[bbllx=0,bblly=55,bburx=288,bbury=233,width=3.0truein]{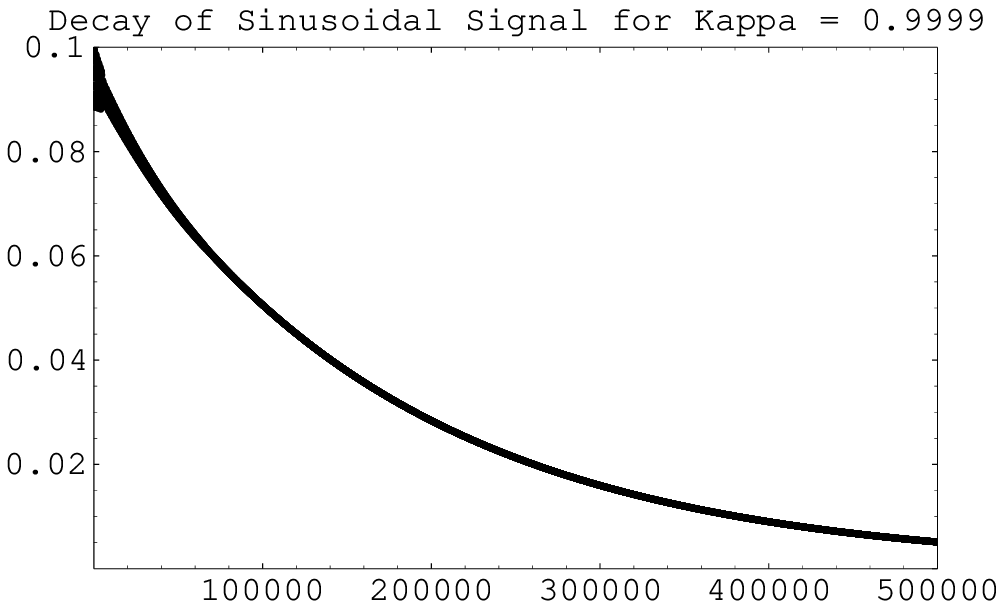}
      \includegraphics[bbllx=0,bblly=57,bburx=288,bbury=231,width=3.0truein]{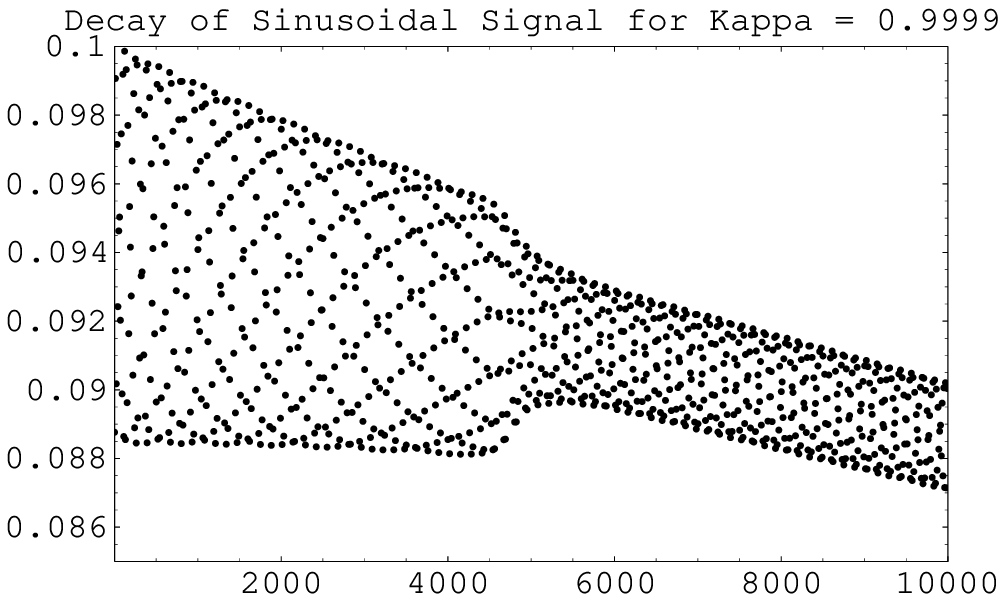}
      }}
  \caption{{\bf Decay of sinusoidal density profile} for $\kappa=0.9999$}
  \label{fig:decay99990}
\end{figure}

Fig.~\ref{fig:decay90000} shows the measured value of $\rho_1(t)$ for
$\kappa=0.9$ and $\kappa=0.99$.
As can be seen, there is an initial transient, due to the inadequacy of
the form used for the local equilibrium $N_-=N_0=N_+=\rho/3$ in the
presence of a spatial gradient.  The left-hand side of
Fig.~\ref{fig:decay99900} shows the measured value of $\rho_1(t)$ for
$\kappa=0.999$, and the right-hand side is an enlargement of the
transient region.  Fig.~\ref{fig:decay99990} shows the same things for
$\kappa=0.9999$.  Note that the transient period lengthens considerably
as $\kappa$ nears unity.  We could have substantially reduced these
transient periods had we used the Chapman-Enskog correction to the local
equilibrium; instead, we simply waited for the transient to die away
before measuring the decay constant $\gamma$.

\begin{figure}
  \center{
    \mbox{
      \includegraphics[bbllx=0,bblly=55,bburx=288,bbury=233,width=3.0truein]{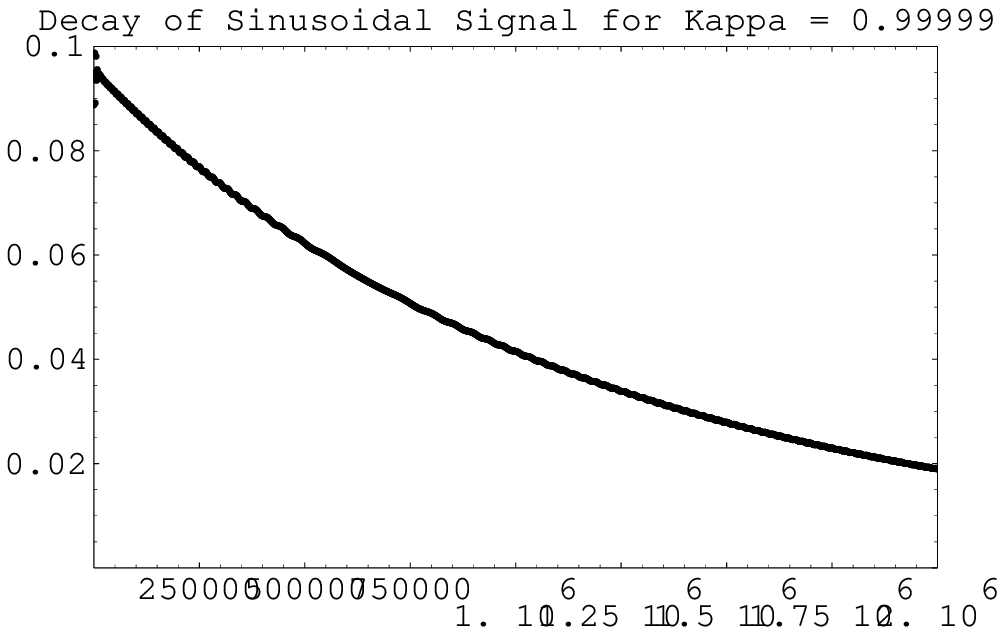}
      \includegraphics[bbllx=0,bblly=59,bburx=288,bbury=229,width=3.0truein]{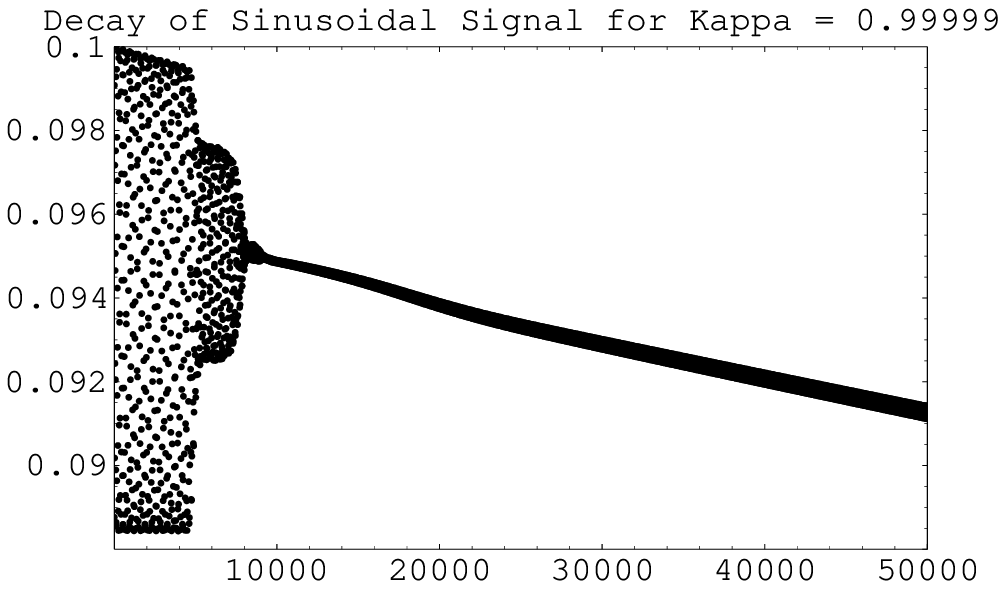}
      }\\
    \mbox{
      \includegraphics[bbllx=0,bblly=59,bburx=288,bbury=229,width=3.0truein]{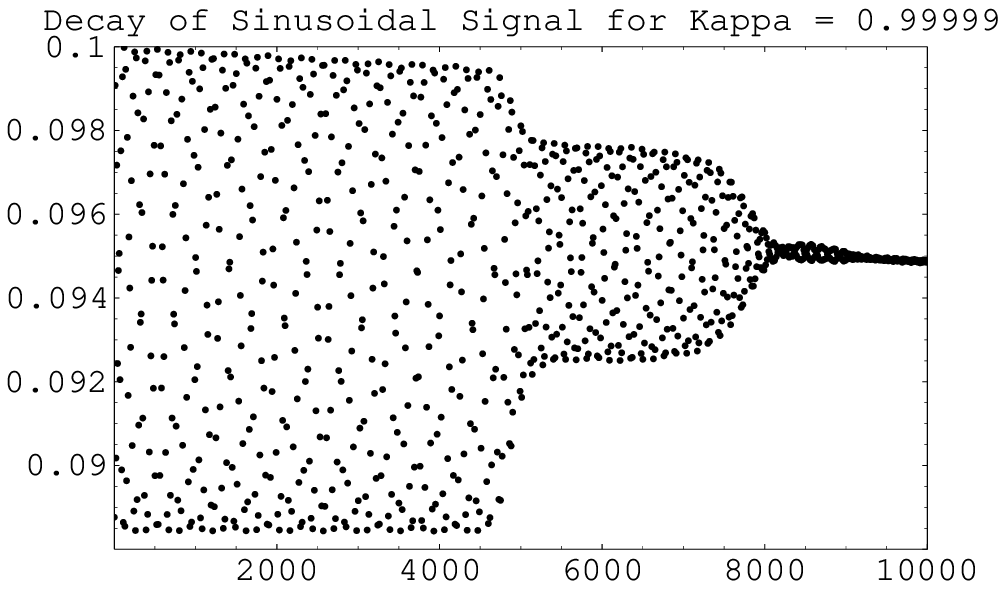}
      }}
  \caption{{\bf Decay of sinusoidal density profile} for $\kappa=0.99999$}
  \label{fig:decay99999}
\end{figure}

\begin{figure}
  \center{
    \mbox{
      \includegraphics[bbllx=0,bblly=55,bburx=288,bbury=233,width=3.0truein]{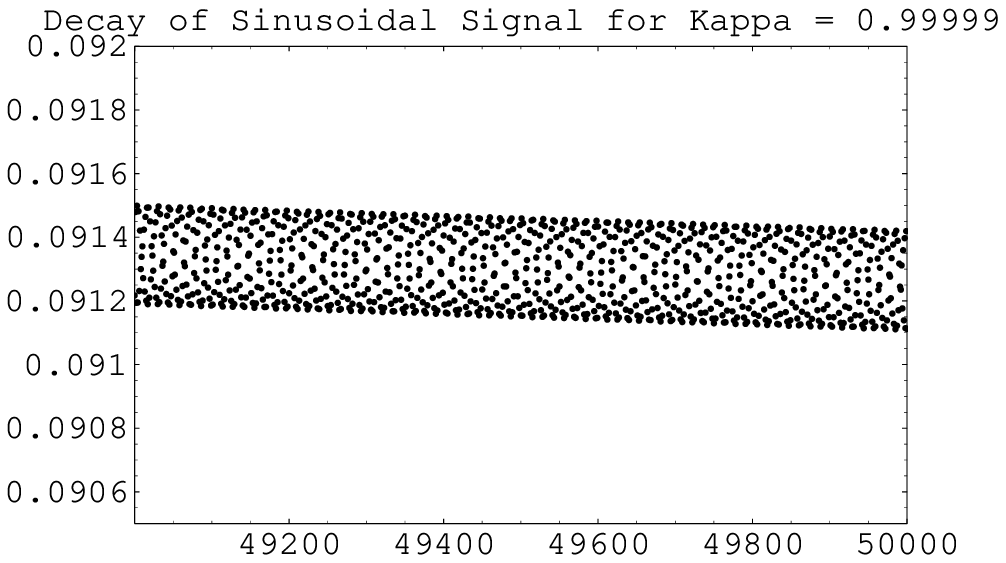}
      \includegraphics[bbllx=0,bblly=59,bburx=288,bbury=229,width=3.0truein]{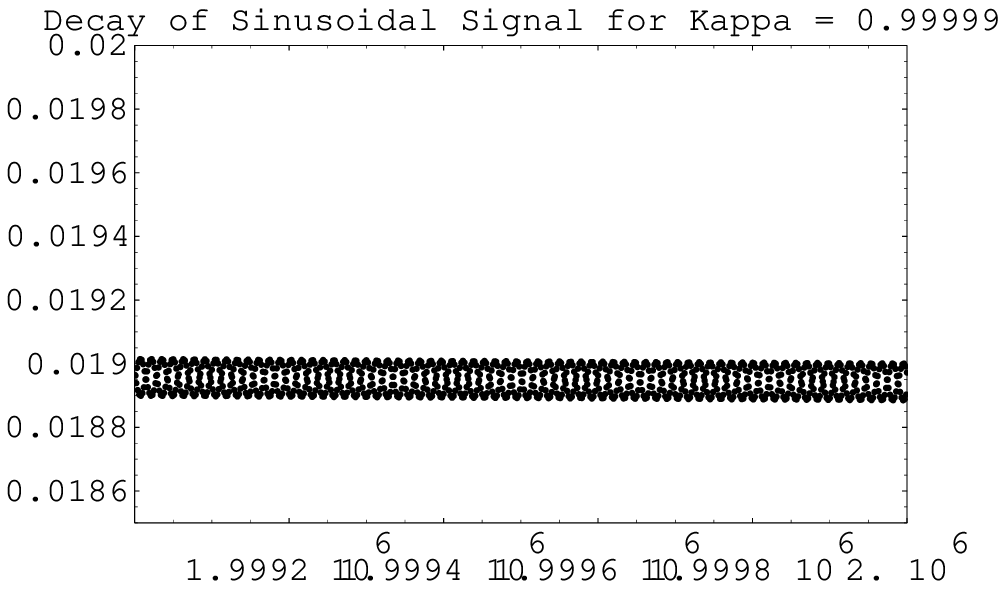}
      }}
  \caption{{\bf Decay of sinusoidal density profile} for $\kappa=0.99999$}
  \label{fig:decayblowup}
\end{figure}

The results for the diffusivity are displayed in Table~\ref{tab:decay}.
Note that we find reasonable agreement until we get to the cases
$\kappa=0.9999$ and $\kappa=0.99999$.  To see what is going wrong for
these cases, the upper-left-hand corner of Fig.~\ref{fig:decay99999}
shows the measured value of $\rho_1(t)$ for $\kappa=0.99999$, and the
upper-right-hand corner shows the same plot with a reduced range for the
ordinate.  In addition to the oddly shaped envelope of the initial
transient period, similar to that seen in Fig.~\ref{fig:decay99990}, and
enlarged in the lower part of this figure, we note that the plot of
$\rho_1(t)$ in the upper right seems to be increasing in thickness, even
after the initial transient.  Upon further enlargement, shown on the
left-hand side of Fig.~\ref{fig:decayblowup}, we see that this thickness
is in fact a subtle transient of extremely long duration.  The
right-hand side of the figure shows this same transient at the longest
time for which the simulation in Fig.~\ref{fig:decay99999} was run.
Though reduced in magnitude (the scales of the ordinates of both graphs
in Fig.~\ref{fig:decayblowup} are equal), the transient is still
present, even though most of the signal itself has diffused away.  We
believe that this transient is causing the anomaly in the measured
diffusivity.  A much longer simulation would be required to test this
assertion.
\begin{table}
  \center{
    \begin{tabular}{l|ll}
    $\kappa$ & $D_{\mbox{\scriptsize theory}}$ & $D_{\mbox{\scriptsize meas}}$\\
    \hline
    $0.9$     & $1.852\times 10^{-2}$ & $1.75364\times 10^{-2}$\\
    $0.99$    & $1.684\times 10^{-3}$ & $1.67419\times 10^{-3}$\\
    $0.999$   & $1.668\times 10^{-4}$ & $1.66674\times 10^{-4}$\\
    $0.9999$  & $1.667\times 10^{-5}$ & $1.66198\times 10^{-5}$\\
    $0.99999$ & $1.667\times 10^{-6}$ & $2.28751\times 10^{-6}$
    \end{tabular}}
  \caption{{\bf Theoretical and measured values of the diffusivity} for
    various values of $\kappa$.}
  \label{tab:decay}
\end{table}

To better understand this loss of separation between the kinetic and
hydrodynamic time scales, note that the time required for the signal to
decay away is
\begin{equation}
\tau_{\mbox{\scriptsize signal}}
\propto
\frac{1}{\gamma}
\propto
\frac{1}{D}
\propto
\frac{\kappa}{1-\kappa}.
\label{eq:tsignal}
\end{equation}
The time required for the transient to decay to $1/e$ of its initial
value, on the other hand, obeys
\begin{equation}
\kappa^{\tau_{\mbox{\scriptsize transient}}}=\frac{1}{e},
\end{equation}
whence
\begin{equation}
\tau_{\mbox{\scriptsize transient}}
\propto
\frac{1}{\ln\kappa}.
\label{eq:ttransient}
\end{equation}
For $\kappa=1-\epsilon$, where $\epsilon$ is small, the characteristic
times in Eqs.~(\ref{eq:tsignal}) and (\ref{eq:ttransient}) both go like
$1/\epsilon$.  Thus, the transients begin to linger for hydrodynamic
time scales, and it becomes impossible to separate the kinetic behavior
from the hydrodynamic behavior.

When the time required for the decay of transients is comparable to
hydrodynamic time scales of interest, the usefulness of the simulation
is questionable.  Thus, while Table~\ref{tab:decay} indicates good
agreement with theory, and the transport coefficient really does tend to zero
without loss of stability, one must be exceedingly careful in the
preparation of the initial condition to exploit this feature of entropic
lattice Boltzmann models.  In particular, transient behavior could be
dramatically reduced by including the Chapman-Enskog correction to
first-order in the gradient.  Indeed, if this problem turns out to be
the only obstacle to stable, conservative, explicit algorithms with
arbitrarily small transport coefficients, higher-order gradient
corrections may also be worth investigating.

\newpage
\section{One-Dimensional Compressible Fluid Model}
\label{sec:fluid}

In this section, we apply the entropic lattice Boltzmann method to a
simple five-velocity model of fluid dynamics in one dimension, first
considered by Renda et al. in 1997~\cite{bib:renda}.  We shall find that
the geometric picture is much richer than that for the diffusion model.
The master polytope for this model is four-dimensional, and we shall
show that the Fourier-Motzkin algorithm is very useful for describing
it.

\subsection{Description of Model}

As a second example, we consider a lattice Boltzmann model for a
one-dimensional compressible fluid with conserved mass, momentum and
energy, first studied by Renda et al.~\cite{bib:renda}.  The velocity
space of this model consists of five discrete values of velocity, namely
$0$, $\pm 1$, and $\pm 2$.  The single-particle distribution at site $x$
with velocity $j\in\{-2,-1,0,+1,+2\}$ is denoted by $N_j$.  As in the
last example, the state vector $\ket{\bfN}$ at a given site $x$ will be
denoted by a ket,
\begin{equation}
  \ket{\bfN}
  =
  \left(
    \begin{array}{l}
      N_{-2}\\
      N_{-1}\\
      N_{ 0}\\
      N_{+1}\\
      N_{+2}
    \end{array}
  \right),
\end{equation}
where we have suppressed the dependence on $x$ for simplicity.

A compressible fluid conserves mass, momentum and kinetic energy, so we
suppose that the corresponding densities
\begin{eqnarray}
  \rho &=& \sum_{j=-2}^{+2} N_j = \braket{\bfrho}{\bfN}\\
  \pi &=& \sum_{j=-2}^{+2} j\; N_j = \braket{\bfpi}{\bfN}\\
  \veps &=& \sum_{j=-2}^{+2} \frac{j^2}{2}\; N_j =
  \braket{\bfveps}{\bfN},
\end{eqnarray}
are conserved by the collisions.  Here we have introduced the bras
\begin{eqnarray}
  \begin{array}{ccc}
    \bra{\bfrho}_j &=& 1\\
    & & \\
    \bra{\bfpi}_j &=& j\\
    & & \\
    \bra{\bfveps}_j &=& j^2/2,
  \end{array}
\end{eqnarray}
or explicitly
\begin{eqnarray}
  \bra{\bfrho} &\equiv&
  \phantom{\frac{1}{2}}
  \left(
    \begin{array}{rrrrr}
      +1 & +1 & +1 & +1 & +1
    \end{array}
  \right)\\
  \bra{\bfpi} &\equiv&
  \phantom{\frac{1}{2}}
  \left(
    \begin{array}{rrrrr}
      -2 & -1 & \phantom{+}0 & +1 & +2
    \end{array}
  \right)\\
  \bra{\bfveps} &\equiv&
  \smallfrac{1}{2}
  \left(
    \begin{array}{rrrrr}
      +4 & +1 & \phantom{+}0 & +1 & +4
    \end{array}
  \right).\\
  \noalign{\hbox{\parbox{6.5truein}{These are the three hydrodynamic
        degrees of freedom in this example.  Because there are a total of
        five degrees of freedom, the other two must be kinetic in nature.
        To span these kinetic degrees of freedom and thereby make the bra basis
        complete, we introduce the linearly independent bras,}}}
  & & \nonumber\\
  \bra{\bfalpha} &=&
  \smallfrac{1}{2}
  \left(
    \begin{array}{rrrrr}
      \phantom{-}0 & +1 & \phantom{-}0 & +1 & \phantom{-}0
    \end{array}
  \right)\\
  \bra{\bfbeta} &=&
  \smallfrac{1}{2}
  \left(
    \begin{array}{rrrrr}
      \phantom{-}0 & -1 & \phantom{-}0 & +1 & \phantom{-}0
    \end{array}
  \right).
\end{eqnarray}
Next, we form a matrix of the five bras and invert it to get the dual
basis of kets,
\begin{eqnarray}
  \lefteqn{
    \left(
      \begin{array}{ccccc}
        \ket{\bfrho} & \ket{\bfpi} & \ket{\bfveps} & \ket{\bfalpha} & \ket{\bfbeta}
      \end{array}
    \right)
    = }\\
  & &
  \left(
    \begin{array}{c}
      \bra{\bfrho}\\ \bra{\bfpi}\\ \bra{\bfveps}\\ \bra{\bfalpha}\\ \bra{\bfbeta}
    \end{array}
  \right)^{-1}
  =
  \left[
    \frac{1}{2}
    \left(
      \begin{array}{rrrrr}
        +2 & +2 & +2 & +2 & +2\\
        -4 & -2 &  0 & +2 & +4\\
        +4 & +1 &  0 & +1 & +4\\
        0 & +1 &  0 & +1 &  0\\
        0 & -1 &  0 & +1 &  0
      \end{array}
    \right)
  \right]^{-1}
  =
  \frac{1}{4}
  \left(
    \begin{array}{rrrrr}
      0 & -1 & +1 & -1 & +2\\
      0 &  0 &  0 & +4 & -4\\
      +4 &  0 & -2 & -6 &  0\\
      0 &  0 &  0 & +4 & +4\\
      0 & +1 & +1 & -1 & -2
    \end{array}
  \right).
  \label{eq:fbasis}
\end{eqnarray}
From Eq.~(\ref{eq:fbasis}) we identify the linearly independent basis
kets, consisting of the hydrodynamic basis kets
\begin{equation}
  \begin{array}{lll}
    \ket{\bfrho} =
    \phantom{\frac{1}{4}}
    \left(
      \begin{array}{r}
        0 \\ 0 \\ +1 \\ 0 \\ 0
      \end{array}
    \right), &
    \ket{\bfpi} =
    \frac{1}{4}
    \left(
      \begin{array}{r}
        -1 \\ 0 \\ 0 \\ 0 \\ +1
      \end{array}
    \right), &
    \ket{\bfveps} =
    \frac{1}{4}
    \left(
      \begin{array}{r}
        +1 \\ 0 \\ -2 \\ 0 \\ +1
      \end{array}
    \right),
  \end{array}
\end{equation}
and the kinetic basis kets
\begin{equation}
  \begin{array}{ll}
    \ket{\bfalpha} =
    \frac{1}{4}
    \left(
      \begin{array}{r}
        -1 \\ +4 \\ -6 \\ +4 \\ -1
      \end{array}
    \right), &
    \ket{\bfbeta} =
    \frac{1}{2}
    \left(
      \begin{array}{r}
        +1 \\ -2 \\ 0 \\ +2 \\ -1
      \end{array}
    \right).
  \end{array}
\end{equation}
Unlike the previous example, there is no obvious correspondence between
the individual bra and ket basis elements.  This is because there is no
natural notion of a metric in this space.  Nevertheless, as noted in
Section~\ref{sec:elbm}, it is always possible to construct an entire ket
basis from an entire bra basis, and that is what we have done here.

We can then expand the state ket in this basis as follows
\begin{equation}
  \ket{\bfN}
  =
  \rho\ket{\bfrho}+\pi\ket{\bfpi}+\veps\ket{\bfveps}+
  \alpha\ket{\bfalpha}+\beta\ket{\bfbeta}
  =
  \rho
  \left(
    \begin{array}{c}
      \smallfrac{1}{4}\left(\vepsbar-\pibar-\alphabar+2\betabar\right) \\
      \alphabar - \betabar \\
      1 - \smallfrac{1}{2}\left(\vepsbar+3\alphabar\right) \\
      \alphabar + \betabar \\
      \smallfrac{1}{4}\left(\vepsbar+\pibar-\alphabar-2\betabar\right)
    \end{array}
  \right),
  \label{eq:fN}
\end{equation}
where we have continued to use an overbar to denote conserved quantities
per unit mass; for example, $\pibar\equiv\pi/\rho$ is the hydrodynamic
velocity.  As with the last example, this representation of the
distribution function $\ket{\bfN}$ has the virtue of making the
conserved quantities manifest.  In this case, we have the inverse
transformation
\begin{eqnarray}
  \rho &=& \braket{\bfrho}{\bfN} = +N_{-2}+N_{-1}+N_{0}+N_{+1}+N_{+2}\nonumber\\
  \pi &=& \braket{\bfpi}{\bfN} = -2N_{-2}-N_{-1}+N_{+1}+2N_{+2}\nonumber\\
  \veps &=& \braket{\bfveps}{\bfN} =
  +2N_{-2}+N_{-1}/2+N_{+1}/2+2N_{+2}\label{eq:finv}\\
  \alpha &=& \braket{\bfalpha}{\bfN} =
  +N_{-1}/2 + N_{+1}/2\nonumber\\
  \beta &=& \braket{\bfbeta}{\bfN} =
  -N_{-1}/2 + N_{+1}/2.\nonumber
\end{eqnarray}

\subsection{Nonnegativity}

We now consider the shape of the master and kinetic polytopes for the
compressible fluid model described in the last subsection.  Here we
shall find a much richer structure than the simple triangular region
that we found for the diffusive example in the last section above.
Referring to Eq.~(\ref{eq:fN}), we see that that nonnegativity of the
components of the single-particle distribution function is guaranteed by
the set of inequalities $\rho\geq 0$ and
\begin{eqnarray}
  0 &\leq & \vepsbar-\pibar-\alphabar+2\betabar\nonumber\\
  0 &\leq & \alphabar - \betabar\nonumber\\
  0 &\leq & 2-\vepsbar-3\alphabar\label{eq:fineqs}\\
  0 &\leq & \alphabar + \betabar\nonumber\\
  0 &\leq & \vepsbar+\pibar-\alphabar-2\betabar\nonumber.
\end{eqnarray}
These define the master polytope in the four dimensional $(\pibar,
\vepsbar, \alphabar, \betabar)$ space.  If we fix the hydrodynamic
parameters, $\pibar$ and $\vepsbar$, the above still constitute a set of
linear inequalities specifying the corresponding kinetic polytope in the
two dimensional $(\alphabar, \betabar)$ space.  Geometrically, these are
the intersections of the master polytope with the (hyper)planes of fixed
hydrodynamic parameters $\pibar$ and $\vepsbar$.  Thus, though the mass
density $\rho$ still scales out of the distribution function, the shape
of the kinetic polytopes in the $(\alphabar, \betabar)$ plane will
depend on the hydrodynamic velocity $\pibar$ and the kinetic energy per
unit mass $\vepsbar$.

\begin{figure}
  \center{
    \mbox{
      \includegraphics[bbllx=72,bblly=280,bburx=540,bbury=525,width=5.0truein]{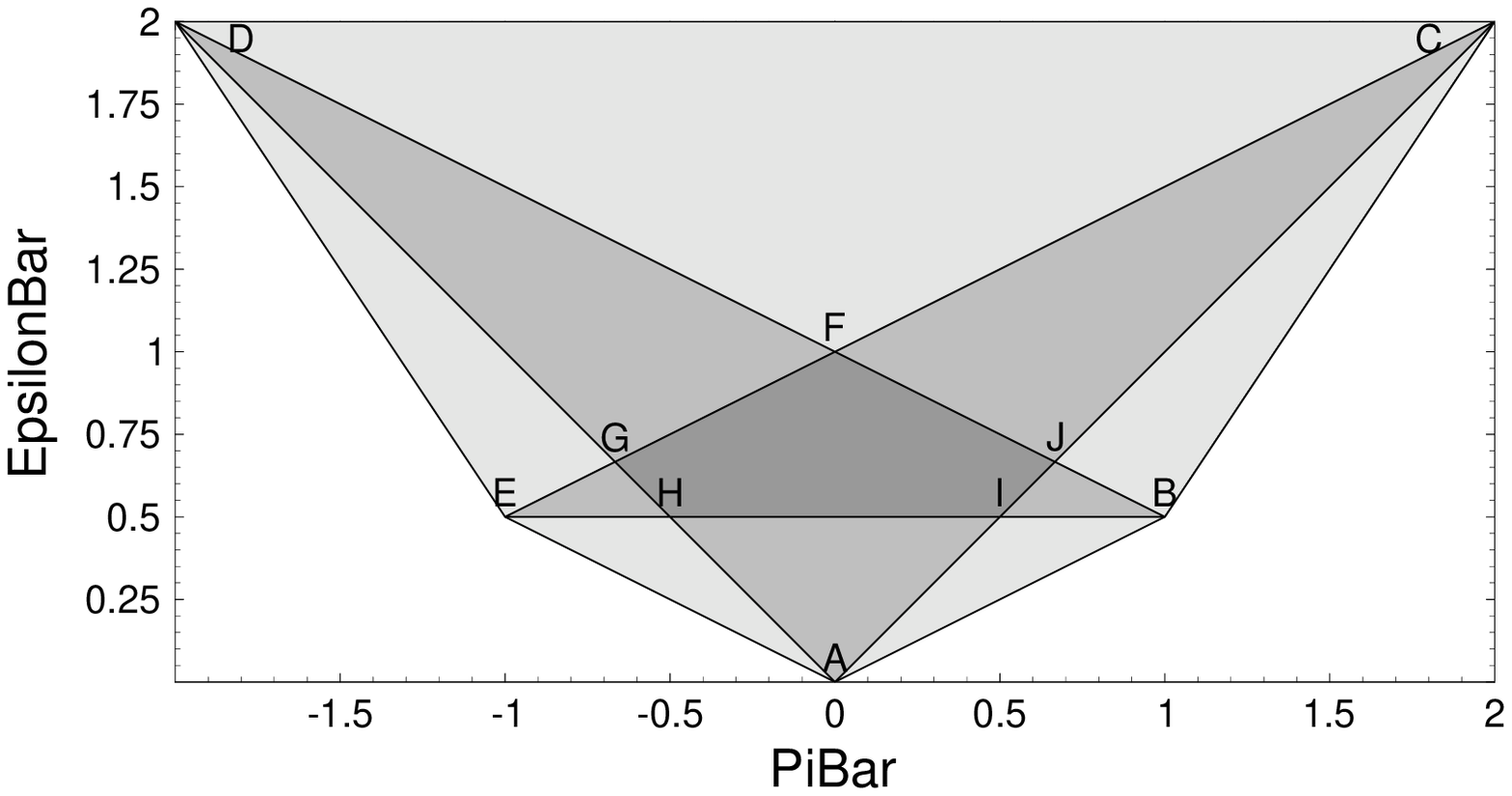}
      }}
  \caption{{\bf Bounds on $\pibar$ and $\vepsbar$:}  The regions with the
    lightest shading correspond to triangular regions in the
    $(\alphabar,\betabar)$ plane, those with intermediate shading correspond to
    quadrilateral regions in the $(\alphabar,\betabar)$ plane, and that with the
    darkest shading corresponds to pentagonal regions in the
    $(\alphabar,\betabar)$ plane.}
  \label{fig:fpolytope}
\end{figure}

It is not particularly easy to visualize the shape of the kinetic
polytope in the $(\alphabar,\betabar)$ plane for given hydrodynamic parameters
$\pibar$ and $\vepsbar$.  In fact, the task is tedious enough that we
have relegated it to Appendix~\ref{app:tour}, which takes the
reader on a detailed tour of the four-dimensional master polytope and
its kinetic polytope cross sections.  To summarize the conclusions of
that Appendix, the projection of the master polytope on the
$(\pibar,\vepsbar)$ plane is illustrated in Fig.~\ref{fig:fpolytope}.
Values of $(\pibar,\vepsbar)$ outside the shaded regions are not
possible.  The shape of the kinetic polytope is then a triangle when
$(\pibar,\vepsbar)$ is in the most lightly shaded regions of
Fig.~\ref{fig:fpolytope}, a quadrilateral when $(\pibar,\vepsbar)$ is in
the intermediately shaded regions of Fig.~\ref{fig:fpolytope}, and a
pentagon when $(\pibar,\vepsbar)$ is in the most darkly shaded region of
Fig.~\ref{fig:fpolytope}.  Illustrations and detailed descriptions of
the kinetic polytopes for particular values of $(\pibar,\vepsbar)$ in
each of the seven distinct regions with $\pibar\geq 0$ are provided in the
seven figures of Appendix~\ref{app:tour}; the chart below
Fig.~\ref{fig:fpolytope} indicates which figure corresponds to each of
its regions.  Finally, since the inequalities are invariant under
$(\pibar,\betabar)\rightarrow (-\pibar,-\betabar)$, the kinetic polytopes for $\pibar\leq 0$
are obtained from those for $\pibar\geq 0$ by simply reflecting in $\betabar$.

\subsection{Fourier-Motzkin Analysis for the 1D Compressible Fluid Model}

To compute the master polytope for the compressible fluid model, we cast
the inequalities of Eqs.~(\ref{eq:fineqs}) into matrix format as
follows,
\begin{equation}
  \left(
    \begin{array}{rrrrr}
      -1 & +2 & -1 & +1 &  0\\
      +1 & -1 &  0 &  0 &  0\\
      -3 &  0 &  0 & -1 & +2\\
      +1 & +1 &  0 &  0 &  0\\
      -1 & -2 & +1 & +1 &  0
    \end{array}
  \right)
  \left(
    \begin{array}{c}
      \alphabar\\
      \betabar\\
      \pibar\\
      \vepsbar\\
      1
    \end{array}
  \right)
  \geq 0.
  \label{eq:fineqsm}
\end{equation}
Here we have put the hydrodynamic parameters $\pibar$ and $\vepsbar$
{\it after} the kinetic parameters $\alphabar$ and $\betabar$ in the column
vector of unknowns, for reasons that will become apparent in a moment.
Performing the Fourier-Motzkin algorithm on this set of inequalities,
we finally arrive at the 23 inequalities,
\begin{equation}
  \left(
    \begin{array}{rrrrr}
      -1 & -2 & +1 & +1 &  0\\
      -1 & +2 & -1 & +1 &  0\\
      -1 &  0 &  0 & -\smallfrac{1}{3} & +\smallfrac{2}{3}\\
      +1 & -1 &  0 &  0 &  0\\
      +1 & +1 &  0 &  0 &  0\\
      0 & -1 & +1 & +1 &  0\\
      0 & -1 & +\smallfrac{1}{3} & +\smallfrac{1}{3} &  0\\
      0 & -1 &  0 & -\smallfrac{1}{3} & +\smallfrac{2}{3}\\
      0 & +1 & -1 & +1 &  0\\
      0 & +1 & -\smallfrac{1}{3} & +\smallfrac{1}{3} &  0\\
      0 & +1 &  0 & -\smallfrac{1}{3} & +\smallfrac{2}{3}\\
      0 &  0 & -1 & +2 &  0\\
      0 &  0 & -1 & +\smallfrac{2}{3} & +\smallfrac{2}{3}\\
      0 &  0 & -1 &  0 & +2\\
      0 &  0 & +1 & +2 &  0\\
      0 &  0 & +1 & +\smallfrac{2}{3} & +\smallfrac{2}{3}\\
      0 &  0 & +1 &  0 & +2\\
      0 &  0 &  0 & -1 & +2\\
      0 &  0 &  0 & +1 & +4\\
      0 &  0 &  0 & +1 & +1\\
      0 &  0 &  0 & +1 & +\smallfrac{1}{4}\\
      0 &  0 &  0 & +1 &  0\\
      0 &  0 &  0 &  0 & +1
    \end{array}
  \right)
  \left(
    \begin{array}{c}
      \alphabar\\
      \betabar\\
      \pibar\\
      \vepsbar\\
      1
    \end{array}
  \right)
  \geq 0.
  \label{eq:fineqsmfm}
\end{equation}
The last of these is a consistency condition which is seen to be
satisfied.  The five inequalities prior to that yield
\[
\max\left(0,-\smallfrac{1}{4},-1,-4\right)
\leq\vepsbar\leq 2,
\]
or
\begin{equation}
  0\leq\vepsbar\leq 2,
\end{equation}
which explains the extent of Fig.~\ref{fig:fpolytope} along the
ordinate.  The six inequalities prior to that yield
\[
\max\left(
  -2,
  -\smallfrac{2}{3}-\smallfrac{2}{3}\vepsbar,
  -2\vepsbar
\right)
\leq \pibar\leq
\min\left(
  +2,
  +\smallfrac{2}{3}+\smallfrac{2}{3}\vepsbar,
  +2\vepsbar
\right),
\]
or
\begin{equation}
  \left| \pibar\right|\leq
  \min\left(
    2,
    \smallfrac{2}{3}+\smallfrac{2}{3}\vepsbar,
    2\vepsbar
  \right),
\end{equation}
which explains the extent of Fig.~\ref{fig:fpolytope} along the
abscissa.  Thus, we have bounded the projection of the master polytope
in the hydrodynamic parameter space.

Given hydrodynamic parameters in the allowed region thus determined,
the first 11 inequalities of Eq.~(\ref{eq:fineqsmfm}) then specify the
kinetic polytopes.  (This is why we put the hydrodynamic parameters
after the kinetic ones in the column vector of unknowns.)  Continuing
our back substitution, the bounds on $\betabar$ are
\begin{equation}
  \max\left(
    -\smallfrac{2-\vepsbar}{3},
    -\vepsbar+\pibar,
    -\smallfrac{\vepsbar-\pibar}{3}
  \right)
  \leq
  \betabar
  \leq
  \min\left(
    +\smallfrac{2-\vepsbar}{3},
    +\vepsbar+\pibar,
    +\smallfrac{\vepsbar+\pibar}{3}
  \right).
  \label{eq:betabounds}
\end{equation}
Finally, the bounds on $\alphabar$ are
\begin{equation}
  \max\left(
    \betabar, -\betabar
  \right)\leq
  \alphabar\leq
  \min\left(
    +2\betabar-\pibar+\vepsbar,
    -2\betabar+\pibar+\vepsbar
  \right).
  \label{eq:alphabounds}
\end{equation}
Eqs.~(\ref{eq:betabounds}) and (\ref{eq:alphabounds}) specify the shape
of the kinetic polytope in the $(\alphabar,\betabar)$ plane, given the
hydrodynamic parameters $\pibar$ and $\vepsbar$.  In fact, they provide
a more succinct way of determining and expressing the bounds of both the
master polytope and the kinetic polytopes than the analysis presented in
Appendix~\ref{app:tour}.

\newpage
\section{Two and Three-Dimensional Fluid Dynamics}
\label{sec:fhpfchc}

In this section, we apply our methodology to much richer models, namely
those most commonly used for two and three-dimensional Navier-Stokes
hydrodynamics.  Thought the Fourier-Motzkin method cannot easily be
applied to the latter, we are able to present a valid collision
representation for both examples.  We indicate how a computer simulation
for such models might be implemented.

\subsection{FHP Model}

The diffusion and compressible fluid models that we considered above had
three and five particle velocities, respectively.  Such models are of
academic and pedagogical interest, but not particularly useful as
simulational tools.  The smallest model that has been used for serious
computational fluid dynamics calculations in two dimensions is called
the FHP model.  This is a six-velocity model on a triangular grid which
has been shown to yield isotropic, incompressible Navier-Stokes flow in
the appropriate scaling limit~\cite{bib:fhp,bib:swolf}.  The six
velocities are given by
\begin{equation}
  \bfc_j =
  \hat{\bfx}\cos\left(\frac{2\pi j}{6}\right) +
  \hat{\bfy}\sin\left(\frac{2\pi j}{6}\right),
\end{equation}
for $0\leq j\leq 5$.  The model conserves mass with density
\begin{equation}
  \rho = \sum_{j=0}^5 N_j
  \label{eq:masscnstrnt}
\end{equation}
and momentum with density
\begin{equation}
  \bfpi = \sum_{j=0}^5 N_j \bfc_j,
  \label{eq:momcnstrnt}
\end{equation}
so the hydrodynamic bras are
\begin{eqnarray}
  \bra{\bfrho} &=&
  \left(
    \begin{array}{cccccc}
      +1 & +1 & +1 & +1 & +1 & +1
    \end{array}
  \right)\label{eq:fhpfirstbra}\\
  \bra{\bfpi_x} &=&
  \left(
    \begin{array}{cccccc}
      +1 & +\smallfrac{1}{2} & -\smallfrac{1}{2} &
      -1 & -\smallfrac{1}{2} & +\smallfrac{1}{2}
    \end{array}
  \right)\\
  \bra{\bfpi_y} &=&
  \left(
    \begin{array}{cccccc}
      0 & +\smallfrac{\sqrt{3}}{2} & +\smallfrac{\sqrt{3}}{2} &
      0 & -\smallfrac{\sqrt{3}}{2} & -\smallfrac{\sqrt{3}}{2}
    \end{array}
  \right)\\
  \noalign{\hbox{\parbox{6.5truein}{and a reasonable choice for the
        kinetic bras might be}}}
  & & \nonumber\\
  \bra{\bfalpha} &=&
  \left(
    \begin{array}{cccccc}
      +1 & -1 & +1 & -1 & +1 & -1
    \end{array}
  \right)\\
  \bra{\bfbeta} &=&
  \left(
    \begin{array}{cccccc}
      -1 & +\smallfrac{1}{2} & +\smallfrac{1}{2} &
      -1 & +\smallfrac{1}{2} & +\smallfrac{1}{2}
    \end{array}
  \right)\\
  \bra{\bfgamma} &=&
  \left(
    \begin{array}{cccccc}
      0 & +\smallfrac{\sqrt{3}}{2} & -\smallfrac{\sqrt{3}}{2} &
      0 & +\smallfrac{\sqrt{3}}{2} & -\smallfrac{\sqrt{3}}{2}
    \end{array}
  \right).\label{eq:fhplastbra}\\
\end{eqnarray}
The kets can then be found as with the preceeding examples; the
hydrodynamic kets are
\begin{equation}
  \begin{array}{lll}
    \ket{\bfrho} = \smallfrac{1}{6}
    \left(
      \begin{array}{r}
        +1\\
        +1\\
        +1\\
        +1\\
        +1\\
        +1
      \end{array}
    \right),
    &
    \ket{\bfpi_x} = \smallfrac{1}{6}
    \left(
      \begin{array}{r}
        +2\\
        +1\\
        -1\\
        -2\\
        -1\\
        +1
      \end{array}
    \right),
    &
    \ket{\bfpi_y} = \smallfrac{\sqrt{3}}{6}
    \left(
      \begin{array}{r}
        0\\
        +1\\
        +1\\
        0\\
        -1\\
        -1
      \end{array}
    \right),
  \end{array}
\end{equation}
and the kinetic kets are
\begin{equation}
  \begin{array}{lll}
    \ket{\bfalpha} = \smallfrac{1}{6}
    \left(
      \begin{array}{r}
        +1\\
        -1\\
        +1\\
        -1\\
        +1\\
        -1
      \end{array}
    \right),
    &
    \ket{\bfbeta} = \smallfrac{1}{6}
    \left(
      \begin{array}{r}
        -2\\
        +1\\
        +1\\
        -2\\
        +1\\
        +1
      \end{array}
    \right),
    &
    \ket{\bfgamma} = \smallfrac{\sqrt{3}}{6}
    \left(
      \begin{array}{r}
         0\\
        +1\\
        -1\\
         0\\
        +1\\
        -1
      \end{array}
    \right).
  \end{array}
\end{equation}

The general representation of the state of the FHP model is then
\begin{eqnarray}
  \ket{\bfN} &=&
  \rho \ket{\bfrho} +
  \pi_x\ket{\bfpi_x} +
  \pi_y\ket{\bfpi_y} +
  \alpha\ket{\bfalpha} +
  \beta \ket{\bfbeta} +
  \gamma\ket{\bfgamma}\nonumber\\
  &=&
  \smallfrac{\rho}{6}
  \left(
    \begin{array}{c}
      1+2\pibar_x                 +\alphabar-2\betabar
      \\
      1+ \pibar_x+\sqrt{3}\pibar_y-\alphabar+ \betabar+\sqrt{3}\gammabar
      \\
      1- \pibar_x+\sqrt{3}\pibar_y+\alphabar+ \betabar-\sqrt{3}\gammabar
      \\
      1-2\pibar_x                 -\alphabar-2\betabar
      \\
      1- \pibar_x-\sqrt{3}\pibar_y+\alphabar+ \betabar+\sqrt{3}\gammabar
      \\
      1+ \pibar_x-\sqrt{3}\pibar_y-\alphabar+ \betabar-\sqrt{3}\gammabar
    \end{array}
  \right).
\label{eq:fhprep}
\end{eqnarray}
In addition, the factors of $\sqrt{3}$ could be absorbed into $\pibar_y$
and $\gammabar$ to further simplify the final expression.

\subsection{FCHC Model}

Early attempts to generalize the FHP model to obtain a three-dimensional
lattice model of the Navier-Stokes equations were thwarted by the
observation that no regular three-dimensional lattice would yield the
required isotropy -- as the triangular lattice did in two dimensions.
The problem was solved by noting that a lattice with the required
isotropy existed in four dimensions, and its three-dimensional
projection could be used to construct an isotropic
model~\cite{bib:fchc}.  The four dimensional lattice is called the
Face-Centered Hypercubic (FCHC) lattice.  It is a self-dual lattice with
coordination number 24.  It can be defined as all sites $(i,j,k,l)$ on a
Cartesian lattice such that $i+j+k+l$ is even.  The 24 lattice vectors
-- or, equivalently, the 24 lattice sites that neighbor the origin
$(0,0,0,0)$ -- are enumerated in Table~\ref{tab:fchcvecs}.
\begin{table}
  \center{
    \begin{tabular}{r|r||cccc}
      \multicolumn{2}{c||}{        } & ${c_j}_w$ & ${c_j}_x$ & ${c_j}_y$ & ${c_j}_z$\\
      \hline
      \hline
      &  1 &  0 & +1 & +1 &  0\\
      &  2 &  0 & +1 & -1 &  0\\
      &  3 &  0 & -1 & +1 &  0\\
      &  4 &  0 & -1 & -1 &  0\\
      &  5 & +1 &  0 &  0 & +1\\
      &  6 & -1 &  0 &  0 & +1\\
      &  7 & +1 &  0 &  0 & -1\\
      &  8 & -1 &  0 &  0 & -1\\
      \cline{2-6}
      &  9 &  0 & +1 &  0 & +1\\
      & 10 &  0 & +1 &  0 & -1\\
      & 11 &  0 & -1 &  0 & +1\\
      $j$ & 12 &  0 & -1 &  0 & -1\\
      & 13 & +1 &  0 & +1 &  0\\
      & 14 & -1 &  0 & +1 &  0\\
      & 15 & +1 &  0 & -1 &  0\\
      & 16 & -1 &  0 & -1 &  0\\
      \cline{2-6}
      & 17 &  0 &  0 & +1 & +1\\
      & 18 &  0 &  0 & +1 & -1\\
      & 19 &  0 &  0 & -1 & +1\\
      & 20 &  0 &  0 & -1 & -1\\
      & 21 & +1 & +1 &  0 &  0\\
      & 22 & -1 & +1 &  0 &  0\\
      & 23 & +1 & -1 &  0 &  0\\
      & 24 & -1 & -1 &  0 &  0
    \end{tabular}
    }
  \caption{Lattice vectors of the FCHC model.}
  \label{tab:fchcvecs}
\end{table}

To project these lattice vectors to three-dimensional space, we simply
ignore the ficticious $l$ coordinate.  When we do this, we note that 6
of the resulting three-dimensional lattice vectors will have two
preimages from the original set of 24 FCHC lattice vectors.  Rather than
maintain these as separate three-dimensional vectors (as is done in
lattice-gas studies), it is possible to combine them in lattice
Boltzmann studies, simply weighting contributions from those directions
by a factor of two.  In fact, this weighting factor can be thought of as
a direction-dependent particle mass $m_j$, so that the three-dimensional
projection of the FCHC model is as presented in
Table~\ref{tab:fchc3dvecs}.
\begin{table}
  \center{
    \begin{tabular}{r|r||c|ccc}
      \multicolumn{2}{c||}{        } & $m_j$ & ${c_j}_x$ & ${c_j}_y$ & ${c_j}_z$\\
      \hline
      \hline
      &  1 &  1 & +1 & +1 &  0\\
      &  2 &  1 & +1 & -1 &  0\\
      &  3 &  1 & -1 & +1 &  0\\
      &  4 &  1 & -1 & -1 &  0\\
      &  5 &  2 &  0 &  0 & +1\\
      &  6 &  2 &  0 &  0 & -1\\
      \cline{2-6}
      &  7 &  1 & +1 &  0 & +1\\
      &  8 &  1 & +1 &  0 & -1\\
      $j$ &  9 &  1 & -1 &  0 & +1\\
      & 10 &  1 & -1 &  0 & -1\\
      & 11 &  2 &  0 & +1 &  0\\
      & 12 &  2 &  0 & -1 &  0\\
      \cline{2-6}
      & 13 &  1 &  0 & +1 & +1\\
      & 14 &  1 &  0 & +1 & -1\\
      & 15 &  1 &  0 & -1 & +1\\
      & 16 &  1 &  0 & -1 & -1\\
      & 17 &  2 & +1 &  0 &  0\\
      & 18 &  2 & -1 &  0 &  0
    \end{tabular}
    }
  \caption{Lattice vectors and masses of the three-dimensional projection of the FCHC model.}
  \label{tab:fchc3dvecs}
\end{table}

The model is required to conserve mass and three components of momentum,
so that the four hydrodynamic bras are given by
\begin{eqnarray}
  \bra{\bfrho}_j  &=& m_j\\
  \bra{\bfpi_x}_j &=& m_j {c_j}_x\\
  \bra{\bfpi_y}_j &=& m_j {c_j}_y\\
  \bra{\bfpi_z}_j &=& m_j {c_j}_z.
\end{eqnarray}
These are shown in the first four rows of Table~\ref{tab:fchc3drows} in
Appendix~\ref{app:fchc}, followed by a choice of twenty linearly
independent kinetic bras.  The corresponding kets are shown in
Table~\ref{tab:fchc3dcols} in Appendix~\ref{app:fchc}, and these can be
used to construct a general analytic form for the state ket,
$\ket{\bfN}$.

\subsection{Attainable Transport Coefficient}

As with the one-dimensional diffusion model, we expect that the entropic
lattice Boltzmann model will allow us to reduce the transport
coefficient of the model.  For fluids, this would mean smaller shear
viscosity and hence higher Reynolds number.  To see if this is possible,
we consider the FHP model in the incompressible limit.  In the
incompressible limit, the Mach number scales like the Knudsen number, so
the Chapman-Enskog method involves only the equilibrium distribution at
zero momentum.  Taking $\zeta_j(x)=x$, it is straightforward to
extremize $h$ to find the equilibrium distribution.  This may be done
perturbatively in Mach (equivalently, Knudsen) number, and the result to
tenth order is
\begin{eqnarray}
\alphabar^{\mbox{\scriptsize eq}} &=&
 2\pi_x
 \left(\pi_x^2-3\pi_y^2\right)
 \left(1-\pi_x^2-\pi_y^2\right)^2
 \left(1+2\pi_x^2+2\pi_y^2\right)+
 \calO\left(\pi^{10}\right)\label{eq:fhpaeq}\\
\betabar^{\mbox{\scriptsize eq}} &=&
 \pi_y^2-\pi_x^2 -
 \left(\pi_x^4 - 6\pi_x^2\pi_y^2 + \pi_y^4\right) +\nonumber\\
 & &
 2\left(\pi_x^6 - 5\pi_x^4\pi_y^2 - 5\pi_x^2\pi_y^4 + \pi_y^6\right) +
 \left(\pi_x^2 + \pi_y^2\right)^2 
 \left(3\pi_x^4 + 6\pi_x^2\pi_y^2 - 5\pi_y^4\right)+
 \calO\left(\pi^{10}\right)\label{eq:fhpbeq}\\
\gammabar^{\mbox{\scriptsize eq}} &=&
 2\pi_x\pi_y 
 \left[
  1 -
  2\left(\pi_x^2 - \pi_y^2\right) +
  4\left(\pi_x^4 - \pi_y^4\right) -
  2\left(\pi_x^2 + \pi_y^2\right)^2\left(3\pi_x^2 + \pi_y^2\right)
 \right]+\calO\left(\pi^{10}\right).\label{eq:fhpceq}
\end{eqnarray} 
Thus the equilibrium values of $\betabar$ and $\gammabar$ are of second
order, while that of $\alphabar$ is actually third order.  For the
computation of viscosity in the incompressible limit, the Chapman-Enskog
analysis requires the linearized collision operator only to zeroth order
in the Mach number.  Thus, for the purposes of working out the
attainable viscosity, we may simply set the equilibrium values of all
three kinetic parameters to zero.  It follows that the outgoing kinetic
parameters are equal to the incoming ones multiplied by $z=1-1/\tau$.

We can now construct an entropic lattice Boltzmann model.  Using
Eq.~(\ref{eq:fhprep}) for guidance, we see that the analog of
Eq.~(\ref{eq:diffes}) for $z_*$ is

{\footnotesize
\begin{eqnarray}
 &\left(1+2\pibar_x                 +\alphabar-2\betabar\right)
  \left(1+ \pibar_x+\sqrt{3}\pibar_y-\alphabar+ \betabar+\sqrt{3}\gammabar\right)
  \left(1- \pibar_x+\sqrt{3}\pibar_y+\alphabar+ \betabar-\sqrt{3}\gammabar\right)
\phantom{=aaaaaaaaaaaaaaaaaai}&\nonumber\\
 &\left(1-2\pibar_x                 -\alphabar-2\betabar\right)
  \left(1- \pibar_x-\sqrt{3}\pibar_y+\alphabar+ \betabar+\sqrt{3}\gammabar\right)
  \left(1+ \pibar_x-\sqrt{3}\pibar_y-\alphabar+ \betabar-\sqrt{3}\gammabar\right)
=\phantom{aaaaaaaaaaaaaaaaaai}&\nonumber\\
 &\left(1+2\pibar_x                 +z_*\alphabar-2z_*\betabar\right)
  \left(1+ \pibar_x+\sqrt{3}\pibar_y-z_*\alphabar+ z_*\betabar+\sqrt{3}z_*\gammabar\right)
  \left(1- \pibar_x+\sqrt{3}\pibar_y+z_*\alphabar+ z_*\betabar-\sqrt{3}z_*\gammabar\right)
\phantom{a}&\nonumber\\
 &\left(1-2\pibar_x                 -z_*\alphabar-2z_*\betabar\right)
  \left(1- \pibar_x-\sqrt{3}\pibar_y+z_*\alphabar+ z_*\betabar+\sqrt{3}z_*\gammabar\right)
  \left(1+ \pibar_x-\sqrt{3}\pibar_y-z_*\alphabar+ z_*\betabar-\sqrt{3}z_*\gammabar\right),
 \label{eq:fhpzsol}
\end{eqnarray}}

where it bears repeating that, just for the purposes of this computation
of viscosity, we have neglected corrections to the equilibrium of order
Mach number squared~\footnote{Note that these terms of order Mach number
  squared are necessary to derive the remainder of the Navier-Stokes
  equations.}.  This is a sixth-order equation for $z_*$; since $z_*=1$
is clearly a root, it can be reduced to fifth order.  In spite of the
fact that fifth order equations generally do not admit solutions in
radicals, this one does happen to do so.  The result, however, is
complicated and uninspiring, and so we omit it here.  The interested
reader can simply type the above equation into, e.g., ${\mbox{\it
    Mathematica}}^\copyright$, and use the {\tt Solve} utility to see
the result.  The important point is that the relevant root for $z_*$
approaches $-1$ as $\alphabar,\betabar,\gammabar\rightarrow 0$.

The rest of the analysis is exactly as it was for the one-dimensional
diffusion model.  We set $z=1-\kappa (1-z_*)$, where $0<\kappa<1$.
Since the usual lattice Boltzmann version of the FHP model has the shear
viscosity proportional to $\tau-1/2$, the entropic version will have it
proportional to $1/\kappa-1$ which goes to zero as $\kappa$ approaches
unity.  Moreover, absolute stability is maintained as this limit is
approached, even though the algorithm is fully explicit and perfectly
conservative.  As with the diffusion model, we are prevented from
achieving arbitrarily small transport coefficient only by considerations
of accuracy (resolution of the smallest eddies), and of the time scale
for the approach to equilibrium.

\subsection{Computer Implementation}

We pause to consider the computer implementation of the entropic lattice
Boltzmann algorithm for the FHP fluid.  The propagation step can be
handled in precisely the same fashion that it is for conventional
lattice Boltzmann simulations.  The principal difference lies in the
implementation of the collision operator, and that is what we describe
here.  Unlike the analysis in the last subsection, a complete computer
implementation needs to solve for the equilibrium exactly -- not just to
within the Mach number squared.

Given the six quantities, $N_j$ for $j=0,\ldots,5$, entering a site, we
first contract them with the bras, Eqs.~(\ref{eq:fhpfirstbra}) through
(\ref{eq:fhplastbra}), to obtain the conservation representation, namely
$\rho$, $\pi_x$, $\pi_y$, $\alpha$, $\beta$ and $\gamma$.  Given $\pi_x$
and $\pi_y$, it is then necessary to solve for the equilibrium values of
$\alphabar$, $\betabar$ and $\gammabar$.  We can use
Eqs.~(\ref{eq:fhpaeq}) through (\ref{eq:fhpceq}) to get excellent
approximations for these, and then use a Newton-Raphson iteration to get
the correct values to machine precision.  Even with an excellent initial
guess, there is some potential for loss of convergence in any
Newton-Raphson iteration, and so it might be worthwhile to investigate
better ways of solving for these values.  Assume that we can do this,
and denote the equilibrium values with an ``eq'' superscript.

Given the equilibrium values, we set
\begin{eqnarray}
\alphabar' &=& \alphabar +
 \frac{1}{\tau}
 \left(\alphabar^{\mbox{\scriptsize eq}} - \alphabar\right)\label{eq:fhpafinal}\\
\betabar'  &=& \betabar  +
 \frac{1}{\tau}
 \left(\betabar^{\mbox{\scriptsize eq}} - \betabar\right)\\
\gammabar' &=& \gammabar +
 \frac{1}{\tau}
 \left(\gammabar^{\mbox{\scriptsize eq}} - \gammabar\right),\label{eq:fhpcfinal}
\end{eqnarray}
and find the value of $\tau$ that solves the nonlinear equation
$h(\bfN')=h(\bfN)$.  This equation will look much like
Eq.~(\ref{eq:fhpzsol}), except that $\alphabar^{\mbox{\scriptsize eq}}$,
$\betabar^{\mbox{\scriptsize eq}}$ and $\gammabar^{\mbox{\scriptsize
    eq}}$ will enter, corresponding to corrections of order Mach number
squared.  Since this nonlinear equation has only one scalar unknown,
namely $\tau$, it can be solved using an iteration method that is
guaranteed to converge, such as regula falsi.  We know that $\tau=1$ is
one bound on the solution; the other bound can be taken as the point at
which the line of outgoing states intersects the boundary of the kinetic
polytope.  This means finding the largest value of $\tau$ between zero
and one for which $N'_j=0$ for some $j\in\{0,\ldots,5\}$.  Using these
two bounds, we iterate the regula falsi algorithm to obtain the solution
$\tau_*$.  We then set $\tau=\tau_*/\kappa$, and use
Eqs.~(\ref{eq:fhpafinal}) through (\ref{eq:fhpcfinal}) to get the
outgoing values of the kinetic parameters.  We finish the collision
procedure by using Eq.~(\ref{eq:fhprep}) to get the outgoing state in
the original representation.

\subsection{Galilean Invariance}
\label{ssec:galileo}

As noted in the Introduction, practitioners of lattice BGK models have
long known that it is necessary to tailor the equilibrium distribution
function in a certain way in order to maintain Galilean invariance.
Recall that the zeroth moment (with respect to the velocity vector) of
the distribution function is determined by the mass density.  Likewise,
the first moment is determined by the momentum density.  For a lattice
BGK model of an ideal gas~\cite{bib:macnam,bib:alex}, the trace of the
second moment is determined by the kinetic energy density.  To obtain
the correct, Galilean-invariant form of the compressible Navier-Stokes
equations, it is also necessary~\cite{bib:ice} to mandate the rest of
the second moment, the third moment, and the trace of the fourth moment.

While it is possible to construct equilibrium distributions with these
moments for various lattices, no lattice model has ever been found for
which these equilibria are guaranteed to be nonnegative.  This means
that the equilibrium toward which we would like to relax in order to
maintain Galilean invariance may itself lie outside of the master
polytope!  There is thus no way that such an equilibrium could be the
maximum of an entropy function that goes to $-\infty$ on the polytope
boundary, such as do the ones that we have been considering.

In such a situation, we may ask whether it is possible to jettison the
requirement of nonnegativity, while still demanding absolute stability.
In fact, our formalism suggests an interesting way of doing just that.
If we no longer demand that the function $h$ go to negative infinity on
the polytope boundary, then we need no longer demand
Eq.~(\ref{eq:zatz}), though Eq.~(\ref{eq:increasing}) is still useful to
ensure that the equilibrium is unique, and Eq.~(\ref{eq:positivity}) is
useful to keep $h$ real-valued in light of Eq.~(\ref{eq:hdef}).  Thus,
we are free to make the choice $\zeta_j(x)$ is $\exp[x^2/(2g_j)]$, where
the $g_j$'s are positive constants.  Indeed, this choice is similar to one made
by Karlin et al.~\cite{bib:karlinprl}.  Eq.~(\ref{eq:equilibrium}) is then
the set of $n$ equations,
\begin{equation}
  \Neqj = g_j\sum_{\sigma=1}^{n_c} Q_\sigma \ket{\bflambda_\sigma}_j,
\label{eq:constq}
\end{equation}
and Eqs.~(\ref{eq:lagmult}) give $n_c$ additional requirements.  One may
then try to solve these $n+n_c$ equations for the $n+n_c$ unknowns,
$Q_\sigma$ and $g_j$, subject to the requirement that $g_j>0$.  In fact,
not all of these unknowns are independent; the scaling $g_j\rightarrow
g_j/x$ and $Q_\sigma\rightarrow Q_\sigma x$, where $x$ is arbitrary,
leaves Eq.~(\ref{eq:constq}) invariant, so there are more requirements
than equations, and hence not all equilibria are derivable from an $h$
function of this form.  Nevertheless, many interesting equilibria are
derivable in this way, and this approach leads immediately to the
construction of an entropically stabilized algorithm for such
equilibria.  Investigation of this approach is work in
progress~\cite{bib:withsauro}.  Its successful resolution may yield an
absolutely stable entropic lattice Boltzmann algorithm for compressible
flow, albeit one that allows for negative $N_j$.

\newpage
\section{Conclusions}
\label{sec:conclusions}

In summary, we have presented a general methodology for constructing
lattice Boltzmann models of fluid dynamics that conserve a given set of
moments of the distribution function.  The method guarantees the
nonnegativity of the distribution function, and allows for convenient
visualization of the dynamics by construction of the polytope of allowed
states using Fourier-Motzkin elimination.  We have shown how such models
can be endowed with a Lyapunov functional, resulting in unconditional
numerical stability, and we presented various choices for the $H$
function for this purpose.  We also described the computer
implementation of such models in some detail.  For lattice Boltzmann
models of diffusion and of incompressible fluid flow, we were able to
present fully explicit, perfectly conservative, absolutely stable
algorithms that appear to work for arbitrarily small transport
coefficient.  Indeed, we showed that the limitations on attainable
transport coefficients for these models arise from considerations of
accuracy, rather than stability.

There are numerous avenues for extension of the current work.  We close
by presenting a list of these.  We hope that the richness and promise of
this style of model will inspire others to take up some of these
questions.
\begin{enumerate}[(i)]
\item We have not discussed the question of boundary conditions in this
  paper.  For a viscous fluid, for example, collisions at a solid
  boundary wall must conserve mass, but not momentum or energy.  The
  latter two quantities can enter or leave the domain through the wall,
  resulting in drag, lift, surface heating, etc.  This results in very
  different polytope structure at the boundaries than in the interior.
  In addition, there is nothing preventing entropy from entering or
  leaving through the wall, so the entire question of how to construct
  $h$ functions at the boundary sites needs to be revisited.
\item While we touched on the question of optimality of conservation
  representations, clearly much more work could be done along these
  lines.  In addition to making it computationally easier to transform
  back and forth between the original and conservation representations,
  one might like to try to optimize the form of the entropically
  stabilized collision operator.
\item We mentioned reversible models, but did not discuss or study them
  in any detail.  Our entropically stabilized models are reversible in
  the limit as $\kappa\rightarrow 1$, but this is precisely the limit in
  which they begin to fail to simulate the parabolic equations of
  interest, namely the diffusion and Navier-Stokes equations.  It would
  be interesting to know if such reversible models are capable of
  simuating time-symmetric partial differential equations, such as
  Euler's equations of inviscid fluid dynamics.
\item The nature of the anomalies encountered as $\kappa\rightarrow 1$
  remains to be completely elucidated.  For example, the lower right
  plot in Fig.~\ref{fig:decay99999} appears to indicate that the
  approach to equilibrium is extremely slow for this model.  Does it
  eventually reach an exponentially decaying equilibrium state for which
  the diffusion coefficient matches that of the theory?  Does roundoff
  error play a role in this regime?  These open questions will require
  substantially more numerical simulation than that presented here.
\item Given that the anomalies as $\kappa\rightarrow 1$ are due to the
  slowness of the approach to the equilibrium state, much effort could
  be expended in trying to start the simulation in as close to a {\it
    global} equilibrium state as possible.  This means taking spatial
  gradient corrections into account.  At a minimum, the Chapman-Enskog
  corrections -- accurate to first-order in the spatial gradient --
  could be incorporated with little effort.
\item The question of how to deal with interaction potentials in this
  formalism could be addressed.  At the level of mean-field theory, this
  would require the introduction of conserved quantities that are
  quadratic in the distribution function.  The allowed regions would
  then no longer be polytopes, convex, or even simply connected.
  Moreover, multiple local equilibria within the allowed region may be
  expected.  For near-equilibrium dynamics, this might give rise to a
  Ginzburg-Landau model with multiple free-energy minima.  This is thus
  a difficult extension of the current theory, but one for which the
  rewards would be great.
\item One could try to construct an adaptive mesh refinement (AMR)
  version of this algorithm for which the nonnegativity of the physical
  densities is the refinement criterion.  Using such an approach, it
  might be possible to relax the criterion of nonnegativity of the
  distribution function.  That is, we could carry negative distribution
  function values, but insist that all physical densities be positive.
  Should the propagation step threaten to create a negative density, we
  would try to refine the lattice locally until that is no longer the
  case.
\item Finally, it would be interesting to consider the potential utility
  of this algorithm for the construction of ``eddy-viscosity'' subgrid
  models of turbulence.  As noted at several points in this paper, while
  entropic lattice Boltzmann algorithms may be stable for arbitrarily
  small transport coefficient, they may begin to lose accuracy.  The
  smallest eddy sizes in three dimensional Navier-Stokes turbulence, for
  example, scale as ${\mbox{Re}}^{-3/4}$, where the Reynolds number
  $\mbox{Re}$ goes inversely with viscosity.  Thus, at sufficiently
  small viscosity, the eddies will be smaller than any fixed grid
  spacing.  If one's goal is an {\it ab initio} simulation of the
  Navier-Stokes equations, this circumstance is grounds for rejecting
  the entropic lattice Boltzmann solution as unphysical.  On the other
  hand, stable, coarse-grained, turbulent flow must obey the same
  conservation laws as the underlying Navier-Stokes equations.  Since
  the entropic lattice Boltzmann collision operators are the most
  general that respect these conservation laws, it is interesting to
  speculate as to whether some physical significance might yet be
  attached to their solutions, even when the smallest eddies are not
  resolved by the lattice.
\end{enumerate}

\section*{Acknowlegements}

This work was presented by BMB at the Seventh International Conference
on the Discrete Simulation of Fluids in Oxford, England in July, 1998.
During the late stages of writing this paper, we became aware that two
other groups~\cite{bib:karlin,bib:hudong} were independently working on
requirements of the form of Eq.~(\ref{eq:entstab}).  As the works were
conducted independently, there are numerous difference in style,
formulation and emphasis, and we believe that the resulting papers,
including this one, are nicely complementary to one another.

It is a pleasure to acknowledge helpful conversations with Frank
Alexander, Bastien Chopard, Nicos Martys, Sauro Succi and Julia Yeomans.
BMB was supported in part by an IPA assignment agreement with the Air
Force Research Laboratory, Hanscom AFB, Massachusetts, and in part by
the United States Air Force Office of Scientific Research under grant
number F49620-95-1-0285.  BMB and PVC would like to thank Schlumberger
Research and the EPSRC's Collaborative Computing Project No. 5 for
making possible BMB's visit to Schlumberger Cambridge Research and the
University of Oxford, where this work was begun in early 1997.  PVC is
grateful to Wolfson College and the Department of Theoretical Physics,
University of Oxford, for a Visiting Fellowship (1996 -- 1999).

\appendix

\newpage
\section{A Tour of the Master Polytope for the Compressible-Fluid Model}
\label{app:tour}

The five inequalities of Eq.~(\ref{eq:fineqs}) can be recast as
\begin{equation}
  0 \leq |\betabar| \leq \alphabar \leq \min\left( \frac{2-\vepsbar}{3}, \vepsbar-|2\betabar-\pibar| \right).
  \label{eq:ineqt}
\end{equation}
This implies
\begin{equation}
  \vepsbar\geq \alphabar+|2\betabar-\pibar|\geq \alphabar\geq |\betabar|\geq 0
\end{equation}
and
\begin{equation}
  \vepsbar\leq 2-3\alphabar\leq 2-3|\betabar|\leq 2,
\end{equation}
and hence $\alphabar\geq 0$ and $\vepsbar\in [0,2]$.  Moreover, we note
that all of the inequalities are invariant under the transformation
\begin{eqnarray*}
  \pibar &\rightarrow& -\pibar\\
  \betabar &\rightarrow& -\betabar,
\end{eqnarray*}
so that we may restrict our attention to $\pibar\geq 0$ without loss of
generality.

\begin{figure}
  \center{
    \mbox{
      \includegraphics[bbllx=72,bblly=170,bburx=540,bbury=640,width=2.75truein]{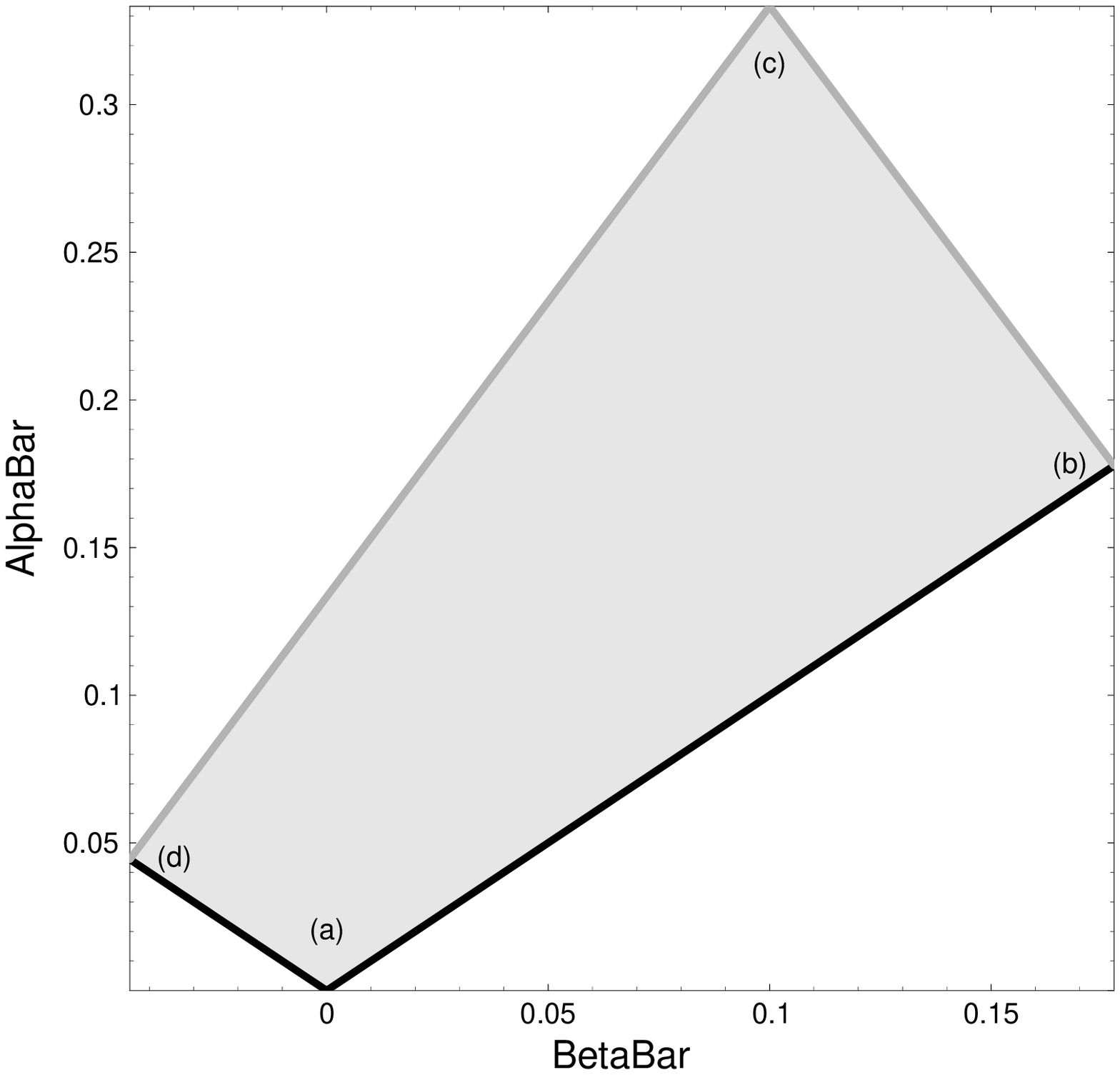}
      }
    \hspace{0.5truein}
    \mbox{
      \includegraphics[bbllx=245,bblly=590,bburx=375,bbury=722,width=2.75truein]{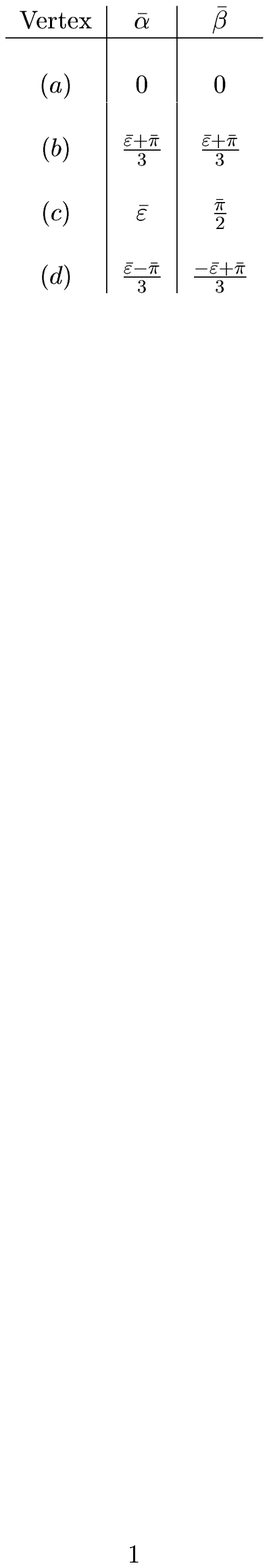}
      }}
  \caption{{\bf Bounds on $\alphabar$ and $\betabar$} for $\pibar=1/5$ and $\vepsbar=1/3$, and the
    coordinates of the vertices given as general functions of $\vepsbar$ and
    $\pibar$.}
  \label{fig:fig1}
\end{figure}

Next note that the second argument of the $\min$ function in
Eq.~(\ref{eq:ineqt}) will be smaller than the first if $0 \leq
\vepsbar < (2-\vepsbar)/3$, or $0 \leq \vepsbar < 1/2$.  In
this case, for sufficiently small $\pibar$, the boundary of the polytope in
the $(\alphabar,\betabar)$ plane will be a quadrilateral, such as that in
Fig.~\ref{fig:fig1}, whose lower bound in $\alphabar$ (shown in black) is
$|\betabar|$ and whose upper bound in $\alphabar$ (shown in gray) is
$\vepsbar-|2\betabar-\pibar|$.  The allowed region of the $(\alphabar,\betabar)$ plane is
the shaded area in between these bounds.  The bottom
vertex~\footnote{Throughout this appendix, we denote vertices in
  parentheses, so as not to confuse them with other variables.} $(a)$ of
this quadrilateral is at the origin, $(\alphabar,\betabar)=(0,0)$, and the upper
vertex $(c)$ is at $(\alphabar,\betabar)=(\vepsbar,\pibar/2)$.  The right vertex
$(b)$ is then at $\betabar=\vepsbar-(2\betabar-\pibar)$ or
$(\alphabar,\betabar)=((\vepsbar+\pibar)/3, (\vepsbar+\pibar)/3)$, and the left vertex
$(d)$ is then at $-\betabar=\vepsbar+(2\betabar-\pibar)$ or
$(\alphabar,\betabar)=((\vepsbar-\pibar)/3,(-\vepsbar+\pibar)/3)$.  These results are
summarized in the table included in Fig.~\ref{fig:fig1}.

We now ask for what range of $\vepsbar$ and $\pibar$ the above-described
quadrilateral boundary in the $(\alphabar,\betabar)$ plane is valid.  We note that
vertices $(a)$ and $(b)$ will be degenerate when $\vepsbar=-\pibar$, and
that vertices $(a)$ and $(d)$ will be degenerate when $\vepsbar=+\pibar$.
Also, it is easy to see that one or the other of these two degeneracies
will occur before any degeneracy involving vertex $(c)$.  It follows
that the above-described quadrilateral boundary in the $(\alphabar,\betabar)$
plane is valid only for $|\pibar| < \vepsbar < 1/2$.  This region of the
$(\pibar,\vepsbar)$ plane is the shaded triangle $AIH$ in
Fig.~\ref{fig:fpolytope}.

\begin{figure}
  \center{
    \mbox{
      \includegraphics[bbllx=72,bblly=170,bburx=540,bbury=640,width=2.75truein]{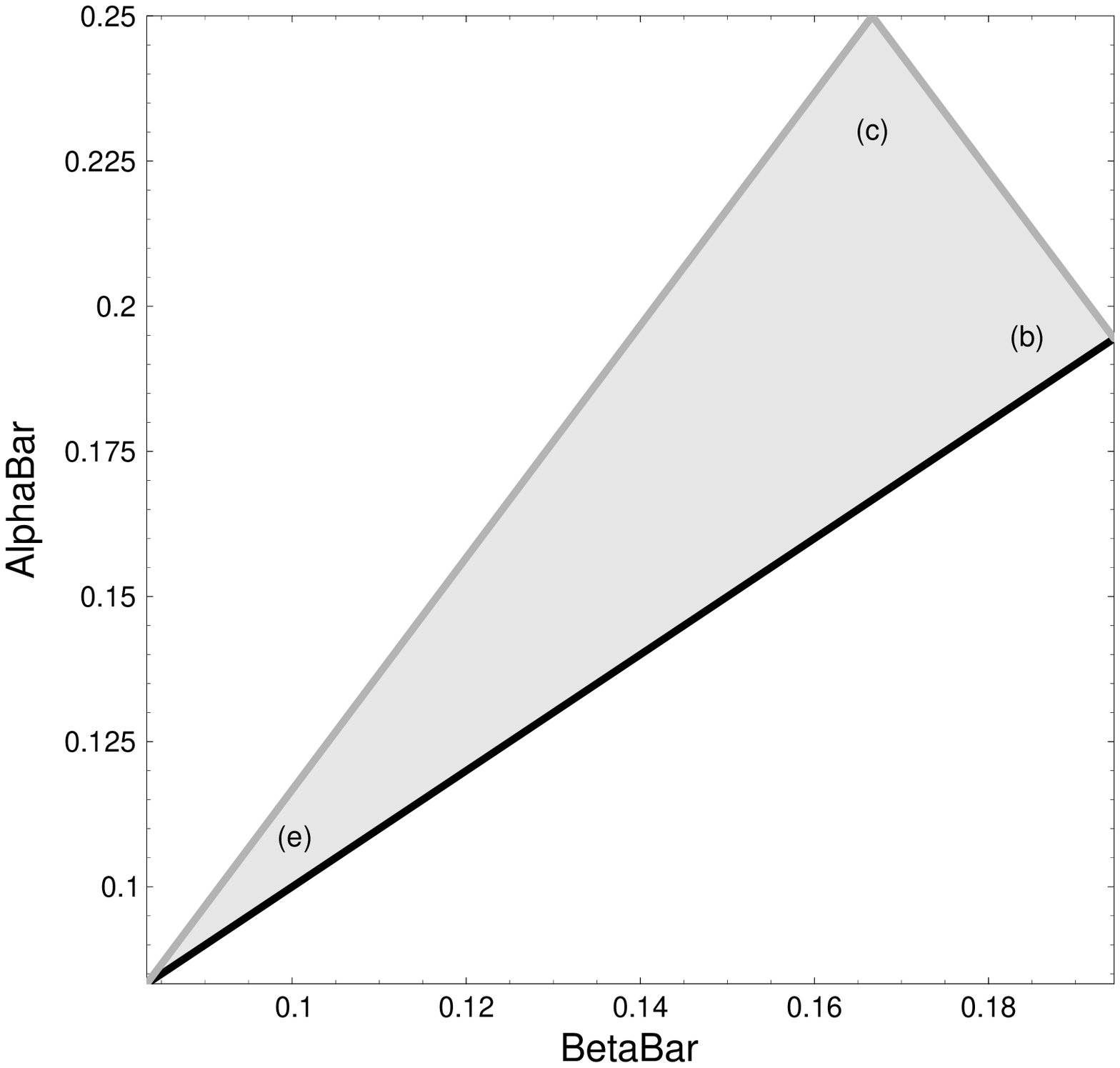}
      }
    \hspace{0.5truein}
    \mbox{
      \includegraphics[bbllx=245,bblly=620,bburx=375,bbury=722,width=2.75truein]{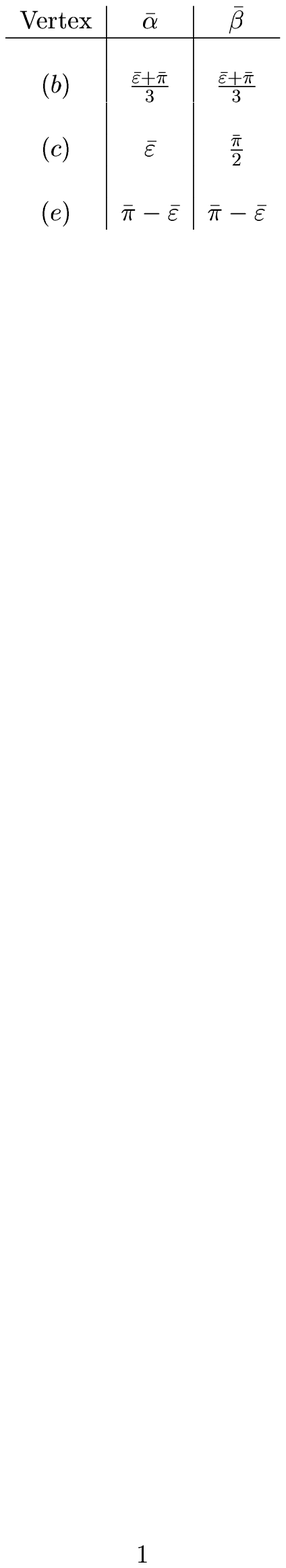}
      }}
  \caption{{\bf Bounds on $\alphabar$ and $\betabar$} for $\pibar=1/3$ and $\vepsbar=1/4$, and the
    coordinates of the vertices given as general functions of $\vepsbar$ and
    $\pibar$.}
  \label{fig:fig2}
\end{figure}

Beginning in triangle $AIH$, boundary $AI$ is encountered when
$\vepsbar=\pibar$ so that vertices $(a)$ and $(d)$ are degenerate.  If we
cross this boundary the allowed region in the $(\alphabar,\betabar)$ plane is no
longer a quadrilateral, but rather becomes a triangle, such as that in
Fig.~\ref{fig:fig2}, whose upper bound in $\alphabar$ (shown in gray) is
$\vepsbar-|2\betabar-\pibar|$ and whose lower bound in $\alphabar$ (shown in black)
is $\betabar$.  The bottom vertex $(e)$ of this triangle is at
$\alphabar=\betabar=\vepsbar+(2\betabar-\pibar)$ or
$(\alphabar,\betabar)=(\pibar-\vepsbar,\pibar-\vepsbar)$; this vertex replaces the
degenerate vertices $(a)$ and $(d)$ of the above-described
quadrilateral.  The expressions for the coordinates of the triangle's
upper vertex $(c)$ and its right vertex $(b)$ are identical to those of
the corresponding vertices of the quadrilateral.  These results are
summarized in the table included in Fig.~\ref{fig:fig2}.

Again, we ask for what range of $\vepsbar$ and $\pibar$ the
above-described triangular boundary in the $(\alphabar,\betabar)$ plane is valid.
We note that vertices $(b)$ and $(c)$ will be degenerate when
$(\vepsbar+\pibar)/3=\vepsbar$ or $\vepsbar=\pibar/2$; for smaller values
of $\vepsbar$ the set of allowed points in the $(\alphabar,\betabar)$ plane is
empty.  We also note that, just as for the quadrilateral of
Fig.~\ref{fig:fig1}, the first argument of the $\min$ function in
Eq.~(\ref{eq:ineqt}) will become the determining factor when
$\vepsbar>1/2$, also invalidating the above argument.  Thus, the
allowed region in the $(\alphabar,\betabar)$ plane will be a triangle with
vertices described in Fig.~\ref{fig:fig2} only if $(\vepsbar,\pibar)$ is
in the shaded triangle $ABI$ in Fig.~\ref{fig:fpolytope}.  Since the
region $AEH$ is the image of $ABI$ under the map $\pibar\rightarrow -\pibar$, it
follows that the allowed region of the $(\alphabar,\betabar)$ plane is also
triangular for $(\vepsbar,\pibar)\in \triangle AEH$, with vertices
identical to those in Fig.~\ref{fig:fig2} with $\betabar\rightarrow -\betabar$.

\begin{figure}
  \center{
    \mbox{
      \includegraphics[bbllx=72,bblly=170,bburx=540,bbury=640,width=2.75truein]{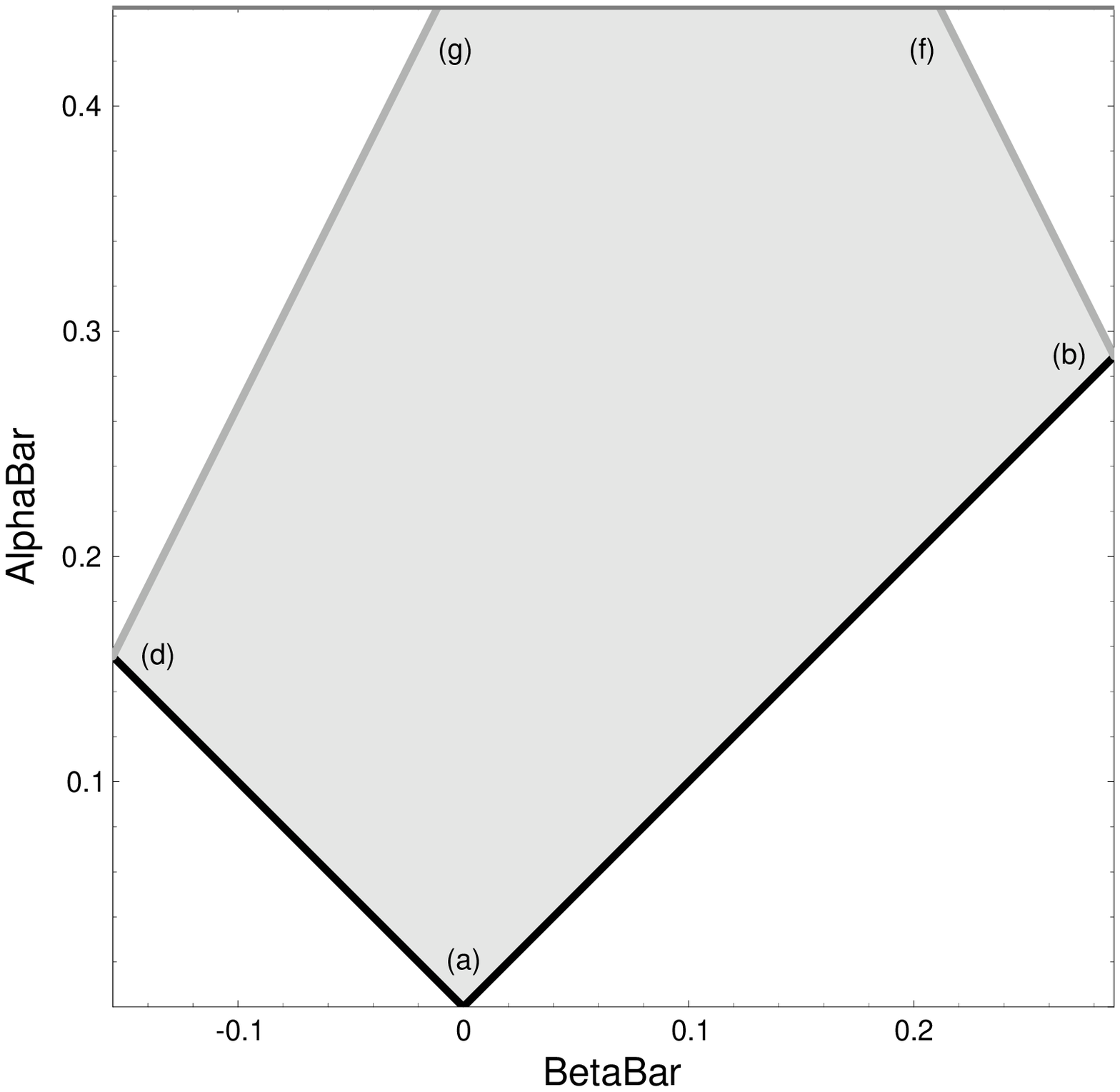}
      }
    \hspace{0.5truein}
    \mbox{
      \includegraphics[bbllx=245,bblly=565,bburx=375,bbury=722,width=2.75truein]{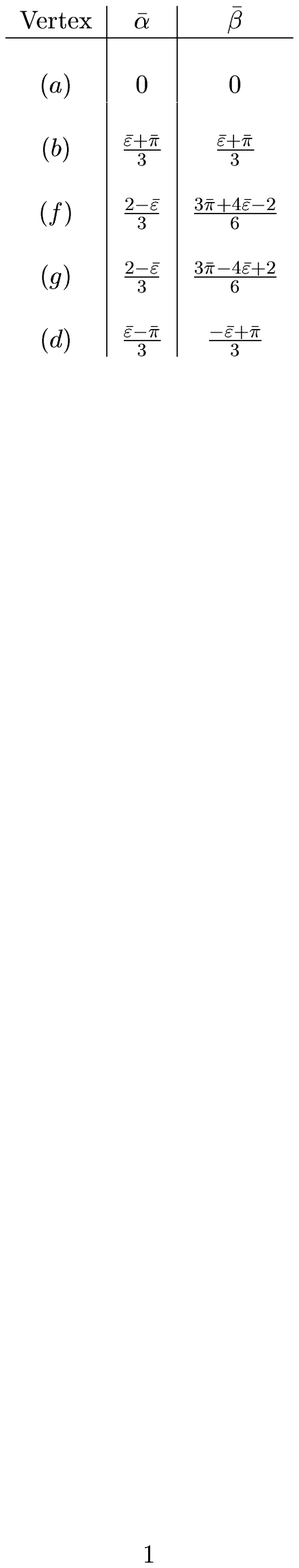}
      }}
  \caption{{\bf Bounds on $\alphabar$ and $\betabar$} for $\pibar=1/5$ and $\vepsbar=2/3$, and the
    coordinates of the vertices given as general functions of $\vepsbar$ and
    $\pibar$.}
  \label{fig:fig3}
\end{figure}

\begin{figure}
  \center{
    \mbox{
      \includegraphics[bbllx=72,bblly=170,bburx=540,bbury=640,width=2.75truein]{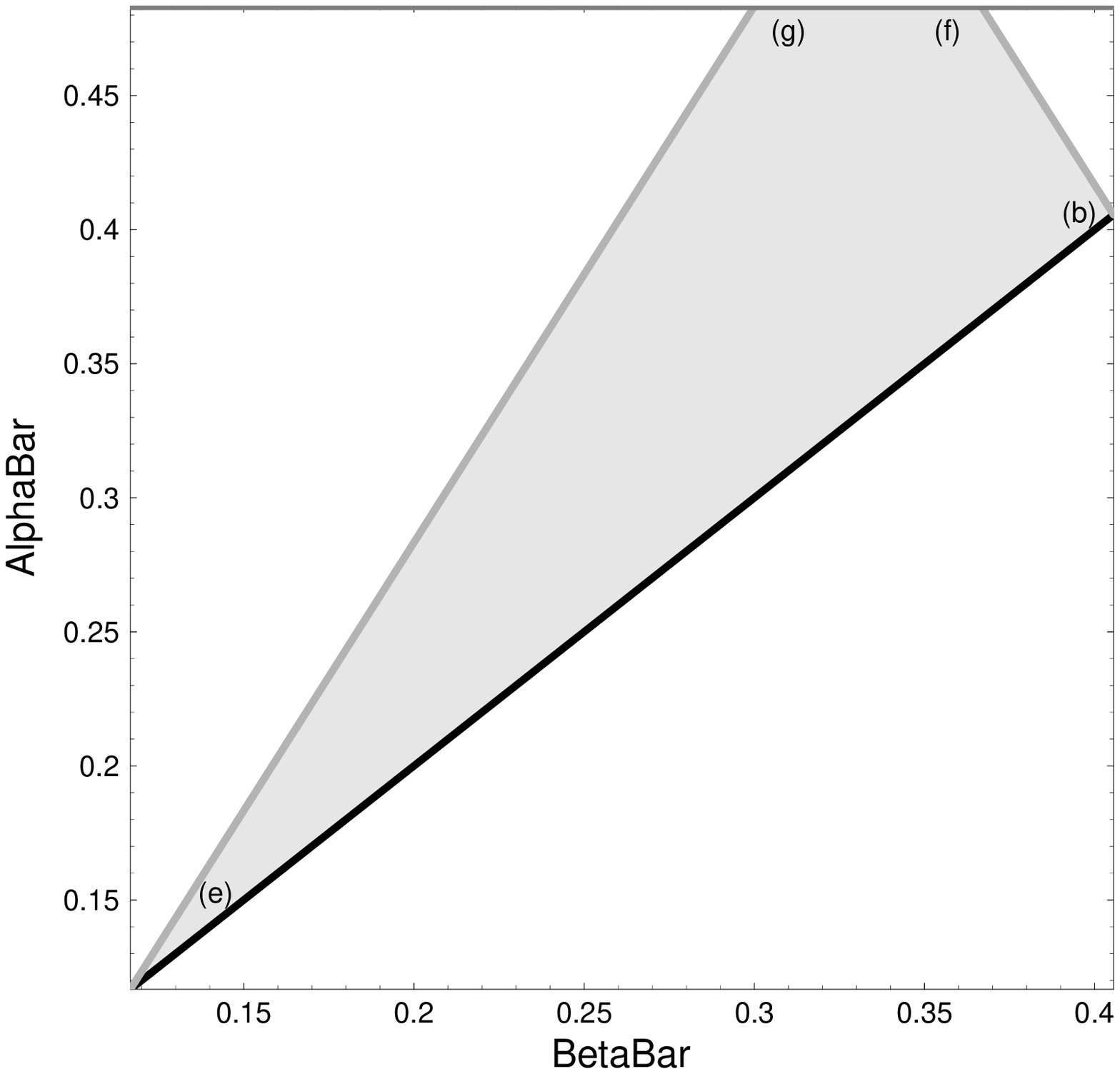}
      }
    \hspace{0.5truein}
    \mbox{
      \includegraphics[bbllx=245,bblly=590,bburx=375,bbury=722,width=2.75truein]{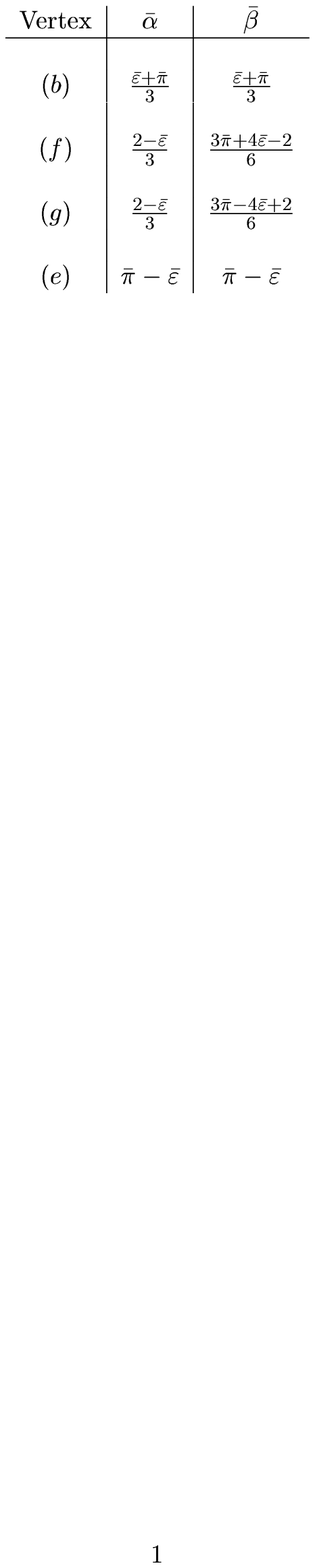}
      }}
  \caption{{\bf Bounds on $\alphabar$ and $\betabar$} for $\pibar=2/3$ and $\vepsbar=11/20$, and the
    coordinates of the vertices given as general functions of $\vepsbar$ and
    $\pibar$.}
  \label{fig:fig4}
\end{figure}

We have now described all of Fig.~\ref{fig:fpolytope} below the line
$\vepsbar=1/2$ (that is, below line $EB$).  We next consider the
situation for $\vepsbar>1/2$.  This means that we have to start
taking into account the inequality
\begin{equation}
  \alphabar\leq\frac{2-\vepsbar}{3}
\end{equation}
of Eq.~(\ref{eq:ineqt}).  This is a constant upper bound on $\alphabar$ which
will become less than the $\alphabar$ coordinate of vertex $(c)$ in
Figs.~\ref{fig:fig1} and \ref{fig:fig2} when $\vepsbar>1/2$.  This
results in a horizontal truncation of the quadrilateral of
Fig.~\ref{fig:fig1} so that it becomes a pentagon, and of the triangle
of Fig.~\ref{fig:fig2} so that it becomes a quadrilateral.  These are
shown in Figs.~\ref{fig:fig3} and \ref{fig:fig4}, respectively.  The top
vertices $(f)$ and $(g)$ in both of these figures are at the upper bound
$\alphabar=(2-\vepsbar)/3$, and $(2-\vepsbar)/3=\vepsbar\mp
(2\betabar-v)$ or $\betabar=(3v\pm 4\vepsbar\mp 2)/6$, respectively; these
vertices replace vertex $(c)$ in Figs.~\ref{fig:fig1} and
\ref{fig:fig2}.  The expressions for the coordinates of vertices $(a)$,
$(b)$, $(d)$ and $(e)$ are identical to those derived previously.  The
vertex coordinates are given in the tables in their corresponding
figures.

\begin{figure}
  \center{
    \mbox{
      \includegraphics[bbllx=72,bblly=170,bburx=540,bbury=640,width=2.75truein]{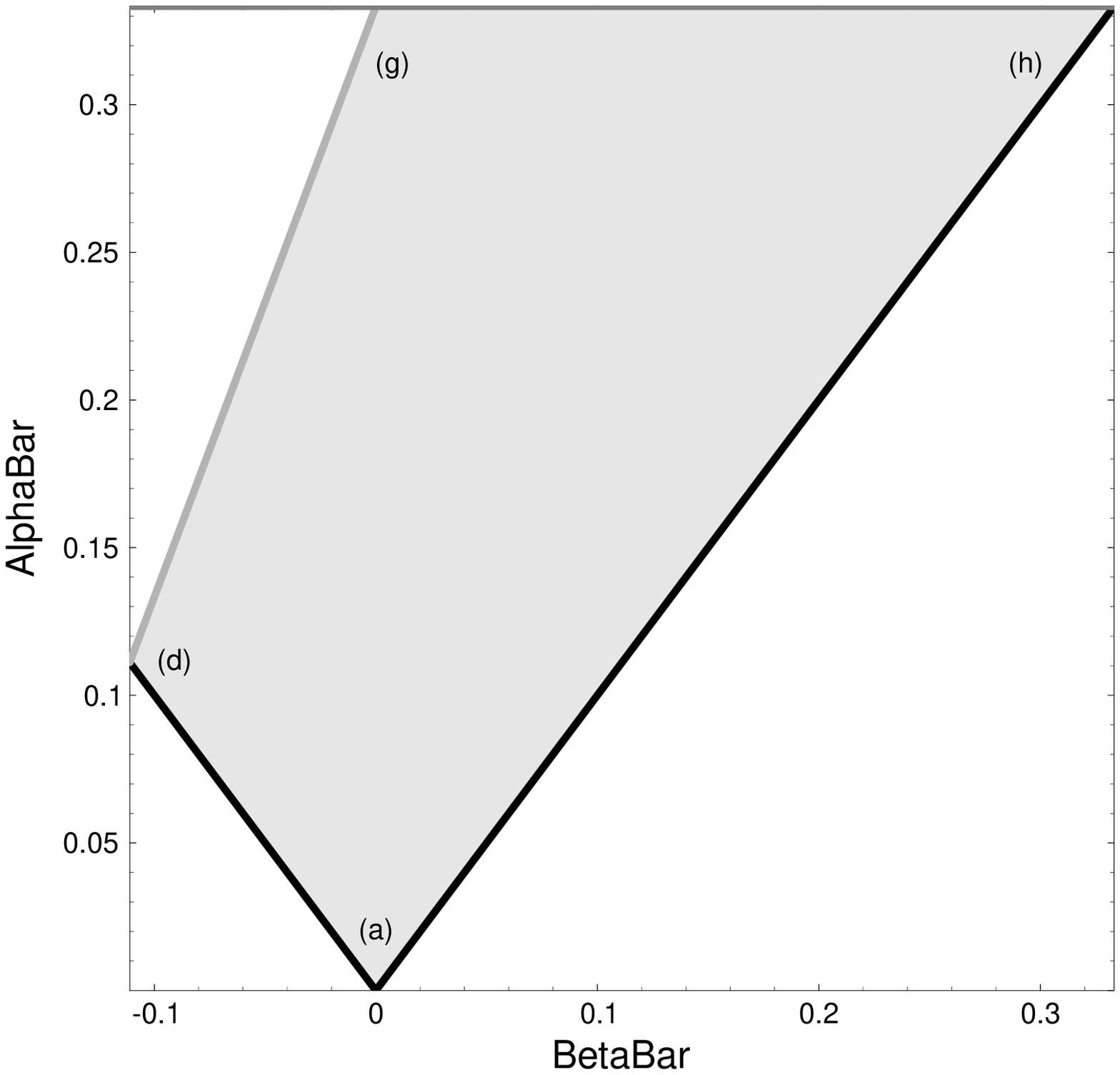}
      }
    \hspace{0.5truein}
    \mbox{
      \includegraphics[bbllx=245,bblly=595,bburx=375,bbury=722,width=2.75truein]{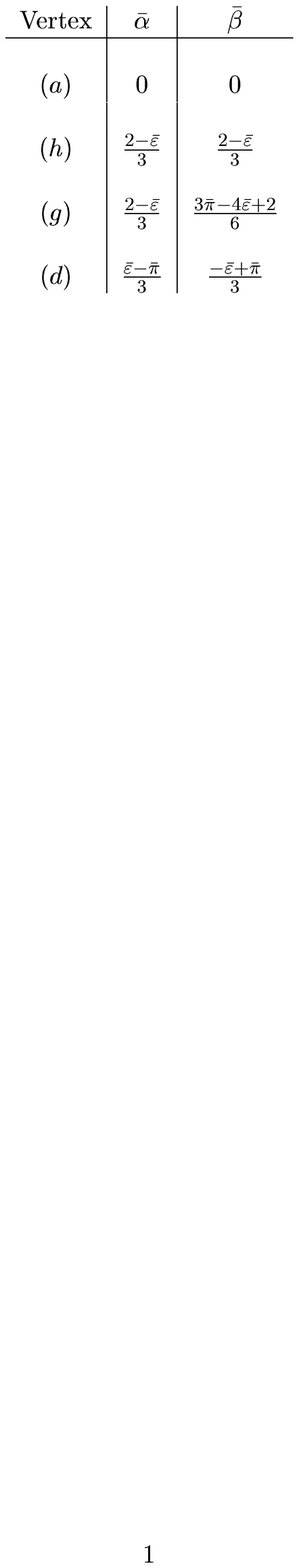}
      }}
  \caption{{\bf Bounds on $\alphabar$ and $\betabar$} for $\pibar=2/3$ and $\vepsbar=1$, and the
    coordinates of the vertices given as general functions of $\vepsbar$ and
    $\pibar$.}
  \label{fig:fig5}
\end{figure}

\begin{figure}
  \center{
    \mbox{
      \includegraphics[bbllx=72,bblly=170,bburx=540,bbury=640,width=2.75truein]{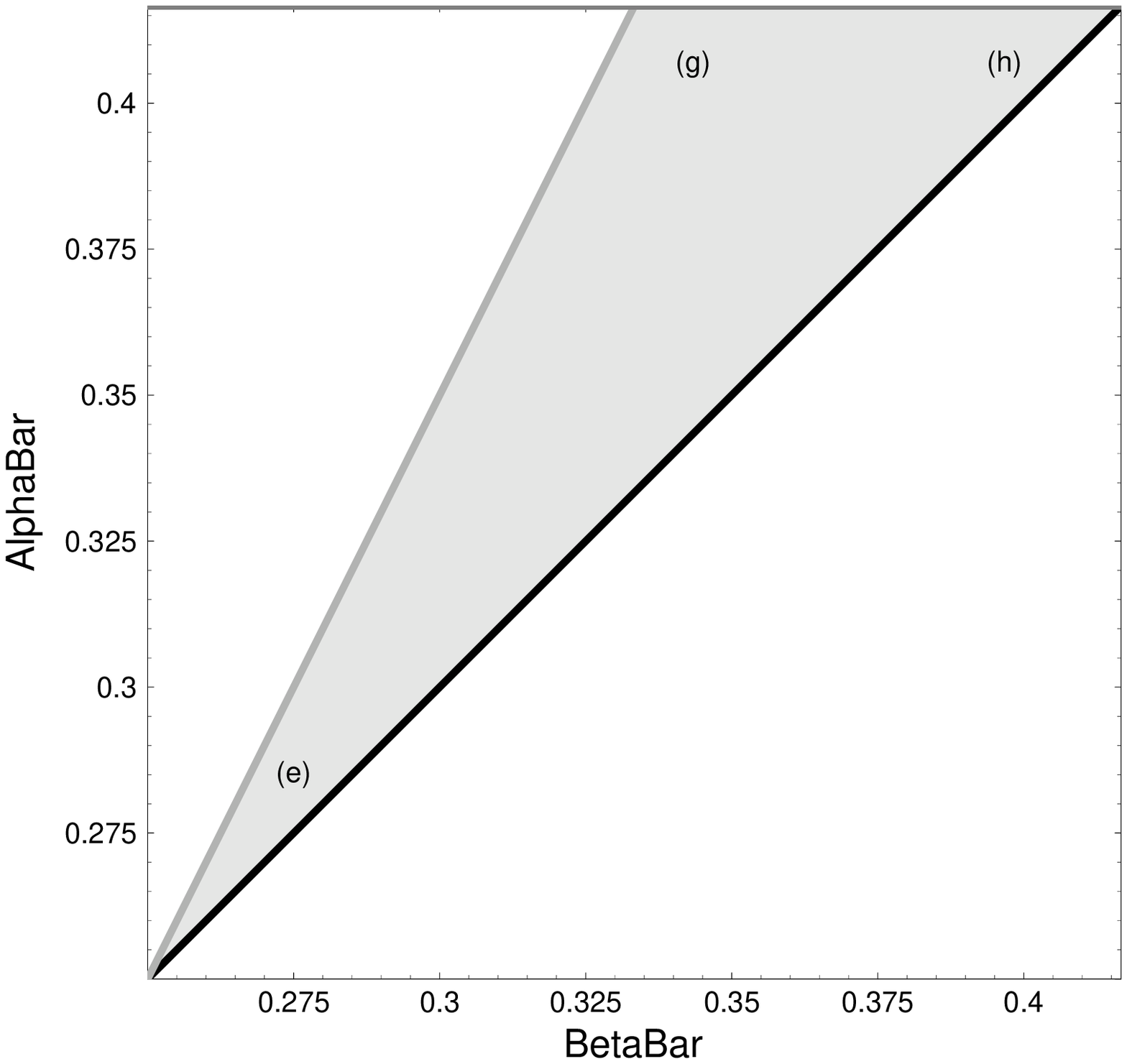}
      }
    \hspace{0.5truein}
    \mbox{
      \includegraphics[bbllx=240,bblly=620,bburx=375,bbury=722,width=2.75truein]{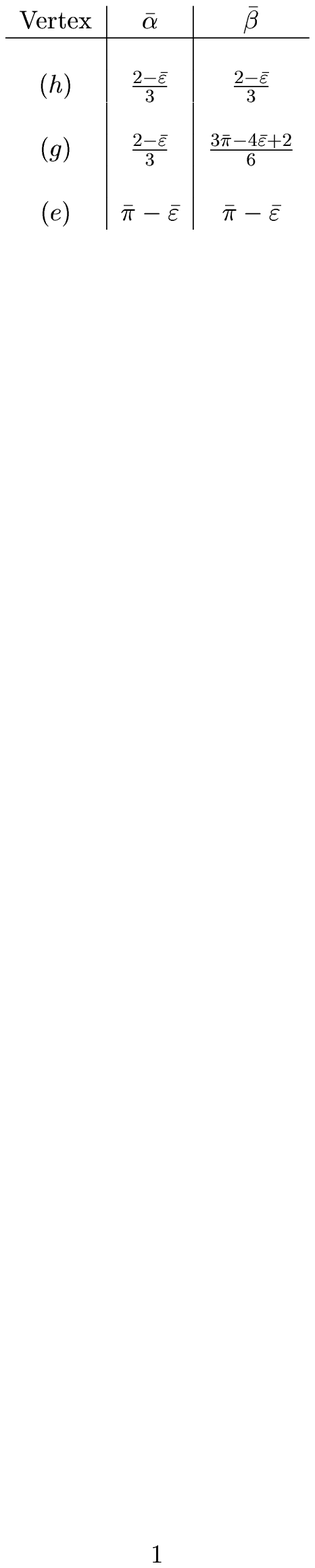}
      }}
  \caption{{\bf Bounds on $\alphabar$ and $\betabar$} for $\pibar=1$ and $\vepsbar=3/4$, and the
    coordinates of the vertices given as general functions of $\vepsbar$ and
    $\pibar$.}
  \label{fig:fig6}
\end{figure}

Yet again, we ask for what range of $\vepsbar$ and $\pibar$ the boundaries
pictured in Figs.~\ref{fig:fig3} and \ref{fig:fig4} are valid.  As
$\vepsbar$ increases, the upper bound on $\alphabar$ decreases until
vertices $(b)$ and $(f)$ coincide.  This happens when
$(\vepsbar+\pibar)/3=(2-\vepsbar)/3$, or $\vepsbar=1-\pibar/2$.  This is
line $FB$ in Fig.~\ref{fig:fpolytope}.  Above this line, the pentagonal
region of Fig.~\ref{fig:fig3} degenerates to a quadrilateral, and the
quadrilateral region of Fig.~\ref{fig:fig4} degenerates to a triangle.
In both cases, vertices $(b)$ and $(f)$ are replaced by a new vertex
$(h)$ whose coordinates are $\alphabar=\betabar=(2-\vepsbar)/3$.  These
situations are shown in Figs.~\ref{fig:fig5} and \ref{fig:fig6}, along
with corresponding tables of vertex coordinates.

The quadrilateral of Fig.~\ref{fig:fig5} will degenerate into the
triangle of Fig.~\ref{fig:fig6} when vertices $(a)$ and $(d)$ merge and
are replaced by vertex $(e)$.  As before, this degeneracy happens when
$\vepsbar=\pibar$, and this is the boundary line $CJ$ in the illustration
of the $(\vepsbar,\pibar)$ plane shown in Fig.~\ref{fig:fpolytope}.  The
triangle of Fig.~\ref{fig:fig6}, in turn, degenerates into the empty set
when $(2-\vepsbar)/3=\pibar-\vepsbar$, or $\vepsbar=3\pibar/2-1$.
Referring to Fig.~\ref{fig:fpolytope}, this is the boundary line $BC$;
thus, we see that the quadrilaterals of Fig.~\ref{fig:fig5} are obtained
when $(\vepsbar,\pibar)\in\triangle CFJ$, and the triangles of
Fig.~\ref{fig:fig6} are obtained when $(\vepsbar,\pibar)\in\triangle BCJ$.

\begin{figure}
  \center{
    \mbox{
      \includegraphics[bbllx=72,bblly=170,bburx=540,bbury=640,width=2.75truein]{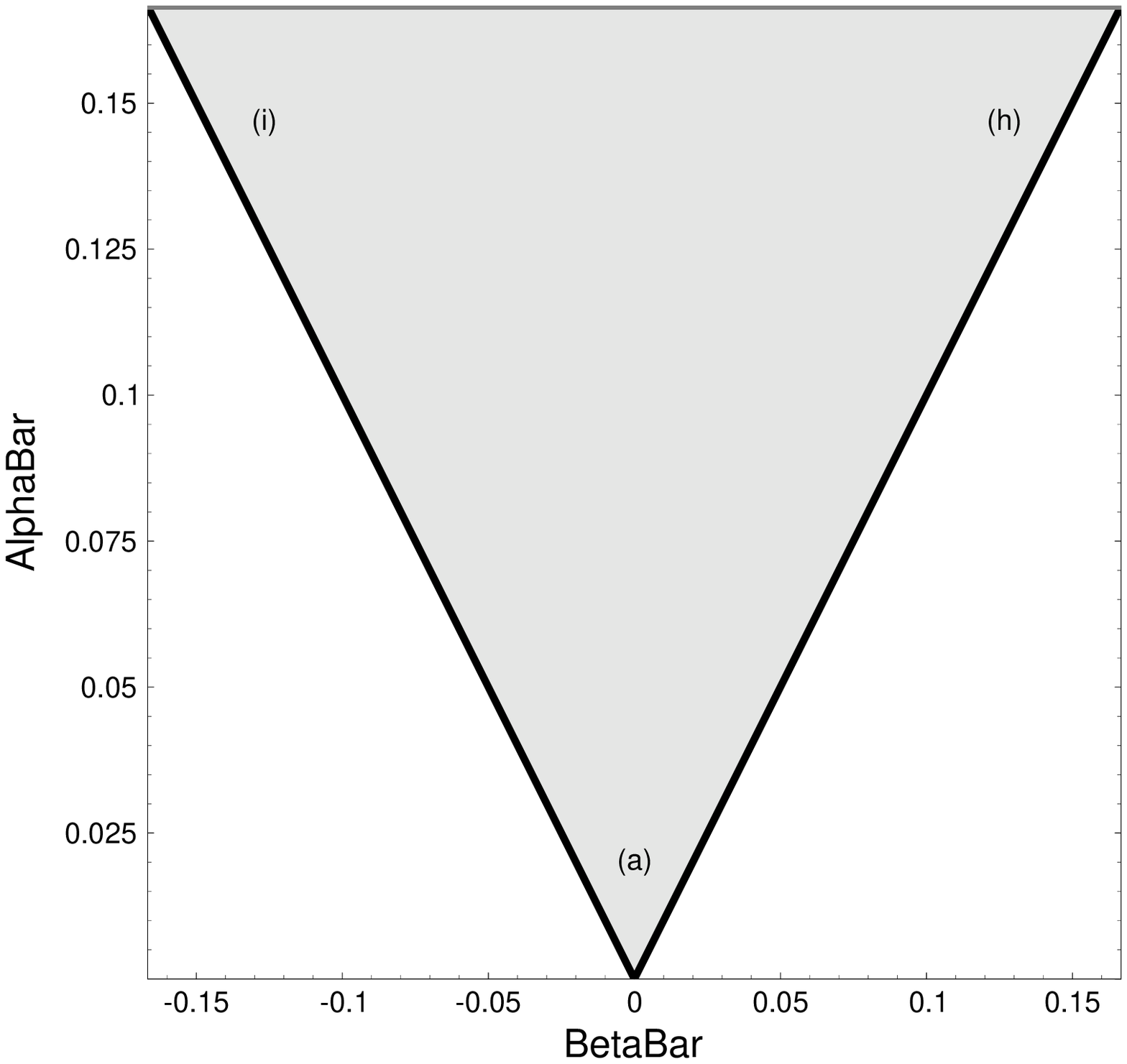}
      }
    \hspace{0.5truein}
    \mbox{
      \includegraphics[bbllx=245,bblly=620,bburx=375,bbury=722,width=2.75truein]{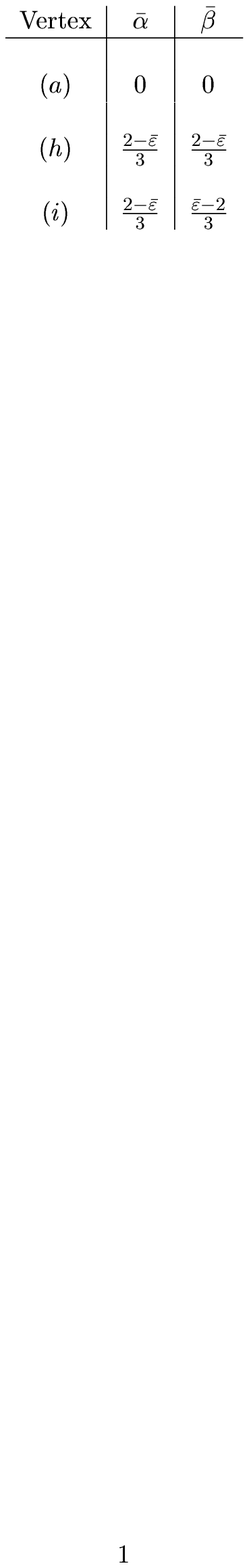}
      }}
  \caption{{\bf Bounds on $\alphabar$ and $\betabar$} for $\pibar=1/2$ and $\vepsbar=3/2$, and the
    coordinates of the vertices given as general functions of $\vepsbar$ and
    $\pibar$.}
  \label{fig:fig7}
\end{figure}

Finally, there is one other way that the quadrilaterals of
Fig.~\ref{fig:fig5} can degenerate.  Vertices $(d)$ and $(g)$ will
coincide if $(2-\vepsbar)/3=(\vepsbar-\pibar)/3$, or
$\vepsbar=1+\pibar/2$.  For values of $\vepsbar$ greater than this,
vertices $(d)$ and $(g)$ are replaced by vertex $(i)$ with coordinates
given by $\alphabar=-\betabar=(2-\vepsbar)/3$.  The resulting isoceles
triangular region is shown in Fig.~\ref{fig:fig7}, along with a
corresponding table of vertex coordinates.  This triangluar region
degenerates to the empty set only when $\vepsbar=2$.  Referring to
Fig.~\ref{fig:fpolytope}, we see that the triangles of
Fig.~\ref{fig:fig7} are obtained when $(\vepsbar,\pibar)\in\triangle CFL$;
in fact, since these regions are symmetric in $\betabar$, they are also
symmetric in $\pibar$, so their description is the same for all
$(\vepsbar,\pibar)\in\triangle CDF$.


\newpage
\section{Basis Bras and Kets for FCHC Model}
\label{app:fchc}

In this appendix, we present one possible choice for the bras and kets
of the FCHC model for three-dimensional fluid dynamics.  The basis bras
are shown in Table~\ref{tab:fchc3drows}, and the basis kets are shown in
Table~\ref{tab:fchc3dcols}.

\begin{table}
  \center{
    {\scriptsize
      $
      \left(
        \begin{array}{rrrrrrrrrrrrrrrrrr}
          1 & 1 & 1 & 1 & 2 & 2 & 1 & 1 & 1 & 1 & 2 & 2 & 1 & 1 & 1 & 1 & 2 & 2 \\
          1 & 1 & -1 & -1 & 0 & 0 & 1 & 1 & -1 & -1 & 0 & 0 & 0 & 0 & 0 & 0 & 2 & -2 \\
          1 & -1 & 1 & -1 & 0 & 0 & 0 & 0 & 0 & 0 & 2 & -2 & 1 & 1 & -1 & -1 & 0 & 0 \\
          0 & 0 & 0 & 0 & 2 & -2 & 1 & -1 & 1 & -1 & 0 & 0 & 1 & -1 & 1 & -1 & 0 & 0 \\
          \hline
          1 & 1 & 1 & 1 & 1 & 1 & 1 & 1 & 1 & -1 & -1 & -1 & -1 & -1 & -1 & -1 & -1 & -1 \\
          -1 & 1 & 1 & 1 & 1 & 1 & 1 & 1 & 1 & 1 & -1 & -1 & -1 & -1 & -1 & -1 & -1 & -1 \\
          1 & 1 & 1 & 1 & 1 & 1 & -1 & -1 & -1 & -1 & -1 & -1 & 1 & 1 & 1 & 1 & 1 & 1 \\
          1 & 1 & 1 & 1 & 1 & 1 & 1 & -1 & -1 & -1 & -1 & -1 & -1 & 1 & 1 & 1 & 1 & 1 \\
          1 & 1 & 1 & 1 & 1 & 1 & 1 & 1 & -1 & -1 & -1 & -1 & -1 & -1 & 1 & 1 & 1 & 1 \\
          1 & 1 & 1 & 1 & 1 & 1 & 1 & 1 & 1 & -1 & -1 & -1 & -1 & -1 & -1 & 1 & 1 & 1 \\
          1 & 1 & 1 & 1 & 1 & 1 & 1 & 1 & 1 & 1 & -1 & -1 & -1 & -1 & -1 & -1 & 1 & 1 \\
          1 & 1 & 1 & 1 & 1 & 1 & 1 & 1 & 1 & 1 & 1 & -1 & -1 & -1 & -1 & -1 & -1 & 1 \\
          1 & 1 & 1 & 1 & 1 & 1 & 1 & 1 & 1 & 1 & 1 & 1 & -1 & -1 & -1 & -1 & -1 & -1 \\
          -1 & 1 & 1 & 1 & 1 & 1 & 1 & 1 & 1 & 1 & 1 & 1 & 1 & -1 & -1 & -1 & -1 & -1 \\
          -1 & -1 & 1 & 1 & 1 & 1 & 1 & 1 & 1 & 1 & 1 & 1 & 1 & 1 & -1 & -1 & -1 & -1 \\
          1 & 1 & 1 & -1 & -1 & -1 & 1 & 1 & 1 & -1 & -1 & -1 & 1 & 1 & 1 & -1 & -1 & -1 \\
          -1 & 1 & 1 & 1 & -1 & -1 & -1 & 1 & 1 & 1 & -1 & -1 & -1 & 1 & 1 & 1 & -1 & -1 \\
          1 & -1 & 1 & -1 & 1 & -1 & 1 & -1 & 1 & -1 & 1 & -1 & 1 & -1 & 1 & -1 & 1 & -1
        \end{array}
      \right)
      $
      }}
  \caption{Bras for the three-dimensional projection of the
    FCHC model.  The first four rows are the hydrodynamic bras, and the
    last 14 present one choice for the kinetic bras.}
  \label{tab:fchc3drows}
\end{table}

\begin{table}
  \center{
    {\tiny
      $
      \smallfrac{1}{48}
      \left(
        \begin{array}{rrrr|rrrrrrrrrrrrrr}
          -16 & 8 & 8 & -16 & 8 & 8 & -16 & 8 & 8 & -16 & 8 & 8 & -16 & 8 & 8 & -16 & 8 & 8\\
          12 & 0 & -24 & 12 & 0 & 0 & 12 & 0 & 0 & 12 & 0 & -12 & 12 & 0 & 0 & 12 & 0 & -12\\
          -24 & 12 & 36 & -24 & 12 & -12 & -24 & 12 & -12 & -24 & 12 & 12 & -24 & 12 & -12 & -24 & 12 & 12\\
          -6 & 6 & 6 & -6 & 18 & -18 & -6 & 6 & -6 & -6 & 0 & 6 & -6 & 6 & -6 & -6 & 0 & 6\\
          -36 & 12 & 12 & 12 & -12 & 12 & -36 & 12 & 12 & -60 & 24 & 12 & -36 & 12 & 12 & -60 & 24 & 12\\
          0 & 0 & 0 & -48 & 36 & 12 & 0 & 0 & 0 & 24 & -18 & -6 & 0 & 0 & 0 & 24 & -18 & -6\\
          29 & -7 & -19 & 5 & 11 & 5 & 5 & -7 & -1 & 29 & -16 & -13 & 29 & -7 & -1 & 29 & -16 & -13\\
          -14 & 10 & 10 & 10 & -14 & -2 & 10 & -14 & -2 & -14 & 10 & 4 & -14 & 10 & -2 & -14 & 10 & 4\\
          -6 & -6 & 18 & -6 & 6 & -6 & -6 & 18 & -18 & -6 & 0 & 6 & -6 & -6 & 6 & -6 & 0 & 6\\
          18 & -6 & -6 & -6 & 6 & -6 & 18 & -6 & 6 & 18 & -12 & -6 & 18 & -6 & -18 & 42 & -12 & -6\\
          -9 & 3 & 15 & 15 & -15 & -9 & -9 & 3 & -3 & -9 & 0 & 9 & -9 & 3 & -3 & -33 & 24 & 9\\
          36 & -12 & -60 & 36 & -12 & 12 & 36 & -12 & 12 & 36 & 0 & -36 & 36 & -12 & 12 & 36 & -24 & -12\\
          22 & -2 & 22 & -26 & 22 & -14 & -2 & -2 & -14 & 22 & -14 & 16 & -2 & -2 & -14 & 22 & -14 & -8\\
          -28 & 20 & 20 & 20 & -28 & -4 & -4 & -4 & -4 & -28 & 20 & 8 & -4 & -4 & -4 & -28 & 20 & 8\\
          24 & -24 & -24 & 24 & 0 & 0 & 24 & 0 & 0 & 24 & -12 & -12 & 24 & 0 & 0 & 24 & -12 & -12\\
          12 & 0 & 0 & -12 & 0 & 0 & 12 & 0 & 0 & 12 & -12 & 0 & 12 & 0 & 0 & 12 & -12 & 0\\
          -8 & 4 & 4 & 16 & -8 & -8 & -8 & 4 & 4 & -8 & 10 & -2 & -8 & 4 & 4 & -8 & 10 & -2\\
          12 & -12 & -12 & 12 & -12 & 12 & 12 & -12 & 12 & 12 & 0 & -12 & 12 & -12 & 12 & 12 & 0 & -12
        \end{array}
      \right)
      $
      }}
  \caption{Kets for the three-dimensional projection of the
    FCHC model, corresponding to the choice of bras in
    Table~\ref{tab:fchc3drows}.  The first four columns are the
    hydrodynamic kets, and the last 14 are the kinetic kets.}
  \label{tab:fchc3dcols}
\end{table}


\begin{thebibliography}{999}
  \parindent=.6em 
\bibitem{bib:fhp} U. Frisch, B. Hasslacher, and Y. Pomeau, {\it
    Phys. Rev. Lett.} {\bf 56} 1505 (1986).
\bibitem{bib:swolf} S. Wolfram, {\it J. Stat. Phys.} {\bf 45}, 471
  (1986).
\bibitem{bib:fchc} U. Frisch et al., {\it Complex Syst.} {\bf 1}, 648
  (1987).
\bibitem{bib:lga} D.H. Rothman and S. Zaleski, {\it Lattice-Gas
    Automata: Simple Models of Complex Hydrodynamics}, (Cambridge
  University Press, 1997).
\bibitem{bib:henon} H\'{e}non, M., {\it Complex Systems} {\bf 1}
  (1987) 763.
\bibitem{bib:me2d3d} See, {\it e.g.}, B.M. Boghosian, P.V. Coveney, and
  A.N.  Emerton, {\em Proc.  R. Soc. Lond. A}, {\bf 452}, 1221 (1996);
  and B.M. Boghosian, P.V. Coveney and P.J. Love, ``Three-dimensional
  lattice-gas model for amphiphilic fluid dynamics'', {\it Proc. R. Soc.
    London A}, in press (1999).
\bibitem{bib:higueraa} F. Higuera and J. Jimenez, {\it Europhys. Lett.}
  {\bf 9} (1989) 663.
\bibitem{bib:higuerab} F. Higuera, S. Succi and R. Benzi,,
  {\it Europhys. Lett.} {\bf 9} (1989) 345.
\bibitem{bib:mcn} G. McNamara and G. Zanetti, {\it Phys. Rev. Lett.}
  {\bf 61} (1989) 2332.
\bibitem{bib:succi} S. Succi, R. Benzi, F. Higuera, {\it Physica D} {\bf
    47} (1991) 219-230.
\bibitem{bib:lbe} R. Benzi, S. Succi, and M.  Vergassola
  {\it Phys. Rep.}, {\bf 222} 145 (1992).
\bibitem{bib:liboff} E. Gross, D. Bhatnager, M. Krook, {\it Phys. Rev.}
  {\bf 94} (1954) 511.
\bibitem{bib:qianbgk} Y.-H. Qian, D. d'Humi\`{e}res, P. Lallemand,
  {\it Europhys. Lett.} {\bf 17} (1992) 479-484.
\bibitem{bib:renda} A. Renda, G. Bella, S. Succi, I.V. Karlin,
  {\it Europhys. Lett.} {\bf 41} (1998) 279-283.
\bibitem{bib:rkpart} B.M. Boghosian, P.V. Coveney, {\it
    Comp. Phys. Comm.} (to appear, 2000).
\bibitem{bib:shanchen} Shan, X., Chen, H., {\it Phys. Rev. E} {\bf 47}
  (1993) 1815-1819.
\bibitem{bib:yeomans} Swift, M.R., Osborn, W.R., Yeomans, J.M., {\it
    Phys. Rev. Lett.} {\bf 75} (1995) 830-833; Swift, M.R., Orlandini,
  S.E., Osborn, W.R., Yeomans, J.M., {\it Phys. Rev. E} {\bf 54} (1996)
  5041-5052.
\bibitem{bib:martys} N. Martys, {\it Int. J. Mod. Phys. C} {\bf 10}
  (1999) 1367-1382.
\bibitem{bib:dhum} D. d'Humi\`{e}res, {\it AIAA Rarefied Gas Dynamics:
    Theory and Applications}, in {\it Progress in Astronautics and
    Aeronautics} {\bf 159} (1992) 450-458.
\bibitem{bib:fm} M. Schechter, {\it Amer. Math. Monthly} {\bf 105}
  (March, 1998) 246-251.
\bibitem{bib:ajw} A.J. Wagner, {\it Europhys. Lett.} {\bf 44} (1998)
  144-149.
\bibitem{bib:karlin} I.V. Karlin, A. Ferrante and H.C. \"{O}ttinger,
  {\it Europhys. Lett.} {\bf 47} (1999) 182-188.
\bibitem{bib:hudong} H. Chen and C. Teixeira, ``H-theorem and Origins of
  Instability in Thermal Lattice Boltzmann Models,'' preprint (1999).
\bibitem{bib:courant} R. Courant, K. Friedrichs, H. Lewy, {\it
    Math. Ann.} {\bf 100} (1928) 32.
\bibitem{bib:cn} J. Crank, P. Nicolson, {\it Proc. Camb. Phil. Soc.}
  {\bf 43} (1947) 50-67.
\bibitem{bib:pr} D.W. Peaceman, H.H. Rachford Jr., {\it
    J. Soc. Ind. Appl. Math.} {\bf 3} (1955) 28.
\bibitem{bib:df} E.C. DuFort, S.P. Frankel, {\it Math. Tables Aids
    Comput.} {\bf 7} (1953) 135-152.
\bibitem{bib:chapensk} See, for example, B.M. Boghosian, ``The
  Chapman-Enskog Method for Lattice Gases,'' in {\it 1989 Lectures in
    Complex Systems}, Santa Fe Institute, E. Jen, ed. (1989).
\bibitem{bib:macnam}  G.R. McNamara, A.L. Garcia, B.J. Alder,
    {\it J. Stat. Phys.} {\bf 87} (1997) 1111-1121.
\bibitem{bib:alex} F.J. Alexander, A.L. Garcia, B.J. Alder,
    {\it Phys. Rev. Lett.} {\bf 74} (1995) 5212-5215.
\bibitem{bib:ice} B.M. Boghosian, P.V. Coveney, {\it Int. J. Mod. Phys.}
  {\bf 9} (1998) 1231-1246.
\bibitem{bib:karlinprl} I. Karlin, {\it Phys. Rev. Lett.} {\bf 81} (1998) 6.
\bibitem{bib:withsauro} B.M. Boghosian, S. Succi, work in progress.
\end{thebibliography}
\end{document}